\documentclass[]{aa}  

\usepackage{graphicx}
\usepackage{natbib}
\usepackage{derivative}
\usepackage{xcolor}
\usepackage{float}
\usepackage{rotating}
\usepackage{booktabs}
\usepackage{mathtools}
\usepackage{placeins}
\usepackage{bm}
\usepackage{longtable}
\usepackage{adjustbox}
\usepackage{supertabular}
\usepackage{comment}
\usepackage{subfigure}
\usepackage{multirow}

\usepackage{txfonts}

\newcommand{\apollinaire}{\texttt{apollinaire} }
\newcommand{\peakbagging}{\texttt{peakbagging} }
\newcommand{\stframework}{\texttt{stellar\_framework} }
\newcommand{\thetak}{\bm{\theta}_k }

\begin{document}

   \title{Deciphering stellar chorus: \texttt{apollinaire}, a \texttt{Python}~3 module for Bayesian peakbagging in helio- and asteroseismology}
   \titlerunning{The \apollinaire package}

   \author{S.N.~Breton\inst{1}
          \and
          R.A.~Garc\'{i}a\inst{1}
          \and 
          J. Ballot\inst{2}
          \and 
          V. Delsanti\inst{3,1}
          \and
          D.~Salabert\inst{4}
            }

    \institute{AIM, CEA, CNRS, Universit\'e Paris-Saclay, Universit\'e de Paris, Sorbonne Paris Cit\'e, F-91191 Gif-sur-Yvette, France
    \\
    \email{sylvain.breton@cea.fr} 
    \and
    IRAP, CNRS, Universit\'{e} de Toulouse, UPS-OMP, CNES 14 avenue Edouard Belin, 31400 Toulouse, France
    \and
    CentraleSup\'{e}lec, 3 Rue Joliot Curie, 91190 Gif-sur-Yvette
    \and
    Universit\'e C\^ote d'Azur, Observatoire de la C\^ote d'Azur, CNRS, Laboratoire Lagrange, France \\
    }

   \date{}

\abstract{Since the asteroseismic revolution, availability of efficient and reliable methods to extract stellar-oscillation mode parameters has been one of the keystone of modern stellar physics. In the helio- and asteroseismology fields, these methods are usually referred as peakbagging. 
We introduce in this paper the \texttt{apollinaire} module, a new \texttt{Python}~3 open-source Markov Chains Monte Carlo  (MCMC) framework dedicated to peakbagging. The theoretical framework necessary to understand MCMC peakbagging methods for disk-integrated helio- and asteroseismic observations are extensively described.
In particular, we present the models that are used to estimate the posterior probability function in a peakbagging framework.
A description of the \texttt{apollinaire} module is then provided. We explain how the module enables stellar background, p-mode global pattern and individual-mode parameters extraction.
By taking into account instrumental specificities, stellar inclination angle, rotational splittings, and asymmetries, the module allows fitting a large variety of p-mode models suited for solar as well as stellar data analysis with different instruments.
After having been validated through a Monte Carlo fitting trial on synthetic data, the module is benchmarked by comparing its outputs with results obtained with other peakbagging codes. An analysis of the PSD of 89 one-year subseries of GOLF observations is performed. Six stars are also selected from the \textit{Kepler} LEGACY sample in order to demonstrate the code abilities on asteroseismic data. The parameters we extract with \texttt{apollinaire} are in good agreement with those presented in the literature and demonstrate the precision and reliability of the module.
}

   \maketitle

\section{Introduction \label{section:apollinaire_peakbagging}}

Models implemented in stellar evolution codes need strong constraints to accurately infer stellar ages \citep{2012A&A...543A..54A,2016MNRAS.456.2183D,2017ApJ...835..172L,2017ApJ...835..173S,2020arXiv201207957T}.
When available, stellar seismic parameters such as pressure-mode (p-mode) frequencies provide excellent constraints on the inputs required for stellar modelling. Reliable tools to infer those parameters from observations are therefore necessary.

The theory behind p modes have been extensively described \citep[e.g.][and references therein]{Unno1989,ChristensenDaalsgardLectureNotes} while solar-like oscillations have been abundantly observed in the Sun through helioseismic instruments such as the Global Oscillations at Low Frequency instrument \citep[GOLF,][]{1995SoPh..162...61G}, the Variability of
solar IRradiance and Gravity Oscillations instrument \citep[VIRGO,][]{1995SoPh..162..101F}, the Solar Oscillations Investigation’s
Michelson Doppler Imager instrument \citep[SOI/MDI,][]{1995SoPh..162..129S}, the Helioseismic Magnetic Imager instrument \citep[HMI,][]{2012SoPh..275..207S}, the Global Oscillations Network Group \citep[GONG,][]{1996Sci...272.1284H}, the  Birmingham Solar Oscillations Network \citep[BiSON,][]{1996SoPh..168....1C} or the solar counterpart of the Stellar Observations Network Group \citep[Solar-SONG,][]{2013JPhCS.440a2051P,2019A&A...623L...9F,2022A&A...658A..27B}, in main-sequence solar-like stars \citep[e.g.][]{2014A&A...566A..20A,2017ApJ...835..172L}, subgiants \citep[e.g.][]{1995AJ....109.1313K}, and red giants, from the red giant branch to the clump \citep[e.g][]{2011Sci...332..205B,2011A&A...532A..86M,2011Natur.471..608B}. After space missions such as the Microvariability and
Oscillations of STars mission \citep[MOST,][]{2000ASPC..203...74M}, the Convection, Rotation and planetary Transit satellite \citep[CoRoT,][]{2009A&A...506..411A} and especially \textit{Kepler}/K2 \citep{2011ApJ...736...19B,2014PASP..126..398H}\footnote{for sake of completeness, the observation of $\alpha$ Ursae Majoris by the Wide-field Infrared Explorer satellite also needs to be mentioned for its precursor role \citep[WIRE,][]{2000ApJ...532L.133B}}, the golden years of solar-like stars asteroseismology are not over with the Transiting Exoplanet Survey
Satellite \citep[TESS,][]{2015JATIS...1a4003R} currently taking place and the launch of the PLAnetary Transits and Oscillations of stars mission \citep[PLATO,][]{2014ExA....38..249R} at the horizon 2026. Ground-based stellar observations provided by networks such as SONG \citep{2007CoAst.150..300G, 2017ApJ...836..142G} also provide asteroseismic data that will benefit from our analysis tools.

Helio- and asteroseismic parameter fitting (usually referred as peakbagging) is a topic which has been extensively discussed in the literature over the last decades. The parameter inference may follow a frequentist approach through the use of Maximum Likelihood Estimators methods \citep[MLE, see e.g.][]{1994A&A...289..649T,1998A&AS..132..107A} or a Bayesian approach, with for example implementation of Maximum A Posteriori algorithms \citep[MAP, see e.g.][]{2009A&A...506....7G}, Monte Carlo Markov Chains methods \citep[MCMC, see e.g.][]{2009A&A...506...15B,2009A&A...506.1043G,2011A&A...527A..56H,2013MNRAS.435..242G,2013MNRAS.432..417G,2015A&A...580A..96D,2016MNRAS.456.2183D,2017ApJ...835..172L,2021AJ....161...62N} or Nested Monte Carlo \citep[see e.g.][]{2014A&A...571A..71C,Corsaro2020}.

The \texttt{apollinaire}\footnote{This paper describes the \texttt{v1.1} of the module, for which additional documentation can be found at: \url{https://apollinaire.readthedocs.io/en/v1.1/}} package aims at providing a ready-to-use, consistent and flexible open-source MCMC peakbagging framework for solar and stellar time series. It has already been used in several publications \citep{Hill2021,2022A&A...658A..27B,Huber2022,Mathur2022,Smith2022}.
The code is fully written in \texttt{Python 3}.
Available solar time series being unequaled in quality and length, their specificities have been fully taken into account during the development phase of the module.
The difficulties of stellar peakbagging have also been deeply considered in order to provide an automated process for p-mode parameter extraction.
Therefore, \apollinaire is able to perform fits globally, by order ($\ell=\{0,1,2,3\}$), by pair ($\ell=\{0,2\}$ or $\{1,3\}$), or for a single mode and to extract parameters such as mode splittings, stellar inclination angles, as well as the choice of symmetric or asymmetric Lorentzian profiles \citep[see e.g.][]{1998ApJ...505L..51N,Korzennik_2005}.
Similar automated frameworks with IDL or \texttt{Python} interface have already been developed in the past few years (see e.g. Fast and AutoMated pEak bagging with \texttt{DIAMONDS} (\texttt{FAMED}) from \citealt{Corsaro2020}, or \texttt{PBJam} from \citealt{2021AJ....161...62N}). With the oncoming of the PLATO mission at the end of the decade, the existence of a wide diversity of peakbagging open source modules can only be an asset for the asteroseismic community, as it will allow non-expert peakbaggers to easily access ready-to-use frameworks for p-mode parameters extraction. 

The layout of the paper is as follows. 
Section~\ref{section:review} presents the principles of parameter fitting in a Bayesian framework and extensively describes the set of models that are implemented in \texttt{apollinaire}.
Section~\ref{section:framework} provides a detailed presentation of how these models are used in the different steps of the \apollinaire framework. Section~\ref{section:benchmark} presents an extended benchmark of the module, with comparison of published results from helioseimic and asteroseismic data. Usual fitting strategies are discussed in Sect.~\ref{section:discussion} while conclusion and perspective for improvement of the current \apollinaire releases are provided in Sect.~\ref{section:conclusion}.

\section{Model spectrum \label{section:review}}

In this paper, we exclusively focus on peakbagging methods for full-disk-integrated time series. The single-sided power spectral density (PSD) is taken as the squared modulus of the Fourier transform of a given time series with the following calibration \citep[e.g.][]{1992nrfa.book.....P,2015EAS....73..193G} verifying the Parseval theorem:
\begin{equation}
    \int_0^{\nu_\mathrm{N}} \mathrm{PSD}(\nu)d\nu = \sigma^2 \;,
\end{equation}
where $\nu_\mathrm{N}$ is the Nyquist frequency of the spectrum and $\sigma$ is the rms value of the temporal signal. 

The goal of the peakbagging process is to extract stellar background parameters and global and individual oscillations mode parameters from the PSD. After having briefly described the principle of mode fitting in a Bayesian framework, the background and p-mode models implemented in \texttt{apollinaire} are presented and extensively described in the following subsections. 

\subsection{Statistics}

The PSD follows a $\chi^2$ distribution with two degrees of freedom \citep{1984PhDT........34W}. 
The likelihood of an ideal spectrum $S$ parametrised by a set $\theta$ of parameters and considered against an observed spectrum $\mathbf{S}_\mathrm{obs}$ at a given set of $k$ frequency bins $\nu_i$ is:

\begin{equation}
    \label{eq:likelihood}
    \mathcal{L} (\mathbf{S}_\mathrm{obs}, \theta) = \prod\limits_{i=1}^{k} \frac{1}{S(\nu_i, \theta)} \exp \left[ - \frac{S_{\mathrm{obs},i}}{S(\nu_i, \theta)} \right]  \;.
\end{equation}

This expression of the likelihood assumes that all frequency bins are
independent, an assumption that is theoretically fulfilled only for uninterrupted, evenly sampled observations. However, in the case of high duty-cycle time series, the independence assumption can be made without introducing any bias in the parameter estimation \citep{Stahn2008,2016MNRAS.456.2183D}. 
The cases of significant gaps will be discussed in Sect.~\ref{sec:obs_window}.


The goal of a Bayesian approach will be to sample the posterior probability $p (\theta | \mathbf{S}_\mathrm{obs})$ defined as:

\begin{equation}
    p (\theta | \mathbf{S}_\mathrm{obs}) = \frac{p (\mathbf{S}_\mathrm{obs} | \theta) p (\theta)}{p (\mathbf{S}_\mathrm{obs})} \;,
\end{equation}
where $p (\mathbf{S}_\mathrm{obs} | \theta)$ is the likelihood $\mathcal{L}$, $p(\theta)$ is the prior probability and $p (\mathbf{S}_\mathrm{obs})$ is a normalisation factor. The prior probability is in fact the heart of the Bayesian approach and contains the information we have before confronting the model to the data. 
In practice, the function that will be sampled is a normalised measure of the posterior distribution, $\mathcal{L} (\mathbf{S}_\mathrm{obs}, \theta) p(\theta)$. The strength of this approach is that it allows the fitter to evaluate the shape of the probability distribution of the model parameters.
Moreover, parameters uncertainties can be extracted directly from it, while the MLE approach only provides a lower bound on the uncertainty, obtained through a Hessian inversion \citep[for more details about the Hessian matrix see][]{1994A&A...289..649T}.   

\subsection{Background model \label{section:background}}

The power distribution in the PSD can be separated between the p-mode contribution and a stellar background \citep[see e.g.][]{2010A&A...511A..46M,2014A&A...570A..41K}. The global spectrum can then be modelled in the following way:

\begin{equation}
    S (\nu) = B(\nu) + P (\nu) \:,
\end{equation}
with $B$ the background contribution and $P$ the p-mode contribution. 

The background is dominated at high frequency by a photon noise term $P_n$. At low frequency, the effects of stellar activity and surface rotation are visible together with long-period instrumental variations \citep[see e.g.][]{2015EAS....73..193G}. 
In presence of stellar activity alone, these low-frequency regions can be modelled through a power law \citep{2010A&A...511A..46M}. However, it is difficult to define a general functional profile both for rotational modulations and instrumental effects and are not taken into account in \texttt{apollinaire}. Therefore, when power excesses due to such effects are identified in the PSD, the minimal frequency chosen for the background analysis should be set sufficiently high to avoid that their contribution biases the fitted profile. In the case of \textit{Kepler}, the typical period of the instrumental effects is $\sim$40-45 days \citep[see e.g.][]{Santos2019,Breton2021} which correspond to frequency regions below $\sim$0.3 $\mu$Hz. The location of the rotational modulations in the PSD depends of course on the surface rotation period of the considered stars, but also of the power distribution between the different harmonics of the signal. The fastest main-sequence solar-like stars with detected surface rotation still exhibit power contribution at a few tenths of $\mu$Hz. Nevertheless, most of the stars are rotating slow enough for their rotational power contribution to influence only the shape of the PSD below a few $\mu$Hz. Concerning the \textit{Kepler} nominal survey, stars exhibiting photometric rotational modulations can be identified with the \citet{Santos2019,Santos2021} reference catalogues.          

Surface convection is the main process shaping the profile of the remaining components of the background. \citet{1985ESASP.235..199H} suggested that these trends could be described through empirical laws (referred as Harvey models or sometimes super-lorentzians, a nomenclature discussion on the term has been provided by \citealt{2014A&A...570A..41K}) of the following form:

\begin{equation}
    \mathcal{H} (\nu) = \frac{A}{1 + \left(\frac{\nu}{\nu_c}\right)^\gamma} \;,
\end{equation}
with A the reduced amplitude component, $\nu_c$ the characteristic frequency and $\gamma$ an exponent which can be linked to the amount of memory in the physical process described by the Harvey model \citep[see e.g.][]{2019LRSP...16....4G}. 


There is no clear consensus in the community on the best number of Harvey models to consider in order to optimally fit the background. \citet{2010A&A...511A..46M} combined one Harvey model and a power law in their model, while \citet{2014A&A...570A..41K} decided to fit the data with two Harvey models after using a Bayesian framework to compare six possible models.

For the sake of generality, the considered limit background is then taken as the sum of $k$ Harvey models, a power law, and a white noise term:
\begin{equation} 
    B (\nu) = \sum\limits_{k} \mathcal{H}_k (\nu) + a\nu^{-b} + P_n \;,
    \label{eq:background}
\end{equation}
where $a$ and $b$ are the power law parameters. 
In the cases of long cadence observations from \textit{Kepler} or TESS, it can be necessary to include a damping factor $\eta$ to take into account the power loss in component of the signals close to the Nyquist frequency \citep{Chaplin2011,2014A&A...570A..41K}
\begin{equation}
    \eta^2 (\nu) = \mathrm{sinc}^2 \left( \frac{\pi \nu}{2 \nu_N} \right) \; . 
\end{equation}
The photon noise is not affected by this effect and Eq.~\ref{eq:background} therefore becomes 
\begin{equation} 
    B (\nu) = \left[ \sum\limits_{k} \mathcal{H}_k (\nu) + a\nu^{-b} \right] \eta^2 + P_n \; .
    \label{eq:background_apodisation}
\end{equation}


\subsection{The Lorentzian model}

In asteroseismology, p-mode profiles are usually described with symmetric Lorentzian profiles. However, helioseismic observations have presented evidence that the p-mode spectral profile was actually asymmetric \citep{1993ApJ...410..829D,1998ApJ...506L.147T}. 
Two solutions were suggested to model p-mode asymmetric Lorentzian profiles. The first one was proposed by \citet{1998ApJ...505L..51N}: 

\begin{equation}
    L (\nu, \nu_0, \Gamma, H, \alpha) = \frac{H}{1 + x^2} \bigg[(1 + \alpha x)^2 + \alpha^2\bigg] \;,
\end{equation}
while the second possibility has been given by \citet{Korzennik_2005}:

\begin{equation}
    L (\nu, \nu_0, \Gamma, H, \alpha) = \frac{H}{1 + x^2} \bigg[1 + \alpha (x - \alpha/2)\bigg] \;.
\end{equation}
In the two previous equations, $H$ is the height of the Lorentzian, $\alpha$ is the asymmetry parameter, and $x$ is the reduced frequency, defined as follows:
\begin{equation}
    x (\nu, \nu_0, \Gamma) = \frac{\nu - \nu_0}{\Gamma/2} \;,
\end{equation}
with $\nu$ the frequency, $\nu_0$ the Lorentzian central frequency, and $\Gamma$ the Lorentzian full width at half maximum (FWHM). If $\alpha=0$, the modelled profile is a standard symmetric Lorentzian. 

It is finally important to note that $H$ and $\Gamma$ can be related to the mode amplitude $A$ through \citep{2006MNRAS.371..935F,2008AN....329..549C,2017ApJ...835..172L}:
\begin{equation}
    H = \frac{2A^2}{\pi \Gamma} \; .
\end{equation}


\subsection{Low resolution and sinc model}

\begin{figure}[ht!]
    \centering
    \includegraphics[width=.49\textwidth]{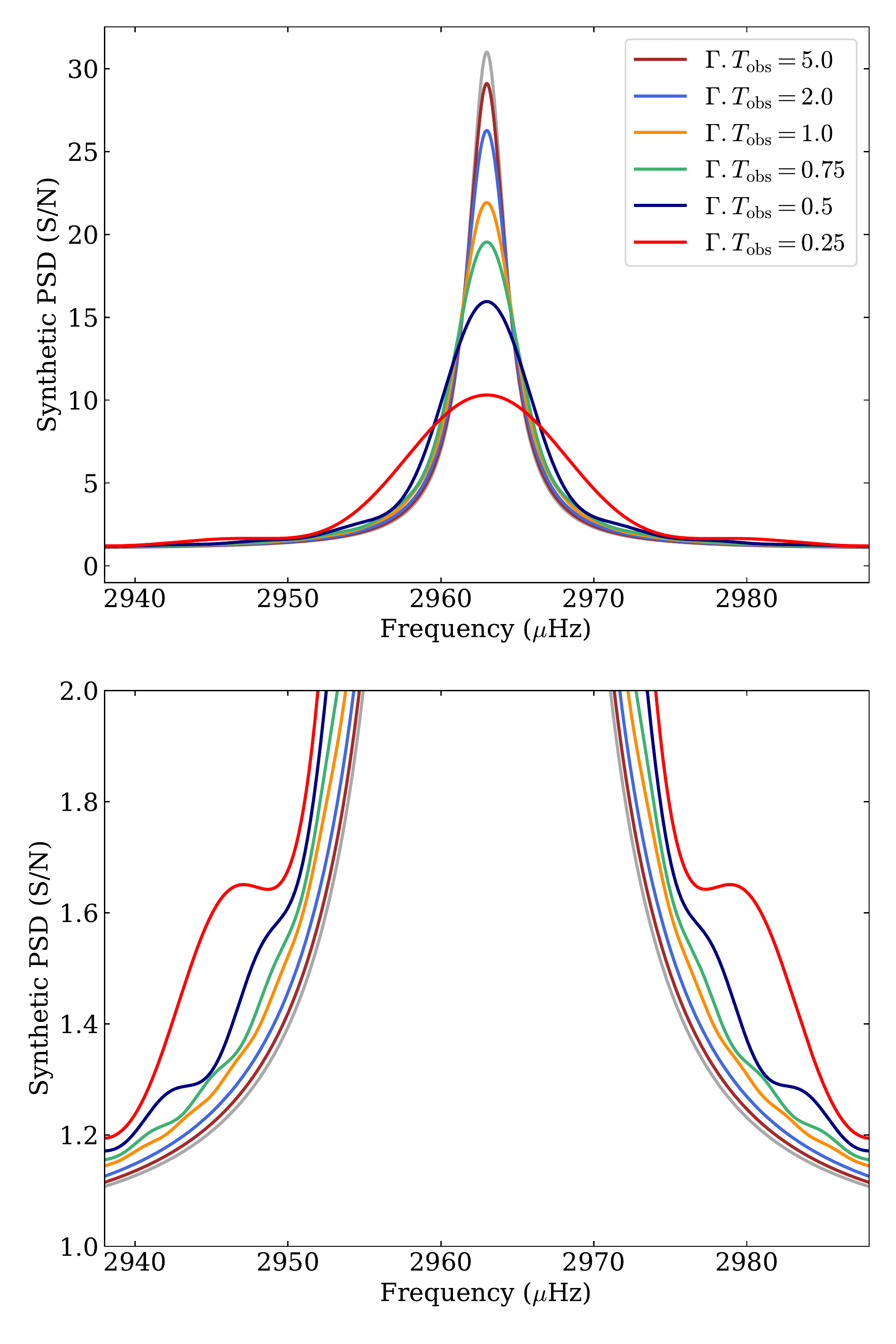}
    \caption{\textit{Top:}effects of the convolution by the Fourier transform of the observational window on an ideal Lorentzian mode profile (grey) in case of continuous observations, with $\Gamma . T_\mathrm{obs}$ = 0.25 (red), 0.5 (dark blue), 0.75 (green), 1 (orange), 2 (light blue), 5 (brown).
    \textit{Bottom:} same as top panel, but with for the y-axis range [1, 2].}
    \label{fig:convol_length}
\end{figure}

In the PSD, the ideal mode profile is in reality convolved by the Fourier transform of the observational window. In case of continuous observations of length $T_\mathrm{obs}$, this Fourier transform has the shape of a sinc function. There is no significant bias in the observed height $H$ (or amplitude $A$) and width $\Gamma$ of the mode in the case $T_\mathrm{obs} \gg 1 / \Gamma$. However, as illustrated in Fig.~\ref{fig:convol_length} when we represent the result of the convolution of the ideal signal by the Fourier transform of the observational window for different $\Gamma . T_\mathrm{obs}$ values, the effect of the convolution appears as soon as $\Gamma . T_\mathrm{obs}$ is of order of several time of unity, and the sinc function profile dominates the Lorentzian profile when $\Gamma . T_\mathrm{obs} < 1$.      
Therefore, if the considered modes have a long lifetime with regards to the duration of observation, the Lorentzian nature of their profile may not appear clearly because there is not enough resolution (frequency bins) to properly characterise the profile. 
In this case, it is more adequate to model the p-mode peaks with squared sinc functions instead of Lorentzians:

\begin{equation}
    L (\nu, \nu_0, \Gamma, H) = H \; \mathrm{sinc}^2 x \;.
\end{equation}

\subsection{Mode description}

Under the effect of slow rotation \citep{1951ApJ...114..373L}, an acoustic mode $M$ of order $n$ and given degree $\ell$ is described as a multiplet of $2\ell+1$ components (modelled with symmetric Lorentzian profiles, asymmetric Lorentzian profiles, or sinc profiles as explained above) according to the following equation:
\begin{equation} \label{eq:mheight-ratio}
M_{n,\ell} (\nu) = \sum\limits_{m=-\ell}^{\ell} L (\nu, \nu_{n,\ell}+ms_{n,\ell}, \Gamma_{n,\ell}, r_{\ell,m} H_{n,\ell}, \alpha)  \;, 
\end{equation}
with $s_{n,\ell}$ being the mode splittings, and $r_{\ell,m}$ the m-height ratio (with $\sum\limits_{m=-\ell}^{m=\ell} r_{\ell,m} = 1$). 
The $r_{\ell,m}$ term is in principle purely geometric and it depends on the stellar inclination angle $i$ \citep[][]{1977AcA....27..203D,1993A&A...268..309T,2003ApJ...589.1009G,2006MNRAS.369.1281B}. 
Typical $r_{\ell,m}$ values considered for different instruments are given in Table~\ref{tab:mheight_ratio}.

\subsection{Mode visibilities \label{section:visibilities_ratios}}

It is often useful to be able to define the mode-height ratio between close-frequency modes. This can be achieved by computing the mode visibility, $V_\ell$, which depends on the stellar limb darkening and links the mode surface amplitude to the disk-integrated mode amplitude.

\begin{equation} \label{eq:mode_visibility}
   V_\ell = \sqrt{(2\ell+1) \pi} \int_0^1 P_\ell (\mu) w (\mu) \, \mu d\mu \;,
\end{equation}
where $P_\ell$ is the $\ell$-th order Legendre polnynomial and $w$ a weighting function. 
Assuming energy equipartition at close frequencies, the height ratio can be computed as the visibility ratio $V_\ell/V_0$.
For some instruments, the dependence is more complex as the instrument response does not depend only on $\mu$. Hence, the relations given by Eq.~\ref{eq:mode_visibility} cannot be directly used \citep[for an example with GOLF, see][]{1999A&A...346..626G,2011A&A...528A..25S}. The values for $V_\ell/V_0$ used in \texttt{apollinaire} are given in Table~\ref{tab:amplitude_ratio}.

\subsection{Modelling the modes for time series with observational gaps \label{sec:obs_window}}

The \texttt{apollinaire} package has been designed to deal with timeseries with large temporal gaps and was used for this purpose in \citet{2022A&A...658A..27B}. The effect of the presence of gaps in the observations is the convolution of the Fourier transform of the ideal time series by the Fourier transform of the observational window. One of the consequences of this convolution is that the hypothesis on the frequency-bin independence is no longer valid \citep[e.g.][]{1994A&A...287..685G}. However, the form of the likelihood that takes this effect into account is computationally challenging and not well suited for an optimised implementation.

The historically considered solution is to ignore the independence loss \citep{1998A&AS..132..107A} and to consider the likelihood given in Eq.~\ref{eq:likelihood}. 
To go further, the model can be corrected to take into account the side lobes generated by the window convolution in the PSD \citep[e.g.][]{2002A&A...390..717S,2004A&A...413.1135S}. We use such an approach in \texttt{apollinaire} to modify the model when fitting PSD from time series with low duty-cycle.
In order to approximate the power redistribution of one peak in the model, we define the observation window $f$ of the time series. The value of $f$ is 1 at the timestamps where data were acquired, 0 otherwise.
The Fourier transform $\tilde{f}$ of $f$ is then computed. The frequency shift $\nu_i$ (relative to the zero-frequency peak of $|\tilde{f}|^2$) and the amplitudes $a_i$ of the $k$ peaks above a given threshold in $|\tilde{f}|^2$ are then stored. The amplitudes are normalised to verify
\begin{equation}
    \sum_{i=0}^k a_i = 1 \;.
\end{equation}
The mode description given in Eq.~\ref{eq:mheight-ratio} is then replaced by:

\begin{equation} \label{eq:mode_with_window}
M_{n,\ell} (\nu) = \sum\limits_{m=-\ell}^{\ell} \sum\limits_{i=0}^{k} L (\nu, \nu_{n,\ell}+ms_{n,\ell} + \nu_i, \Gamma_{n,\ell}, a_i r_{\ell,m} H_{n,\ell}, \alpha)  \;. 
\end{equation}

\begin{figure}[ht!]
    \centering
    \includegraphics[width=.49\textwidth]{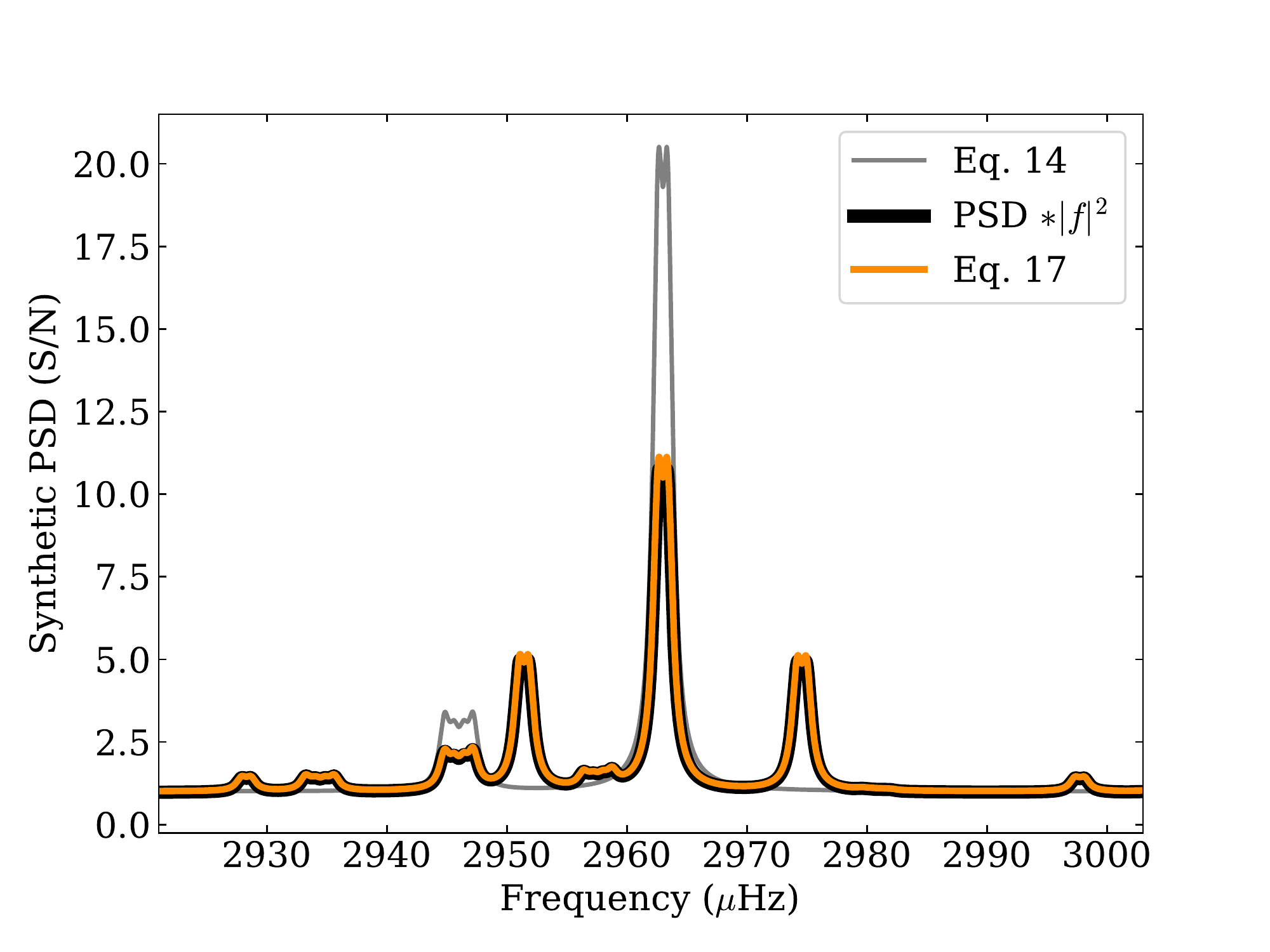}
    \caption{Model for a $\ell=1,3$ pair with inclination angle $i = 90^o$ and splittings $s = 0.4$ $\mu$Hz, without taking the observational window into account (grey) and modified model with the observational window effect applied following Eq.~\ref{eq:mode_with_window} (orange). The actual convolution of the synthetic PSD by $|f|^2$ is represented in black for comparison.}
    \label{fig:window_effect}
\end{figure}

The side lobes power redistribution of the mode is illustrated in Fig.~\ref{fig:window_effect}, where we consider an observational window with a regular observation cycle of 720 minutes with observations followed by 720 minutes without observations. We consider a $\ell=1,3$ pair with $i = 90^o$ and $\nu_s = 0.4$ $\mu$Hz and we represent the model with 100\% duty cycle for comparison. The power redistribution of the modes appears clearly when the formula from Eq.~\ref{eq:mode_with_window} is applied to compute the mode profiles, and we see that the modified model, approximating $|\tilde{f}|^2$ as the sum of $k+1$ Dirac functions, is extremely close to the profile we obtain when we actually perform the convolution operation between the PSD and the $|f|^2$. 

\section{Description of the framework \label{section:framework}}

\begin{figure}[ht!]
    \centering
    \includegraphics[width=.48\textwidth]{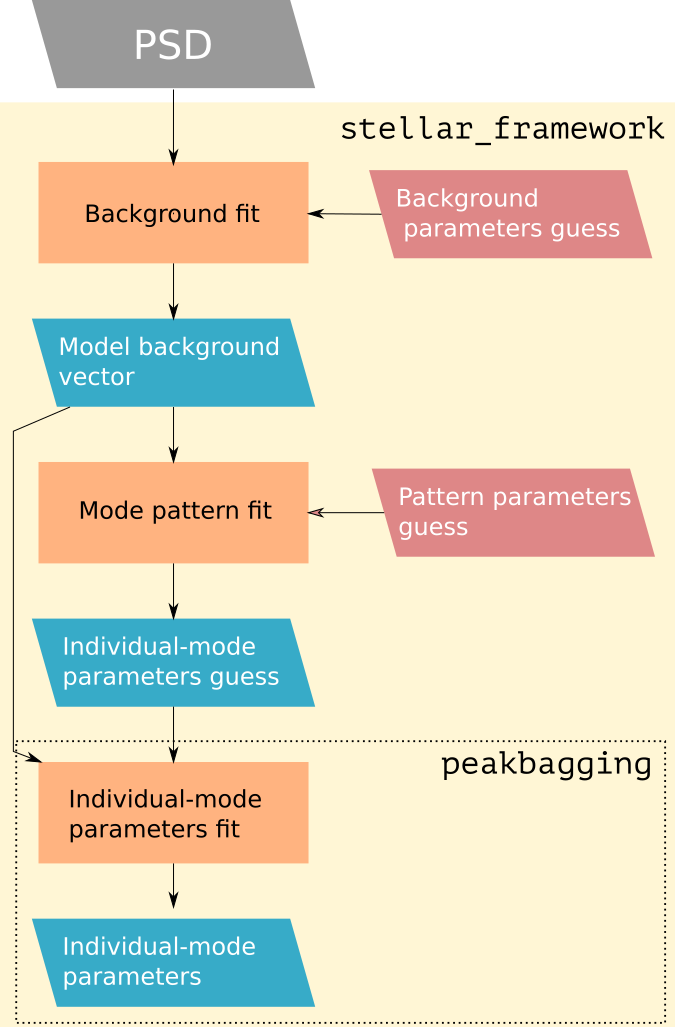}
    \caption{Simplified representation of the \apollinaire flow chart.}
    \label{fig:apollinaire_flowchart}
\end{figure}

The \apollinaire package has been designed to perform MCMC samplings for each step of the peakbagging procedure: background fit, asymptotic parameters fit to determine the global modes pattern, and extraction of individual mode parameters. 
Those three operations are performed one after another when using the \stframework function but can also be performed independently. The global flowchart of an analysis performed with the \stframework function is represented in Fig.~\ref{fig:apollinaire_flowchart}. It should be noted that both the background and the global pattern fits relies on two steps: a MLE algorithm is applied  with rough guesses in order to find good values to initialise the MCMC sampling, the MCMC sampling itself is then performed. The user can also override the automatically computed guesses for theses two steps. 

\subsection{Sampling the posterior probability with MCMC \label{section:sampling_MCMC}}

The MCMC are implemented with the Python package \texttt{emcee} \citep{2013PASP..125..306F}. The sampling strategy follows \citet{2010CAMCS...5...65G}. The sampler, which can be seen as an improvement of the single site Metropolis scheme \citep{Sokal1997,2009_book_liu}, is designed as an affine invariant ensemble of walkers. Walkers positions are updated one after another by the algorithm. The proposal for the new position of a given walker is created by taking into account the positions of other walkers. Acceptance or rejection of each move is assessed through the Metropolis-Hastings rule \citep{1953JChPh..21.1087M,1970Bimka..57...97H}. In every part of the framework (background, global mode pattern, individual mode parameters), the user is free to choose the number of walkers and the number of steps to perform in order to sample the distribution, as well as the number of steps discarded as part of the burn-in phase. 

The results yielded by a Bayesian approach strongly rely on the choice of prior functions made for the different parameters. At the same time, the prior function reflects the level of knowledge that the Bayesian fitters already possesses concerning the model for which they want to sample the posterior distribution. From this point of view, the choice of the prior functions to use in order to sample the posterior has necessarily something arbitrary. Discussing considerations over the expected mode profile, several possibilities have been suggested in the Bayesian peakbagging literature.
Following \citet{Benomar2012}, \citet{2016MNRAS.456.2183D} considered for example a smoothness condition for the mode frequencies while using uniform priors for every other parameters. On the contrary, 
\citet{2017ApJ...835..172L} considered uniform priors for frequencies and inclination angle while using a modified Jeffrey's priors \citep{2011A&A...527A..56H} for amplitudes and widths. \citet{Corsaro2020} adopted uniform priors for all free parameters, justifying this choice by the fast computation that it allowed in their framework. 

In \texttt{apollinaire}, prior functions are taken to be uniform distributions within two bounds with the exception of the inclination angle, $i$, the heights $H$ (or amplitudes $A$), and the widths $\Gamma$.  
It should be underlined that the uniform priors confine the sampled distribution to the compact support defined by the bounds and represent in this sense a strong constraint on the posterior. However, this can be justified by the fact that it is for example reasonable to suppose that, for a given mode, the actual mode frequency cannot be located outside a small frequency window around the mode power excess.    

Most of the usual priors defined on a compact support can actually be linked to a uniform prior through a change of variable. For $H$ (or $A$) and $\Gamma$, we therefore consider a uniform prior distributions in $\log H$ (or $\log A$) and $\log \Gamma$. This is equivalent with constraining $H$ (or $A$) and $\Gamma$ with Jeffrey's priors. This way, in the limit of the fixed boundaries, the prior does not contain information on the parameter scaling. For $i$, the prior distribution is $p(i) = \sin i$ in order to fulfill the \textit{a priori} isotropic distribution of stellar rotation axes \citep{2019LRSP...16....4G}.

In the \texttt{stellar\_framework} function, the automatically generated priors have been set in order to cover a range wide enough in order to ensure that the sampled distribution is not biased by boundary effects. Additional difficulties that may arise in the specific case of frequencies will be discussed in Sect.~\ref{section:pattern}.       

When a MCMC is sampled, the code automatically extracts summary statistics from it. For a given parameter, it returns the median of the marginalised sampled distribution, $y$. 16th and 84th centiles $y_{16}$ and $y_{84}$ are also extracted to obtain the uncertainties $\sigma_-$ and $\sigma_+$ over $y$. In case of a Gaussian distribution $\sigma_- = \sigma_+ = \sigma$ with $\sigma$ the standard deviation of the distribution. Some \apollinaire output files (see Appendix~\ref{section:format} and online documentation for more details) only provide a $\sigma_\mathrm{sym}$ over $y$:   
\begin{equation}
    \sigma_\mathrm{sym} = \max \, (y - y_{16}, \, y_{84} - y) \; .
    \label{eq:sigma}
\end{equation}
In this case, if it is the natural logarithm of the parameters that has been sampled, median, 16th and 84th centiles are transformed again before computing $\sigma_\mathrm{sym}$ and being returned. 

\subsection{Inputs and outputs}

The guess and priors for background and global pattern fits are automatically generated by \texttt{apollinaire}. The user can also manually provides guess and priors if needed. 
The inputs for individual mode-parameter extractions is more complex and \apollinaire uses for them text files with a specific syntax: the a2z files, which were originally developed as part of the A2Z pipeline \citep{2010A&A...511A..46M}. 
The syntax of the file is dedicated to provide a simple and straightforward way to specify nature, extent, initial value and bounds for each parameter to fit.
These files will be auto-generated by the \texttt{stellar\_framework} function but can also be manually created in order to directly use the \texttt{peakbagging} function.
They can be read as a \texttt{pandas}\footnote{\url{https://pandas.pydata.org/}} DataFrame through the auxiliary function \texttt{read\_a2z}.

The chains are stored as Hierarchical Data Format version 5 (hdf5) files. 
The code provides functions to read these files for the user that would need to perform a more thorough analysis on the chains than the extraction of summary statistics described in Sect.~\ref{section:sampling_MCMC}. 
Using these files, it is for example straightforward to obtain the marginalised distribution for each parameter or the covariance matrix between two parameters.
Corner plots visually summarising the sampled distributions can be saved as pdf or png files if filenames are specified. They represent both the marginalised distribution for each parameter and the covariance distribution for each pair of parameters.

Parameters fitted by the \texttt{perform\_mle\_background}, \texttt{perform\_mle\_pattern}, \texttt{explore\_distribution\_background} and \texttt{explore\_distribution\_pattern} functions are returned as \texttt{numpy} arrays. 
When these functions are called by the \texttt{stellar\_framework} functions, the results they provide are stored in text files with dedicated headers. For convenience, the \texttt{peakbagging} function returns both an a2z DataFrame which can be saved to an a2z file with the auxiliary function \texttt{save\_a2z} and a so-called pkb\footnote{originally developed inside the Kepler Asteroseismic Science Operations Center (KASOC) working package 6} array which can be saved with the \texttt{save\_a2z} (when the function is called by \texttt{stellar\_framework}, this is automatically done). The returned a2z DataFrame is useful to perform new MCMC samplings with modified input values while the pkb array provides a mode-by-mode summary statistics and allow to simply reconstruct the best-fit p-mode model computed by \texttt{apollinaire}.

More details about a2z files, a2z DataFrame, pkb files and pkb arrays can be found in Appendix~\ref{section:format}. Examples for a2z and pkb files are also provided.

\subsection{Background fit}

Besides the PSD, the only additional inputs needed by \apollinaire to automatically compute initial guesses and priors for the background fit are the stellar effective temperature, $T_\mathrm{eff}$, the asymptotic large spacing $\Delta\nu$ and the frequency at maximum power $\nu_\mathrm{max}$.
If no previous estimation of $\Delta\nu$ and $\nu_\mathrm{max}$ are available, it is possible to provide the code with the stellar mass, $M$, and radius, $R$, in order to compute an estimate of $\Delta\nu$ and $\nu_\mathrm{max}$ through the scaling laws \citep{1995A&A...293...87K}.
\begin{equation}
\begin{aligned}
    &\Delta \nu \approx \Delta \nu_\odot \left( \frac{M}{\mathrm{M}_\odot} \right)^{1/2} \left( \frac{R}{\mathrm{R}_\odot} \right)^{3/2} \, \\
    &\nu_\mathrm{max} \approx \left( \frac{\mathrm{R}_\odot}{R} \right)^{2} \frac{M}{\mathrm{M}_\odot} \left( \frac{\mathrm{T}_{\mathrm{eff}, \odot}}{T_\mathrm{eff}} \right)^{1/2} \,
\end{aligned}
\end{equation}
with $\rm \Delta \nu_\odot$, $\rm M_\odot$, $\rm R_\odot$, $\rm T_{\mathrm{eff}, \odot}$ the reference solar values for the asymptotic large spacing, mass, radius, and effective temperature, respectively. These values are set to 135 $\mu$Hz, 1 $\rm M_\odot$, 1 $\rm R_\odot$ and 5770 K in \apollinaire. 

The limit spectrum fitted on the data is the sum of the background term given by Eq.~\ref{eq:background} and a Gaussian p-mode envelope term:
\begin{equation}
    S_\mathrm{B} (\nu) = B(\nu) + H_\mathrm{max} \exp \left[ - \left( \frac{\nu - \nu_\mathrm{max}}{W_\mathrm{gauss}}\right)^2 \right] \; ,
\end{equation}
with $H_\mathrm{max}$ being the maximal height of the p-mode Gaussian envelope and $W_\mathrm{gauss}$ its width or standard deviation. 

When only one Harvey model is considered, the initial $\gamma$ value is fixed to 2, to 4 when two Harvey models are considered. 
This parameter can be fixed \citep{2014A&A...570A..41K} or set to vary. In order to fit more than two Harvey models, the user should manually provide the guesses. Otherwise, input profile guesses are automatically generated. If the spectrum has been acquired with photometric observations, the values provided by Table~2 of \citet{2014A&A...570A..41K} are used. If the code has to deal with a PSD obtained from a solar radial-velocity time series, initial guess values have also been implemented using GOLF as a reference. 

In order to obtain a significant gain of computing time, it is possible to resample the PSD to a version with much less data points. The resampling techniques are described in Appendix~\ref{appendix:background}. 
As it will be shown in Sect.~\ref{section:benchmark_legacy}, this method can be reliably used in order to extract mode frequencies but the main caveat is that the background parameters obtained by fitting the resampled PSD should be considered with caution. In particular, the resampling will filter out the power in the p-mode region, and $H_\mathrm{max}$, $\nu_\mathrm{max}$ and $W_\mathrm{env}$ obtained with this method will be significantly biased.


\subsection{Limit p-mode spectrum}

To fit the p modes, \texttt{apollinaire} divides the observed spectrum by the fitted background profile $B$ in order to get the signal-to-noise (S/N) spectrum:
\begin{equation}
    \mathbf{S_{\mathrm{obs,SN}}} = \frac{\mathbf{S}_\mathrm{obs}}{B} (\nu) \;.
\end{equation}

This way, the background contribution fitted at the previous step is removed.
The limit spectrum, $S_\mathrm{SN}$, that is then adjusted to $\mathbf{S_{obs,\mathrm{SN}}}$ is the following:
\begin{equation} \label{eq:limit_spectrum_sn}
 S_\mathrm{SN} (\nu) = \sum\limits_{n} \sum\limits_{\ell} \frac{M_{n,\ell}}{B} (\nu) + b \;, 
\end{equation}
with $M_{n,\ell}$ given by Eq.~\ref{eq:mheight-ratio} and $b$ corresponding to an additive factor to locally adjust the background. 

\subsection{Pattern fit \label{section:pattern}}

In order to constrain the priors of the mode main individual parameters, $\nu_{n,\ell}$, $H_{n,\ell}$, and $\Gamma_{n,\ell}$, \texttt{apollinaire} performs a global pattern fit on the orders located around $\nu_\mathrm{max}$. \citet{1980ApJS...43..469T} presented the following asymptotic relation for mode frequencies, within the approximation $n \gg \ell$:
\begin{equation}
    \nu_{n,\ell} \approx \left(n + \frac{\ell}{2} + \epsilon \right) \Delta \nu \;,
\end{equation}
where $\epsilon$ is a phase constant. 
Slightly modifying the formalism adopted in \citet{2017ApJ...835..172L}, this relation can be approximated to, in the $\nu_\mathrm{max}$ neighbourhood:
\begin{equation}
    \nu_{n,\ell} \approx \left(n + \frac{\ell}{2}  + \epsilon \right) \Delta \nu - \delta \nu_{0\ell} - \beta_{0\ell} (n - n_\mathrm{max}) + \frac{\alpha}{2} (n - n_\mathrm{max})^2 \; ,
    \label{eq:tassoul_2nd}
\end{equation}
where the small separations $\delta\nu_{0\ell}$ are given by: 
\begin{equation}
\begin{split}
    &\delta\nu_{00} = 0 \; , \\ 
    &\delta\nu_{01} = \left<\frac{1}{2}(\nu_{n,1} - \nu_{n+1,0}) - \nu_{n,1}\right>_n \; , \\ 
    &\delta\nu_{02} = \left<\nu_{n,0} - \nu_{n-1,2}\right>_n \; , \\
    &\delta\nu_{03} = \left<\frac{1}{2}(\nu_{n,3} - \nu_{n+1,0}) - \nu_{n,3}\right>_n \; ,
\end{split}
\end{equation}
while $\alpha$ and $\beta_{0\ell}$ are respectively the curvature terms on $\Delta\nu$ and $\delta\nu_{0\ell}$. The parameter $n_\mathrm{max}$, which is not an integer, follows the relation:
\begin{equation}
    n_\mathrm{max} = \frac{\nu_\mathrm{max}}  {\Delta\nu} - \epsilon \; .
\end{equation}

Mode heights in the limit spectrum can be approximated through the p-mode envelope parameters $H_\mathrm{max}$ and $W_\mathrm{gauss}$, considering:
\begin{equation}
H_{n,\ell} = \frac{H_\mathrm{max}}{B (\nu_{n,\ell})} \exp \left[ - \left( \frac{\nu_{n,\ell} - \nu_\mathrm{max}}{W_\mathrm{gauss}}\right)^2 \right] \; ,
\label{eq:height_glob}
\end{equation}
while $\Gamma$ is taken as a FWHM value common to all modes. 

Using Eq.~\ref{eq:tassoul_2nd} and \ref{eq:height_glob}, the pattern fit step aims at approximating the mode pattern around $\nu_\mathrm{max}$ with a given set $\theta$ of global parameters: $\epsilon$, $\alpha$, $\Delta\nu$, $\nu_\mathrm{max}$, $H_\mathrm{max}$, $W_\mathrm{gauss}$, $\Gamma$, $\delta\nu_{02}$, $\beta_{02}$, $\delta\nu_{01}$, $\beta_{01}$, $\delta\nu_{13}$ (with $\delta\nu_{13} = \delta\nu_{03} - \delta\nu_{01}$), $\beta_{03}$. The last four parameters can be ignored, for example if the star to fit presents $\ell=1$ mixed modes: only pairs $M_{n-1,2}$, $M_{n,0}$ will then be fitted (see Appendix~\ref{appendix:subgiant_fit}). It is also possible to ignore just the $\ell=3$ mode when the signal-to-noise ratio of the considered PSD is not good enough. Indeed, in this situation, the $\delta\nu_{13}$ and $\beta_{03}$ parameters will be difficult to constrain and the sampled distribution will be prior dominated.

Guesses for $\epsilon$, $\alpha$, $H_\mathrm{max}$, $\Gamma$,  $\delta\nu_{02}$, $\delta\nu_{01}$, $\delta\nu_{13}$ can be manually provided, otherwise they will be automatically computed. If a guess is given for a $\delta\nu_{0k}$, the initial value for the corresponding $\beta_{0k}$ will be set to 0. To determine initial automated guesses for parameters, we adopted prescriptions from \citet{2012A&A...537A...9C} for stars with $\Delta\nu < 14$ $\mu$Hz and $\nu_\mathrm{max} < 450$ $\mu$Hz. For main sequence stars, we derived well performing initial values from the results obtained by \citet{2017ApJ...835..172L}. Proper guesses for stars with $450 \; \mu\mathrm{Hz} < \nu_\mathrm{max} < 1000 \; \mu\mathrm{Hz}$ are not implemented in the current version of \texttt{apollinaire} but a recipe on how to proceed with these stars is given in Appendix~\ref{appendix:subgiant_fit}.

The pattern fit step will fit $\theta$ by considering the $k$ orders closest to $\nu_\mathrm{max}$. By default, the \stframework function uses $k=3$.
The bounds of the fitting window are set $0.2 \, \Delta\nu$ below and above the smallest and largest mode frequency included in the pattern. 
In the standard procedure, the MCMC exploration is directly performed from the initial rough guess computed by \texttt{apollinaire}. 
It is possible to use a fast MLE run to quickly refine this initial guess and to use the values yielded by the MLE as a starting point for the MCMC sampling. This can save some computing time by reducing the number of discarded steps at the beginning of the MCMC exploration, but may also reduce the walkers opportunities to explore different region of the distribution to sample in the case of multimodal distribution.   


As underlined by \citet{2017ApJ...835..172L}, the discrepancies between the actual position of the modes and the frequencies yielded by the relation given by Eq.~\ref{eq:tassoul_2nd} can be physically explained by acoustic glitches \citep[e.g.][]{Mazumdar2014,Houdayer2021}.   
Moreover, when optimised only on the central orders, the formula given by Eq.~\ref{eq:tassoul_2nd} should be extrapolated with caution for modes located far from $\nu_\mathrm{max}$. Indeed, in this case, there can be significant discrepancies between the mode frequencies predicted by the formula and the actual position of the mode in the PSD.
This can be straightforwardly checked using an échelle diagram and it will be illustrated in particular in Sect.~\ref{section:benchmark_legacy}. It is of course possible to include more orders when sampling the pattern parameters distribution, but this will require more computing time, on one hand because the likelihood will be computed on more data points, on the other hand because the chain will take more steps to converge. In this situation, the results obtained with the summary statistics alone should also be considered with caution as the inclusion of low S/N modes can strongly influence the multi-modality of the sampled distribution.      

An échelle diagram may also be particularly useful in order to check that the code correctly identified the $\ell = 0$ and the $\ell = 1$ mode. When this is not the case, the failure in the sampling is generally related to an improper prior for $\epsilon$. In this situation, the solution is to manually impose new initial values and prior bounds for $\epsilon$.    

\subsection{Selecting orders to fit \label{sec:order_selection}}

The result of the pattern fit is used to obtain guess values for $\nu_{n,\ell}$, $H_{n,\ell}$, and $\Gamma_{n,\ell}$ for each mode. These values are provided inside an a2z DataFrame. There are two ways to determine the modes for which the function will provide guess values. With the first method, the function will simply build the a2z DataFrame for a given number of order symmetrically distributed around $\nu_\mathrm{max}$. The second method performs a H0 screening \citep[e.g.][]{2009A&A...506....1A,2010MNRAS.406..767B,2012A&A...543A..54A,2016MNRAS.456.2183D,2017ApJ...835..172L} on the p-mode region to assess the strong ($\ell=0$ or 1) mode detectability. Following the procedure presented in \citet{2016MNRAS.456.2183D}, the PSD is rebinned over $t$ bins and a $\chi^2_{2t}$ statistics is considered. For each mode, the rebinning is performed considering an odd number of bins, the central bin being the one with the frequency closest to the mode frequency estimated with the fitted global parameters. The adopted threshold for the rejection of the null hypothesis H0 is $p=0.001$. The maximal value $t_\mathrm{max}$ considered for the rebinning is 99, or the $t$ value corresponding to a bin width of 5 $\mu$Hz, respectively. If the H0 hypothesis is rejected for more than a third of the considered rebinning, we consider the mode as detectable. A guess for this mode will be added in the a2z DataFrame along with the guess for the closest $\ell=\{2,3\}$ mode. It should be stated that it is also possible to manually provide a a2z DataFrame that will override the guess automatically generated by the function.

\subsection{Individual parameters extraction: the \peakbagging function \label{section:individual_parameter}}

\begin{figure}[ht!]
    \centering
    \includegraphics[width=.48\textwidth]{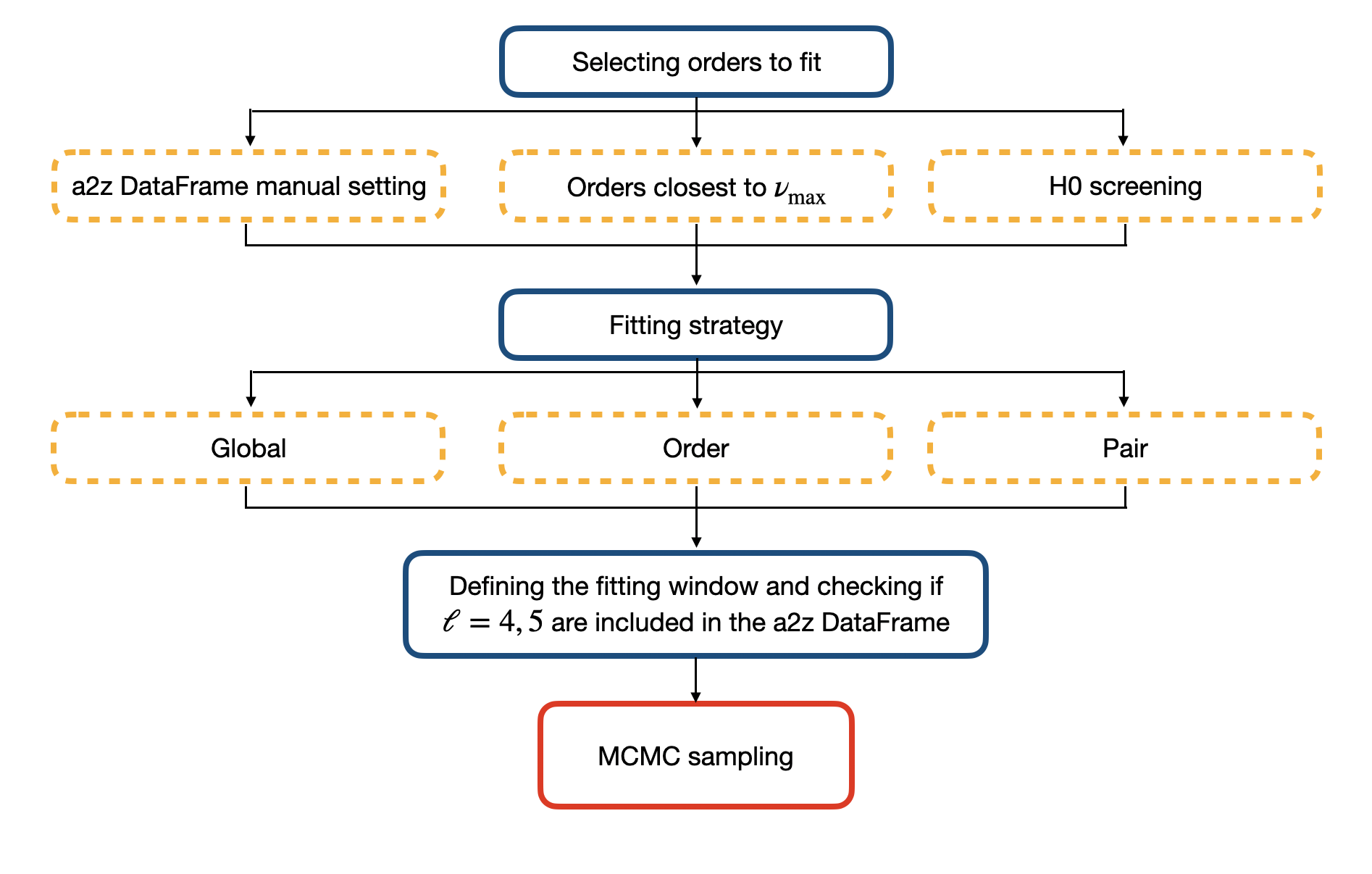}
    \caption{Summary of the available options for the p-mode individual parameter extraction step.}
    \label{fig:peakbagging_diagram}
\end{figure}

In what follows, we refer to the $M_{n-1,2}$, $M_{n,0}$, $M_{n-1,3}$, and $M_{n,1}$ group of modes as a peakbagging order to avoid any confusion. 
Mode parameters can be fitted globally, by peakbagging order, or by pair. In the latter case, modes $M_{n-1,2}$, $M_{n,0}$ or $M_{n-1,3}$ and $M_{n,1}$ are then fitted together, respectively. The reader should note that the pair fitting strategy is also suited to fit any individual mode if the parameters of the pair corresponding mode is not specified in the a2z input.
The model used to compute the likelihood and posterior probability follows Eq~\ref{eq:limit_spectrum_sn}. The initial value for the local signal-to-noise background term $b$ is taken to 1 and set to vary between $10^{-6}$ and 5. 

The minimal set of parameters that will be fitted by the code are mode frequencies, $\nu_{n,\ell}$, heights, $H_{n,\ell}$ (or amplitudes, $A_{n,\ell}$) and FWHM, $\Gamma_{n,\ell}$. Additional parameters that can be fitted are splittings, $s_{n,\ell}$, inclination angle, $i$ and asymmetries, $\alpha_{n,\ell}$. It is possible to fit the projected splittings, $s_\star = s \sin i$ \citep{2006MNRAS.369.1281B,2008A&A...486..867B}, instead of $s$. Except for mode frequencies, every input parameter can be set to parametrise all modes, a whole order, a pair, or only one mode. In other words, it is possible to impose each mode of a given peakbagging order to have, for example, the same FWHM. 

If the user does not provide a fitting window size, the data to fit are restricted to an adaptive fitting window: only PSD elements within $[\nu_-, \nu_+]$ are considered, with:
\begin{equation}
\begin{split}
    \nu_- &= \nu_\mathrm{low} - \frac{\nu_\mathrm{up} - \nu_\mathrm{low}}{d} \; , \\
    \nu_+ &= \nu_\mathrm{up} + \frac{\nu_{\mathrm{up}} - \nu_\mathrm{low}}{d} \; , 
\end{split}
\end{equation}
where $\nu_\mathrm{low}$ and $\nu_\mathrm{up}$ are respectively the minimal and maximal guess frequencies of the modes to fit. The value of $d$ is then determined by the considered method: if the fit is made by order, $d=3$, if the fit is made by pair, $d=1$. 

For modes above 800 $\mu$Hz and if the fit is made by pair, another possibility offered by the code is to use the $\Delta\nu$ value to constrain the fitting window. 
The 800 $\mu$Hz value has been chosen to ensure that the considered mode does not exhibit avoided crossings, which might add additional difficulties due to the potential presence of mixed modes in the fitting window.
In this case, the window bounds are:

\begin{equation}
\begin{split}
    \nu_- &= \nu_\mathrm{center} - w_w \frac{\Delta\nu}{\Delta\nu_\odot} \; , \\
    \nu_+ &= \nu_\mathrm{center} + w_w \frac{\Delta\nu}{\Delta\nu_\odot} \; , 
\end{split}
\end{equation}
where $\nu_\mathrm{center}$ is the centre of the window given by the mean of the guess frequencies of the two modes to fit (or simply the guess frequency of the mode to fit if only parameters for one mode are specified),
and the reference value $\Delta\nu_\odot = 135$~$\mu$Hz. The possible values of $w_w$ are summarised in Table~\ref{tab:ww}.

\begin{table}[ht]
    \centering
    \caption{Possible values of $w_w$.}
    \begin{tabular}{c|ccc}
        \hline \hline
        Interval ($\mu$Hz)  & 1200-2000 & 2000-2500 & > 2500 \\
        \hline
        $w_w$ ($\mu$Hz)  & 35 & 55 & 75 \\
        \hline
    \end{tabular}
    
    \label{tab:ww}
\end{table}

As underlined in section~\ref{section:review}, the mode visibilities and m-height ratios have instrumental dependencies. There are two ways to deal with mode visibilities: on one hand, the user can choose to fit amplitude parameters individually for each mode, on the other hand, it is possible to specify the amplitude ratios $V_\ell/V_0$ in the a2z input file. Usual ratios \citep{2011JPhCS.271a2053S} are reminded in Table~\ref{tab:amplitude_ratio}.

\begin{table}[ht]
    \centering
    \caption{Mode visibilities ratios $V_\ell/V_0$}
    \begin{tabular}{c|ccc}
        \hline \hline
        Ratio  & GOLF & VIRGO & \textit{Kepler}/K2/TESS \\
        \hline
        $V_1/V_0$  & 1.69 & 1.53 & 1.5 \\
        $V_2/V_0$  & 0.81 & 0.59 & 0.7 \\
        $V_3/V_0$  & 0.17 & 0.09 & 0.2 \\
        $V_4/V_0$  & 0.0098 & - & - \\
        $V_5/V_0$  & 0.001 & - & - \\
        \hline
    \end{tabular}
    \label{tab:amplitude_ratio}
\end{table}

The m-height ratios to use are selected through the \textit{instrument} argument of the \texttt{peakbagging} function. The implemented ratios for each instrument are given in Table~\ref{tab:mheight_ratio}.  

\begin{table}[ht]
    \centering
    \caption{m-height ratios $r_{\ell,m}$. The parameter $i$ is the stellar inclination angle.}
    \begin{tabular}{c|ccc}
        \hline \hline
        $r_{\ell,m}$  & GOLF & VIRGO & \textit{Kepler}/K2/TESS \\
        \hline
        $r_{1,0}$  & 0 & 0 & $\cos^2 i$ \\
        $r_{1,\pm 1}$  & 0.5 & 0.5  & $\frac{1}{2}\sin^2 i$  \\
        $r_{2,0}$  & 0.65/2.65 & 0.75/2.75  & $\frac{1}{4} (3 \cos i - 1)^2$  \\
        $r_{2,\pm 1}$  & 0 & 0 & $\frac{3}{8}\sin^2 2i$ \\
        $r_{2,\pm 2}$  & 1/2.65 & 1/2.75 & $\frac{3}{8}\sin^4 i$ \\
        $r_{3,0}$  & 0 & 0 & $\frac{1}{64} (5 \cos 3i + 3 \cos i)^2$ \\
        $r_{3,\pm 1}$  & 0.41/2.82 & 0.63/3.26 & $\frac{3}{64} \sin^2 i (5 \cos 2i + 3)^2$ \\
        $r_{3,\pm 2}$  & 0 & 0  & $\frac{15}{8} \sin^4 i \cos^2 i$ \\
        $r_{3,\pm 3}$  & 1/2.82 & 1/3.26 &  $\frac{5}{16} \sin^6 i$ \\
        $r_{4, 0}$  & 0.1/2.7 & - &  - \\
        $r_{4,\pm 1}$  & 0 & - &  - \\
        $r_{4,\pm 2}$  & 0.3/2.7 & - &  - \\
        $r_{4,\pm 3}$  & 0 & - &  - \\
        $r_{4,\pm 4}$  & 1/2.7 & - &  - \\
        $r_{5, 0}$  & 0 & - &  - \\
        $r_{5,\pm 1}$  & 0.117 & - &  - \\
        $r_{5,\pm 2}$  & 0 & - &  - \\
        $r_{5,\pm 3}$  & 0.137 & - &  - \\
        $r_{5,\pm 4}$  & 0 & - &  - \\
        $r_{5,\pm 4}$  & 0.246 & - &  - \\
        \hline
    \end{tabular}

    \label{tab:mheight_ratio}
\end{table}

It should also be noted that for $H$ and $\Gamma$, it is the natural logarithms of the parameters that are expected to have a Gaussian distribution \citep{1994A&A...289..649T,2019LRSP...16....4G}. For parameters of these types, the \peakbagging function thus samples the natural logarithm. However, it should be remembered that it is straightforward to obtain the sampled distribution of the parameter itself by a simple transformation on the MCMC elements. 

\subsection{Dealing with $\ell=4$ and $\ell=5$ leaks in solar spectra \label{sec:intermediate_degree_leak}}

Due to the very high signal-to-noise ratio of radial-velocity helioseismic data, some $\ell=4$ and $\ell=5$ modes (referred as intermediate-degree modes) may arise above noise. It has been shown that omitting their contribution to the power distribution of the spectrum could introduce a bias in the fitted frequencies up to three times the uncertainty \citep{2008MNRAS.389.1780J}. If guesses for intermediate-degree modes are specified in the input a2z file (for example using theoretical frequency value for those modes), the code will check their presence inside the defined window. 

If, when reading the a2z DataFrame, it appears that an intermediate-degree mode is present in the fitting window, its frequency will be added to the parameters to fit. Heights and FWHMs of intermediate-degree mode are computed considering the ratio presented in Table~\ref{tab:amplitude_ratio} and using the closest $\ell=0$ mode as a reference. If no $\ell=0$ is fitted at this time, the closest $\ell=1$ is used instead. Power is distributed between the m-components of the mode following Table~\ref{tab:mheight_ratio}. The splittings are set to 400 nHz and do not vary. No asymmetry is considered for these modes.

The diagram shown in Fig.~\ref{fig:peakbagging_diagram} summarises the possible options described in Sect.~\ref{sec:order_selection}, \ref{section:individual_parameter} and \ref{sec:intermediate_degree_leak}.

\subsection{Quality assurance \label{section:quality_assurance}}

Several frequentist and Bayesian metrics for peakbagging quality assurance have been suggested over the years \citep[see e.g.][]{2012A&A...543A..54A,2016MNRAS.456.2183D,2017ApJ...835..172L}. The quality assurance metric implemented in \texttt{apollinaire} is inspired from the Bayesian machinery described in \citet{2016MNRAS.456.2183D} but avoids resampling a MCMC to obtain the probability of the different models to compare, which obviously saves a large amount of computing time. 

The \texttt{apollinaire} package implements a Bayesian quality assurance computing tool where three possibilities are assumed for each fitted pair of modes (odd or even):
\begin{enumerate}
    \item The strong ($\ell=0$ or $1$) and the weak ($\ell=2$ or $3$) modes are both detected (model $M_\mathrm{sw}$ with associated probability $p_\mathrm{sw}$).
    \item Only the strong mode is detected (model $M_\mathrm{s}$ with associated probability $p_\mathrm{s}$).
    \item Neither of the two modes is detected (model $M_0$ with associated probability $p_0$).
\end{enumerate}

We have assumed that a weak mode could not be detected if the strong mode was not also detected. Assuming that a sufficient number of initial steps has been discarded, the MCMC contains $n$ sets of parameters $\bm{\theta}$. A subset of $k$ ($k \leq n$) elements still representative of the MCMC distribution can be selected by thinning the chain. 
Indeed, this sub-ensemble of parameters should follow the same distribution that the full MCMC if enough elements are considered.  
For each of those $k$ sets of parameters $\thetak$, likelihoods $p (D|M_\mathrm{sw}, \thetak)$, $p (D|M_\mathrm{s}, \thetak)$, $p (D|M_0, \thetak)$ corresponding to the three hypothesises are then compared (considering the data $D$ within the same frequency window that was used for the actual fit). For the $M_0$ model, as the frequency window is narrow enough, we consider a flat background which is computed as the median of the power distribution of the frequency bins inside the window.  

The estimated probability $p_\alpha$ (marginalised over the parameter distribution) that a model $M_\alpha$ is the most likely considering the data is then computed as follows:
\begin{equation}
    \label{eq:quality_assurance}
    p_\alpha = \frac{\# \;  \{ \thetak  \; | \; p (D|M_\alpha, \thetak) = \max\limits_{i \in \{ \alpha, \beta, \gamma\} } p (D|M_i, \thetak) \}  }{k} \; ,
\end{equation}
where given an ensemble $E$, $\#E$ denotes its cardinal. The indexes $\alpha$, $\beta$, and $\gamma$ in Eq.~\ref{eq:quality_assurance} should be properly replaced by sw, s, and 0 depending on the considered case. We use Eq.~\ref{eq:quality_assurance} to estimate the probability $p_\alpha$ as the fraction of explored parameter sets $\theta_k$ for which the model $\alpha$ is the most likely among the three models. 
If we consider the specific case where $n = k$, Eq.~\ref{eq:quality_assurance} gives the exact proportion of cases in the sampled distribution where the $M_\alpha$ model is the most likely. The probability $p_\alpha$ is therefore the likelihood of $M_\alpha$ for this distribution.
The thinning step to reduce to $k$ samples allows a significant gain in computing time to obtain $p_\alpha$ while conserving the sampled distribution properties.

$p_0$ is the probability of the null hypothesis H0 for the detection of the strong mode while $p_\mathrm{s} + p_0$ is the H0 probability for the weak mode.
For the considered pair, the natural logarithm of the Bayes factor $K$ for the detection of a mode of degree $\ell$ is then given by:
\begin{equation}
\begin{split}
   \ln K = \ln \; (p_\mathrm{sw} + p_\mathrm{s}) - \ln p_0 \quad & ; \quad \ell \in \{0, 1\} \\
   \ln K = \ln p_\mathrm{sw} - \ln \; (p_\mathrm{s} + p_0) \quad & ; \quad \ell \in \{2, 3\} \; ,
\end{split}
\end{equation}

We remind here the interpretation of the $\ln K$ value for the evidence against H0, as outlined by \citet{1995_kass_raftery} :
\begin{equation*}
    \ln K = \left\{ 
    \begin{aligned}
& < 0 \quad & \text{favours H0} \\
& 0 \text{ to } 1 \quad & \text{not worth more than a bare mention}  \\
& 1 \text{ to } 3 \quad & \text{positive} \\
& 3 \text{ to } 5 \quad & \text{strong} \\
& > 5 \quad & \text{very strong} \\
    \end{aligned}
    \right.
\end{equation*}

It is easy to see that if the model $p_\mathrm{sw}$, for example, is favoured in any case, this will correspond to $p_\mathrm{sw} = 1$ and therefore $\ln K > 5$. If we have $p_\mathrm{sw} = 2/3$ and $p_0 = 1/3$, we will have $\ln K \approx 0.69$, the $M_\mathrm{sw}$ model is only barely favoured compared to the H0 hypothesis. Modes with $\ln K < 1$ should be cautiously considered when exploiting the peakbagging results for modelling purposes.

\section{Benchmark with Monte Carlo synthetic spectra and published peakbagging results \label{section:benchmark}}

In this section, in order to assess the reliability of the code, we present the results of a Monte Carlo benchmark with synthetic spectra.  
We then compare \texttt{apollinaire} results with results obtained with other peakbagging codes. As the method has been designed to perform helioseismic and asteroseismic analysis, we decided to perform our benchmark with both solar and stellar data\footnote{Full results, \textit{Kepler} light curves and data analysis tools used to perform the benchmark can be accessed through the following repository: \url{https://gitlab.com/sybreton/benchmark_peakbagging_apollinaire}.}.

\subsection{Monte Carlo trial \label{section:monte_carlo}}

\begin{figure}[ht!]
    \centering
    \includegraphics[width=0.49\textwidth]{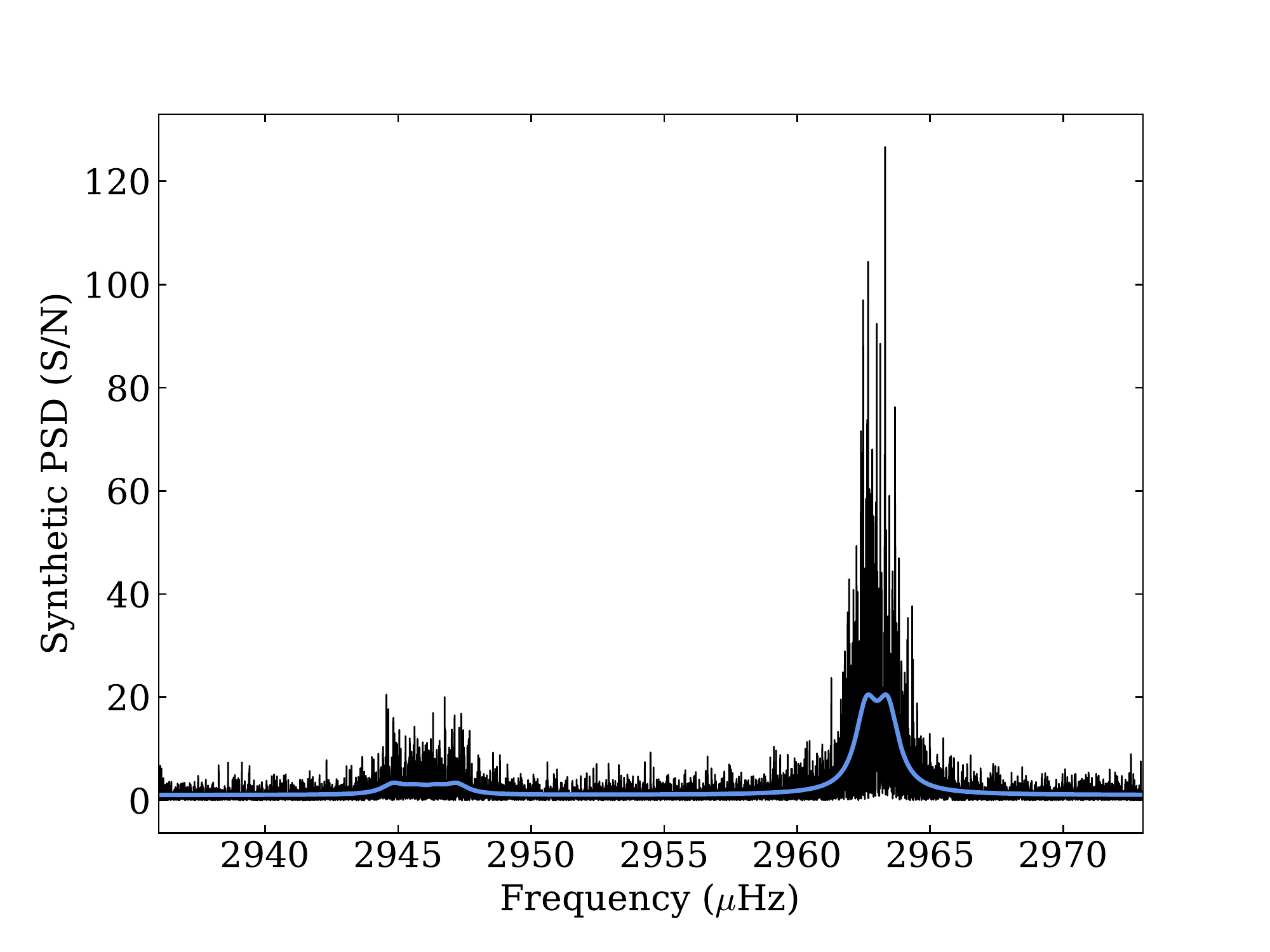}
    \caption{Example of a synthetic spectrum generated with the pair 2 set of parameters and a frequency vector with 1460-day resolution (black). The ideal spectrum is overplotted in blue.}
    \label{fig:mode_example_monte_carlo}
\end{figure}

Considering three different sets of parameters (see Table~\ref{tab:monte_carlo_pairs}), 
we generate a limit spectrum for three pairs of mode. 
This limit spectrum is then multiplied by a noise vector following with a distribution following a $\chi^2$ with two degrees of freedom. 
The pairs are generated for two frequency vectors, the first with one-year resolution and the second with four-year resolution. An example of such a synthetic pair is shown in Fig.~\ref{fig:mode_example_monte_carlo}. Each pair is then fitted using the \texttt{peakbagging} function. For each pair and each resolution, we repeat this process 200 times. For each considered pair and resolution, Tables~\ref{tab:monte_carlo_results_strong_mode} and \ref{tab:monte_carlo_results_weak_mode} summarise the percentage of times for which the true value, considering the uncertainties, is outside the 68\% and 99.7\% credible intervals, defined by the one and three $\sigma$ departure from the fitted value, respectively, with the fitted value taken as the median of the sampled distribution. 
The percentage of fits for which we find the true value to be in the 68\% and 99.7\% credible intervals is therefore close to expectations, the standard deviation for the frequency of success being 3.3~\% in the case of 200 draws following a binomial law of parameters $(n_\mathrm{draw}=200, p=0.68)$, and 0.4~\% for $(n_\mathrm{draw}=200, p=0.997)$, with $n_\mathrm{draw}$ the number of draws. When we consider the 2400 fitted frequencies at once, we find the uncertainties obtained with \texttt{apollinaire} to be conservative in this experiment. Indeed, 71.5\% of the fitted frequencies are in the 68\% credible interval. With a standard deviation of 0.95\% for 2400 independent experiments, this is approximately four standard deviations away from the 68\% expected value. 

\begin{figure}[ht!]
    \centering
    \includegraphics[width=0.49\textwidth]{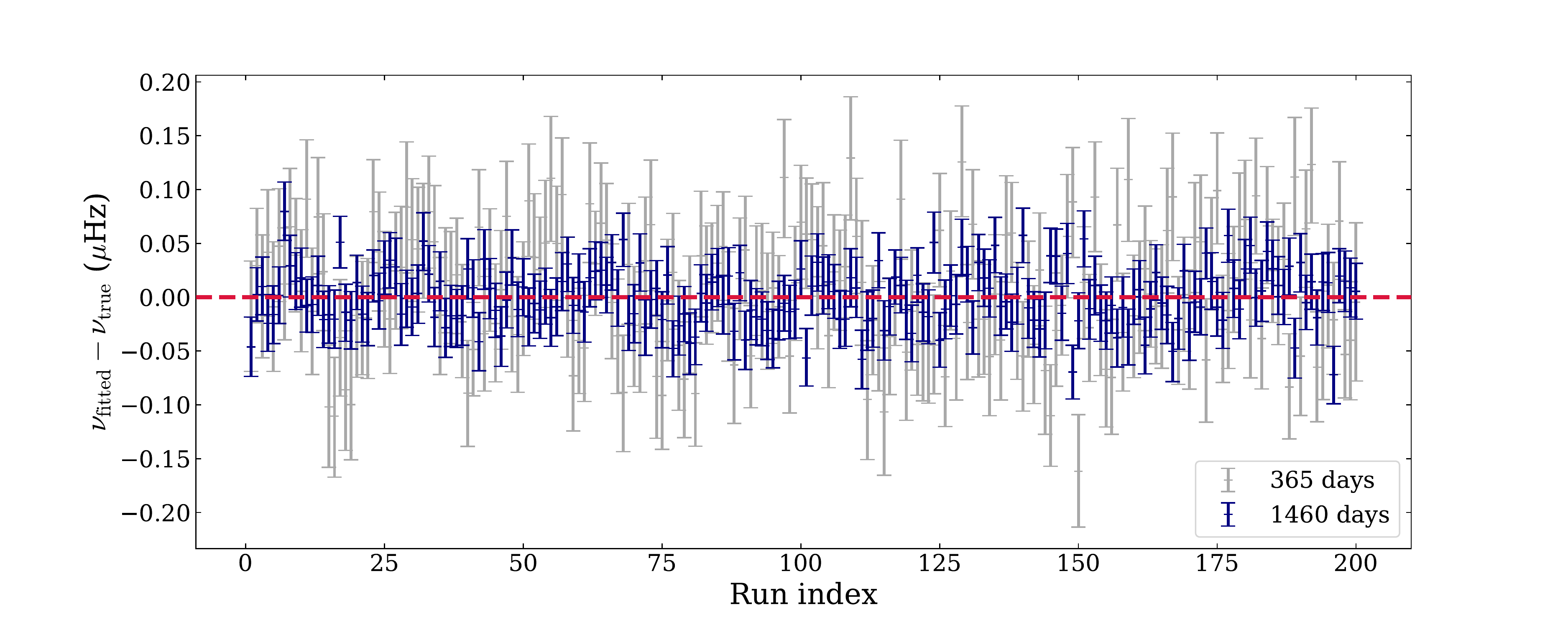}
    \caption{Error $\nu_\mathrm{fitted} - \nu_\mathrm{true}$ on the $\ell=1$ mode from the pair 2 and corresponding uncertainties for every Monte Carlo run with 365-day (grey) and 1460-day (blue) resolution. The zero vertical coordinate is highlighted by the red dashed line.}
    \label{fig:frequency_error}
\end{figure}

\begin{figure}[ht!]
    \centering
    \includegraphics[width=0.49\textwidth]{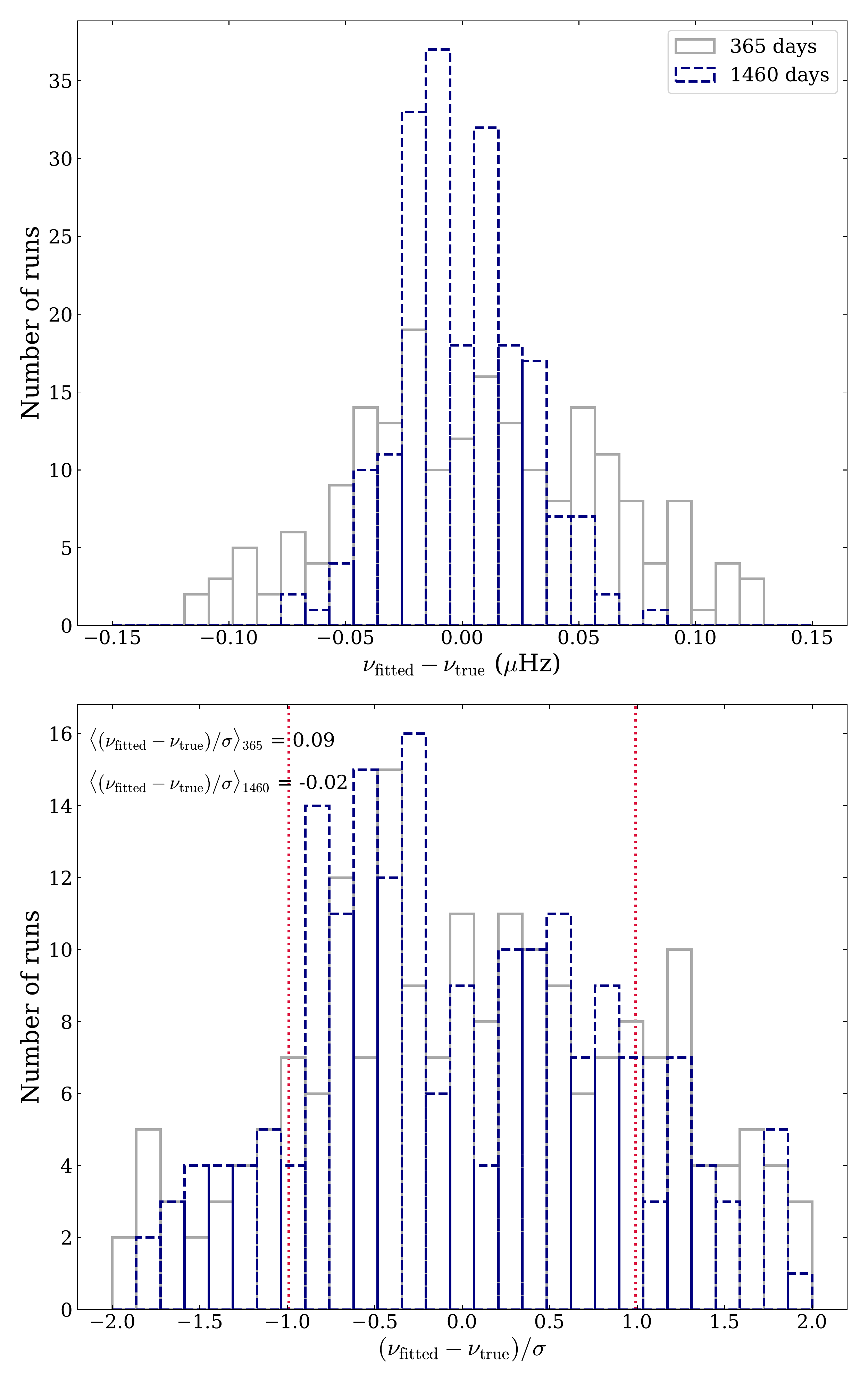}
    \caption{\textit{Top:} $\nu_\mathrm{fitted} - \nu_\mathrm{true}$ distribution on the $\ell=1$ mode from the pair 2 for every Monte Carlo run with 365-day (grey) and 1460-day (blue) resolution.
    \textit{Bottom:} $(\nu_\mathrm{fitted} - \nu_\mathrm{true}) / \sigma$ distribution on the $\ell=1$ mode from the pair 2 for every Monte Carlo run with 365-day (grey) and 1460-day (blue) resolution. The mean value $\langle (\nu_\mathrm{{fitted}} - \nu_\mathrm{{true}})/\sigma \rangle$ for both distribution is specified and the standard deviation of the 1460-day distribution is shown in red.}
    \label{fig:frequency_hist}
\end{figure}

We note that the uncertainties obtained for the 1460-day pairs are significantly smaller than for the 365-day pairs. An example of comparison between the true value and \texttt{apollinaire} fitted value is shown in Fig.~\ref{fig:frequency_error} and \ref{fig:frequency_hist}, where the frequency error $\nu_\mathrm{fitted} - \nu_\mathrm{true}$ between the fitted value and the true value is represented for the $\ell=1$ mode of the pair 2. 
The spread reduction of the error distribution appears clearly in the histograms. We also represent in Fig.~\ref{fig:frequency_hist} the $(\nu_\mathrm{fitted} - \nu_\mathrm{true})/\sigma$ distribution where we specify the mean value $\langle (\nu_\mathrm{fitted} - \nu_\mathrm{true})/\sigma \rangle$. As expected, the standard deviation for the $(\nu_\mathrm{fitted} - \nu_\mathrm{true})/\sigma$ distribution is close to 1 in both cases.   
We find no systematic bias in the fitted parameters, except for the splittings in the pair 3 which are systematically underestimated due to the large mode width. 

\begin{table}[h]
    \centering
    \caption{Mode parameters of the three pairs used for the Monte Carlo trial.}
    \begin{tabular}{c|cc|cc|cc}
    \hline \hline
        & \multicolumn{2}{c}{pair 1} & \multicolumn{2}{c}{pair 2} & \multicolumn{2}{c}{pair 3} \\
    \hline
      $\ell$ & 2 & 0 & 3 & 1 & 2 & 0 \\
      $\nu$ ($\mu$Hz) & 1810 & 1822 & 2946 & 2963 & 3703 & 3710 \\
      $H$ (S/N) & 14 & 20 & 6 & 30 & 21 & 30 \\
      $\Gamma$ ($\mu$Hz) & 0.35 & 0.35 & 1 & 1 & 4 & 4 \\
      $s$ ($\mu$Hz) & 0.4 & 0.4 & 0.4 & 0.4 & 0.4 & 0.4 \\
      $i$ ($^{\circ}$) & 90 & 90 & 90 & 90 & 90 & 90 \\
    \hline
    \end{tabular}
    \label{tab:monte_carlo_pairs}
\end{table}

\begin{table}[h]
    \centering
    \caption{Percentage of values in the 68\% and 99.7\% credible intervals for the strong ($\ell={0,1}$) mode in the Monte Carlo trial.}
    \begin{tabular}{cc|cc|cc|cc}
    \hline \hline
        & & \multicolumn{2}{c}{pair 1} & \multicolumn{2}{c}{pair 2} & \multicolumn{2}{c}{pair 3} \\
    \hline
      \multicolumn{2}{c|}{Resolution (day)} & 365 & 1460 & 365 & 1460 & 365 & 1460 \\
      \multirow{2}{*}{$\nu$}    & 68\%   & 68.5 & 70   & 64.5 & 71.5 & 75   & 67  \\
                                & 99.7\% & 99.5 & 99.5 & 99.5 & 100  & 100  & 100  \\
      \multirow{2}{*}{$H$}      & 68\%   & 74.5 & 69.5 & 76   & 72.5 & 65   & 72  \\
                                & 99.7\% & 99.5 & 99.5 & 100  & 100  & 100  & 99.5 \\
      \multirow{2}{*}{$\Gamma$} & 68\%   & 71.5 & 73.5 & 68.5 & 72   & 72   & 70   \\
                                & 99.7\% & 100  & 99.5 & 100  & 100  & 100  & 99.5 \\
      \multirow{2}{*}{$s$}      & 68\%   & -    & -    & 77.5 & 71   & -    & - \\
                                & 99.7\% & -    & -    & 100  & 100  & -    & - \\
    \hline
    \end{tabular}
    \label{tab:monte_carlo_results_strong_mode}
\end{table}

\begin{table}[h]
    \centering
    \caption{Percentage of values in the 68\% and 99.7\% credible intervals for the weak ($\ell={2,3}$) mode in the Monte Carlo trial.}
    \begin{tabular}{cc|cc|cc|cc}
    \hline \hline
        & & \multicolumn{2}{c}{pair 1} & \multicolumn{2}{c}{pair 2} & \multicolumn{2}{c}{pair 3} \\
    \hline
      \multicolumn{2}{c|}{Resolution (day)} & 365 & 1460 & 365 & 1460 & 365 & 1460 \\
      \multirow{2}{*}{$\nu$}    & 68\%   & 77.5 & 72   & 76   & 73.5 & 71.5 & 74   \\
                                & 99.7\% & 100  & 99.5 & 99.5 & 100  & 100  & 100  \\
      \multirow{2}{*}{$H$}      & 68\%   & 75.5 & 70.5 & 70.5 & 68.5 & 71.5 & 72   \\
                                & 99.7\% & 100  & 99   & 100  & 100  & 99.5 & 100  \\
      \multirow{2}{*}{$\Gamma$} & 68\%   & 74.5 & 73   & 74.5 & 74.5 & 74.5 & 76.5 \\
                                & 99.7\% & 99.5 & 100  & 99.5 & 100  & 100  & 99.5 \\
      \multirow{2}{*}{$s$}      & 68\%   & 68   & 73   & 67.5 & 68   & 86   & 77.5 \\
                                & 99.7\% & 99   & 99.5 & 99.5 & 99.5 & 100  & 100  \\
    \hline
    \end{tabular}
    \label{tab:monte_carlo_results_weak_mode}
\end{table}

\subsection{GOLF solar PSD analysis}


To study the p-mode frequency shifts induced by the the magnetic solar cycle, \citet[][]{2015A&A...578A.137S} performed an analysis of 69 one-year subseries of the GOLF instrument, spanning from 1996 April 11 to 2014 March 5, with 91.25 days overlap. The modes were fitted using a MLE method \citep{2007A&A...463.1181S}. In order to compare the results from this method with \texttt{apollinaire} fits, we use the same MLE code to perform a similar analysis by considering subseries of one-year spanning from 1996 April 11 to 2020 July 6, with a 91.25 days overlap\footnote{The GOLF time series used for this work can be downloaded at \url{https://irfu.cea.fr/dap/LDEE/Phocea/Vie_des_labos/Ast/ast_visu.php?id_ast=3842}}. We remove 5 subseries which have a duty-cycle below 90\% and we consider therefore 89 subseries. For each of these subseries, we fit modes $\ell={0,1,2,3}$. However, as $\ell=3$ have a low signal-to-noise ratios in the GOLF data, we compare in the following only the results obtained for mode $\ell={0,1,2}$. The lowest frequency considered mode is $n=11$, $\ell=2$ while the highest frequency considered mode is $n=26$, $\ell=1$. This means we compare fit results between \texttt{apollinaire} and the MLE method for 4005 modes. 
The procedure to fit the series with \texttt{apollinaire} is the following.
Frequency bins below 50 $\mu$Hz are not considered to fit the background. 
The signal-to-noise spectrum is then computed by dividing the PSD by the fitted background model. Modes are fitted by pair. Heights $H_{n,\ell}$, FWHMs $\Gamma_{n,\ell}$ and splittings $s_{n,\ell}$ are fitted independently for each mode. As asymmetries are expected to depend on frequency only, one asymmetry value is fitted for each pair. Finally, $\ell=\{4,5\}$ power leakages are accounted for during the fit. 
For the background fit, the MCMC are sampled with 500 walkers iterated over 500 steps, with the 100 first steps discarded as burn-in. For the individual mode fit, the MCMC are sampled using 500 walkers iterated over 1000 steps. The 400 first steps are discarded to take the burn-in phase into account. 

\begin{figure}[ht!]
    \centering
    \includegraphics[width=0.48\textwidth]{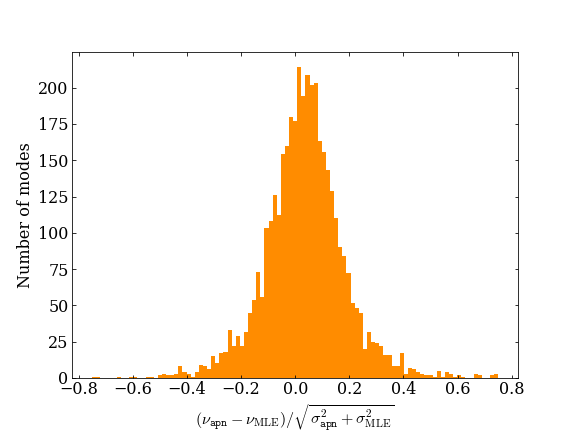}
    \caption{$(\nu_\mathtt{apn} - \nu_\mathrm{MLE}) / \sqrt{\sigma_\mathtt{apn}^2 + \sigma_\mathrm{MLE}^2}$ distribution for mode frequencies fitted in the 89 considered GOLF time series.}
    \label{fig:golf_benchmark}
\end{figure}

We consider the $1\sigma$ and $3\sigma$ intervals relative to the values and corresponding uncertainties $\sigma$ fitted with \texttt{apollinaire}.
Among the 4005 fitted modes, the frequencies fitted by the MLE method, $\nu_\mathrm{MLE}$, lay outside the $1\sigma$ interval for 38 modes (0.9\%) and outside the $3\sigma$ interval for 10 modes (0.2\%). The MLE heights $H$ lay outside the $1\sigma$ interval for 464 modes (11.5\%) and outside the $3\sigma$ interval for 23 modes (0.6\%). The MLE FWHMs $\Gamma$ lay outside the $1\sigma$ interval for 185 modes (4.6\%) and outside the $3\sigma$ interval for 18 modes (0.4\%). The relative high number of $H$ and $\Gamma$ values laying outside the $1\sigma$ interval can be explained by the strong anticorrelation that exists between these two parameters, especially at high frequency. We also identify at least two modes which have not been correctly fitted by the MLE method. Finally, we represent the $(\nu_\mathtt{apn} - \nu_\mathrm{MLE}) / \sqrt{\sigma_\mathtt{apn}^2 + \sigma_\mathrm{MLE}^2}$ distribution in Fig.~\ref{fig:golf_benchmark} in order to show the comparison of the two methods exhibits no systematic bias.

\subsection{The \textit{Kepler} main-sequence LEGACY catalogue \label{section:benchmark_legacy}}

In order to extract strong modelling constraints to be used in stellar evolution codes \citep{2017ApJ...835..173S}, \citet[][referred as L17]{2017ApJ...835..172L} performed a thorough peakbagging analysis of the so-called \textit{Kepler} LEGACY sample, which contains 66 main-sequence stars with stochastically excited p modes\footnote{The mode frequency, amplitude and width tables are publicly available on VizieR at: \url{http://cdsarc.u-strasbg.fr/viz-bin/Cat?J/ApJ/835/172}}. We select six targets to analyse within the LEGACY sample, KIC 5184732, 6106415, 6225718, 6603624, 12069424, and 12069449, and we compare the \texttt{apollinaire} results to the reference values provided by L17. Some fundamental stellar properties of the selected targets are given in Table~\ref{tab:target_presentation}, taken from the \textit{Kepler} data release 25 \citep[DR25,][]{2017ApJS..229...30M}. 

\begin{table}[ht]
\centering
\caption{KIC and fundamental stellar properties of LEGACY sample selected targets. 
}
\begin{tabular}{lcrrrr}
\hline \hline
KIC & Name & $T_\mathrm{eff}$ (K) &  $\log g$ &  $R$ ($R_\odot$) &   $M$ ($M_\odot$) \\
\hline
5184732  & -      & 5835 &   4.257 &    1.326 &  1.159 \\
6106415  & -      & 6028 &   4.295 &    1.202 &  1.039 \\
6225718  & -      & 6320 &   4.316 &    1.207 &  1.096 \\
6603624  & -      & 5671 &   4.319 &    1.162 &  1.027 \\
12069424 & 16CygA & 5775 &   4.294 &    1.165 &  0.973 \\
12069449 & 16CygB & 5745 &   4.359 &    1.069 &  0.952 \\
\hline
\end{tabular}
\label{tab:target_presentation}
\end{table}

As we want to show how \texttt{apollinaire} behaves when used blindly without any previous knowledge of the seismic values, effective temperatures $T_\mathrm{eff}$, masses $M$, and radii $R$ are used to estimate $\nu_\mathrm{max}$ and $\Delta \nu$ values from the global seismic scaling laws. We could of course have used as input the $\nu_\mathrm{max}$ and $\Delta \nu$ values from L17.
The PSD we analyse have been obtained from KEPSEISMIC-calibrated\footnote{KEPSEISMIC data are available at MAST via \url{http://dx.doi. org/10.17909/t9-mrpw-gc07}} light curves \citep{2011MNRAS.414L...6G}. 
The background profile is fitted considering two Harvey models and a flat noise contribution (see Sect.~\ref{section:background}). 
We fit the background with the complete PSD, but, in order to assess the effect on the mode frequencies of a background fit performed on a rebinned PSD, we also sample the posterior probability of our background model on a PSD resampled following the method described in Appendix~\ref{appendix:background}.
Frequency bins below 50 $\mu$Hz are not considered. We use 500 walkers iterated over 5000 steps, with the 100 first iteration discarded as burn-in. 
We deliberately choose a large number of steps to ensure the convergence of the chains.
The global mode pattern is adjusted on the signal-to-noise spectrum obtained by dividing the PSD by the background model, according to the strategy described in Sect.~\ref{section:pattern} and considering $\ell = \{0,1,2,3\}$ of the three orders closest to $\nu_\mathrm{max}$. 
We use 500 walkers iterated over 5000 steps and we discard the 250 first drawn points. We visually check that the fitted pattern had a satisfying profile in order to provide correct frequency guesses for the individual mode-parameter extraction.
For the individual mode parameter fit, we limit for each target our analysis to the seven orders closest to $\nu_\mathrm{max}$.
All the chosen targets have a p-mode S/N ratio sufficient enough to fit mode parameters farther from $\nu_\mathrm{max}$, but we make this choice in order to ensure that the frequency estimate yielded by Eq.~\ref{eq:tassoul_2nd} are correct and to avoid a manual redefinition of the priors for low and high-order mode before the mode parameters sampling.
The MCMC are sampled using 500 walkers iterated over 5000 steps. The 200 first steps are discarded to take the burn-in phase into account.

\begin{figure}[ht!]
    \centering
    \includegraphics[width = 0.48 \textwidth]{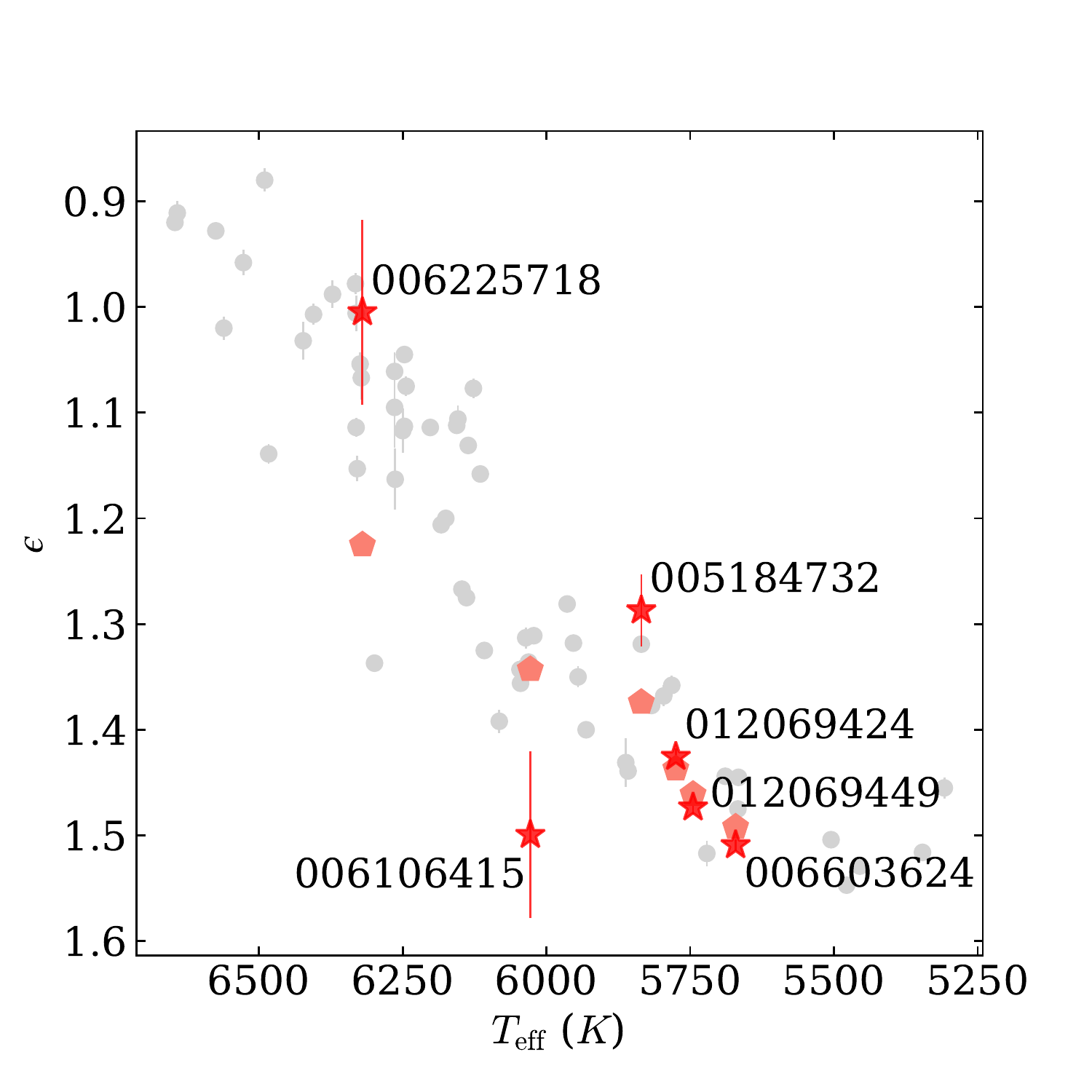}
    \caption{L17 $\epsilon$ values (\textit{pentagons}) compared to \apollinaire $\epsilon$ values (\textit{stars}) for the six considered targets. Corresponding KIC are specified next each red dot. The epsilon values obtained by L17 for the other stars of the LEGACY sample are represented in grey. Some of the uncertainties are too small to be represented on the figure.}
    \label{fig:epsilon}
\end{figure}

\begin{figure*}
 \centering
  \subfigure{\includegraphics[width=.48\textwidth]{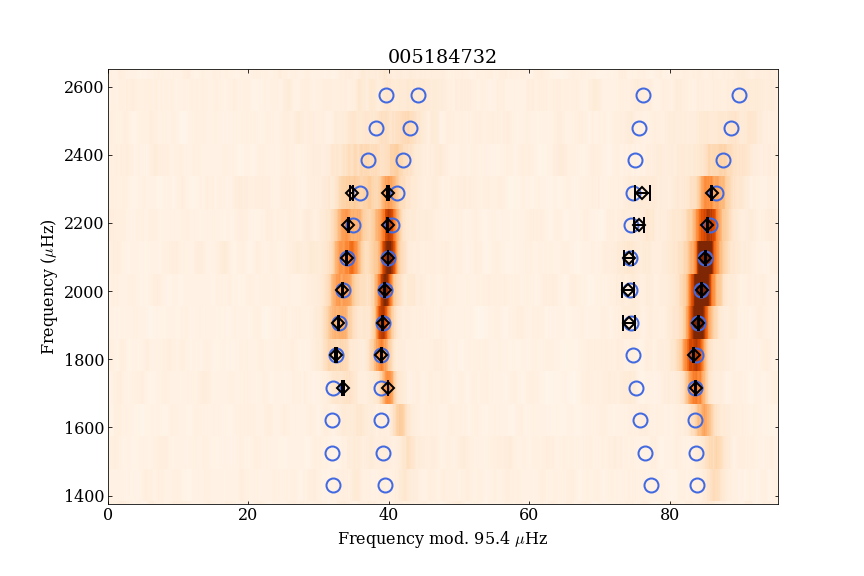}}
  \subfigure{\includegraphics[width=.48\textwidth]{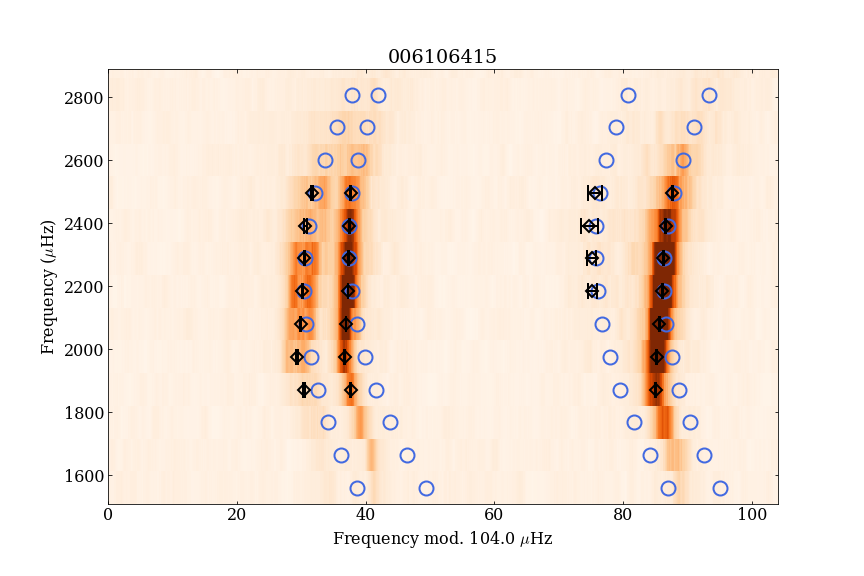}} 
  \subfigure{\includegraphics[width=.48\textwidth]{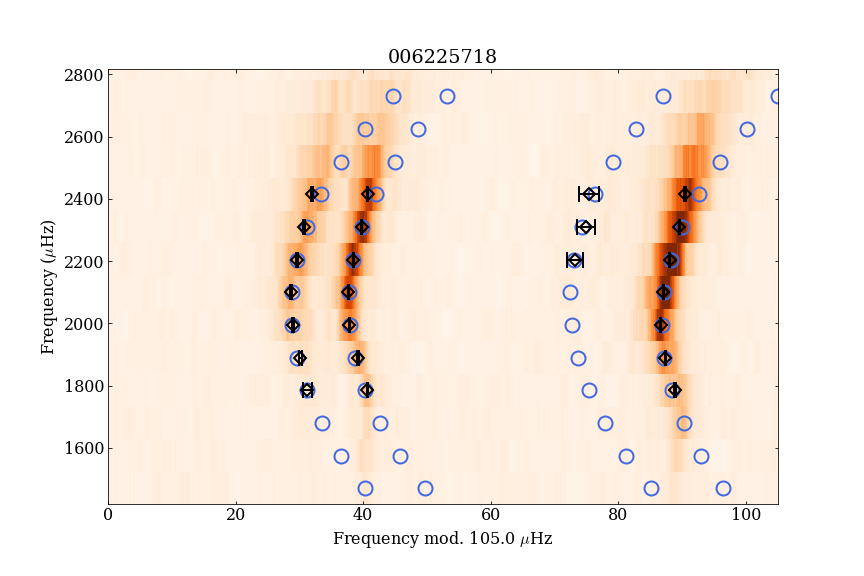}}
  \subfigure{\includegraphics[width=.48\textwidth]{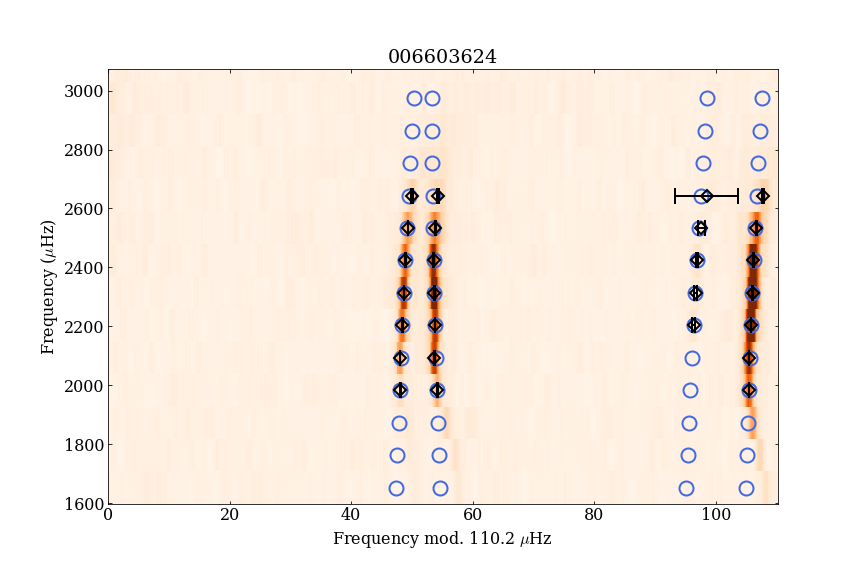}} 
  \subfigure{\includegraphics[width=.48\textwidth]{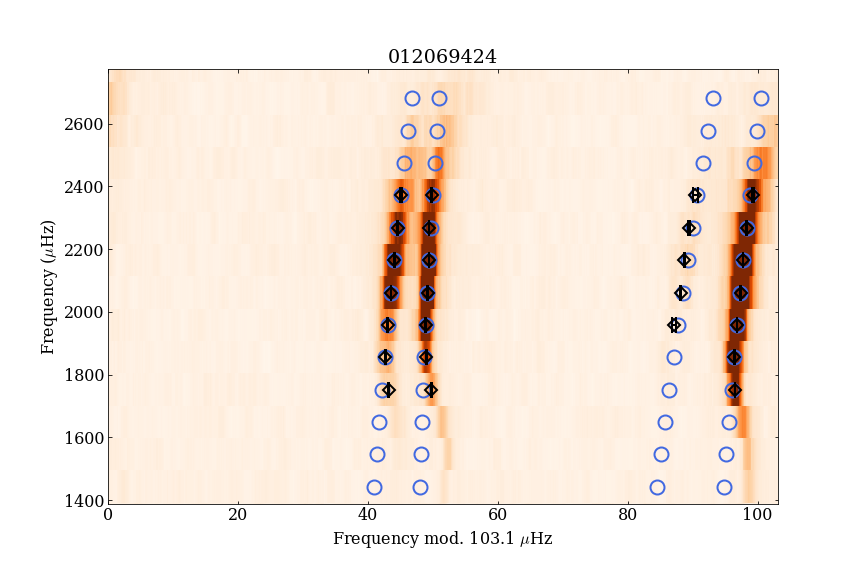}}
  \subfigure{\includegraphics[width=.48\textwidth]{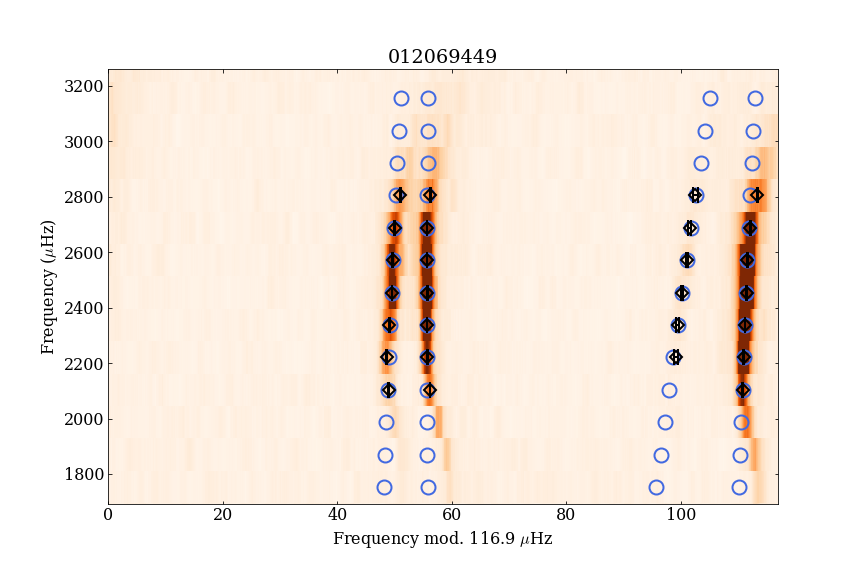}}
\caption{Échelle diagram of the six selected targets \apollinaire fitted p-mode frequencies (black diamonds) with corresponding errorbars. The blue circles signal the frequency initial guesses inferred from Eq.~\ref{eq:tassoul_2nd}. 
}
\label{fig:ed_stellar_benchmark}
\end{figure*}

\begin{figure*}[ht!]
 \centering
  \subfigure{\includegraphics[width=.48\textwidth]{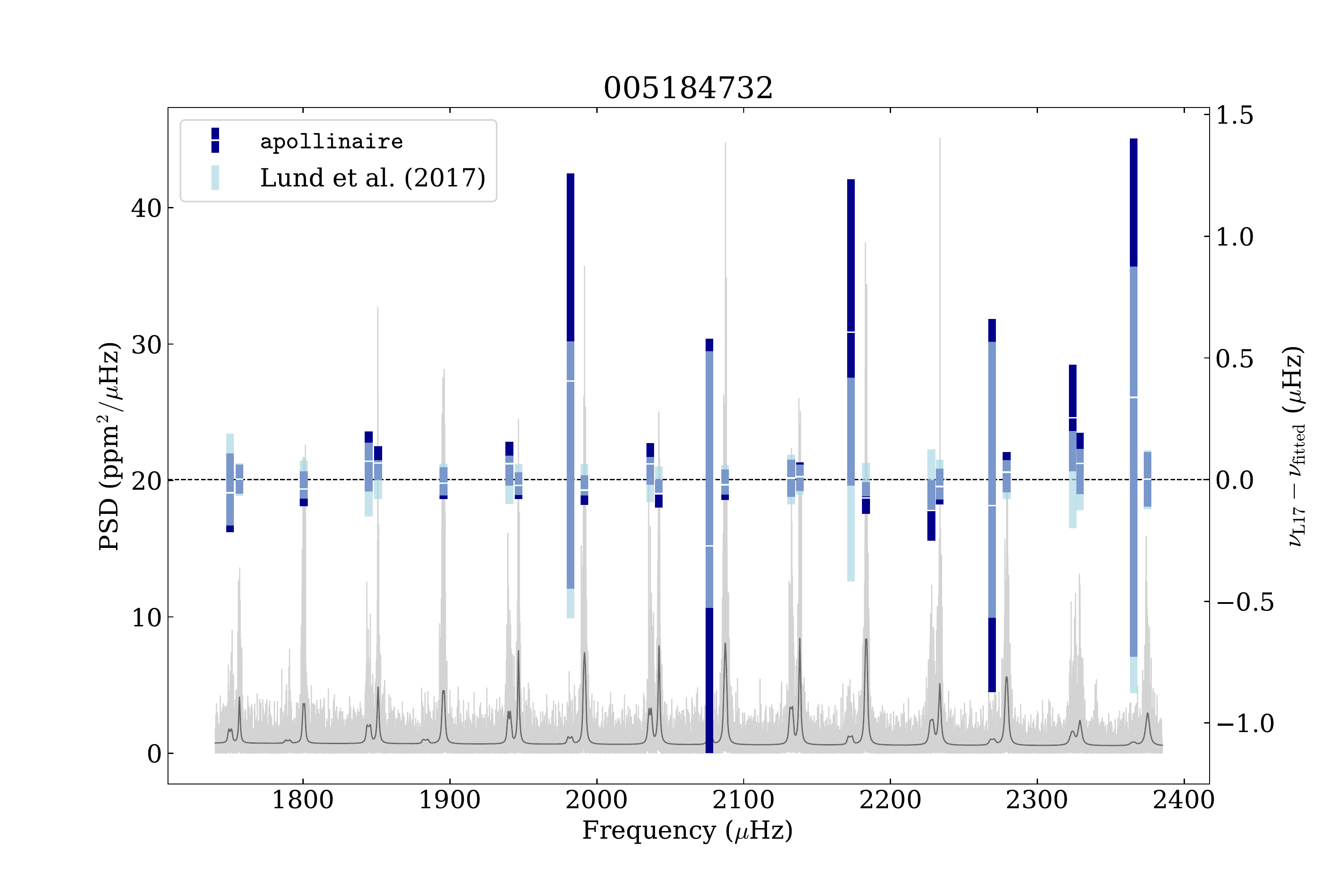}}
  \subfigure{\includegraphics[width=.48\textwidth]{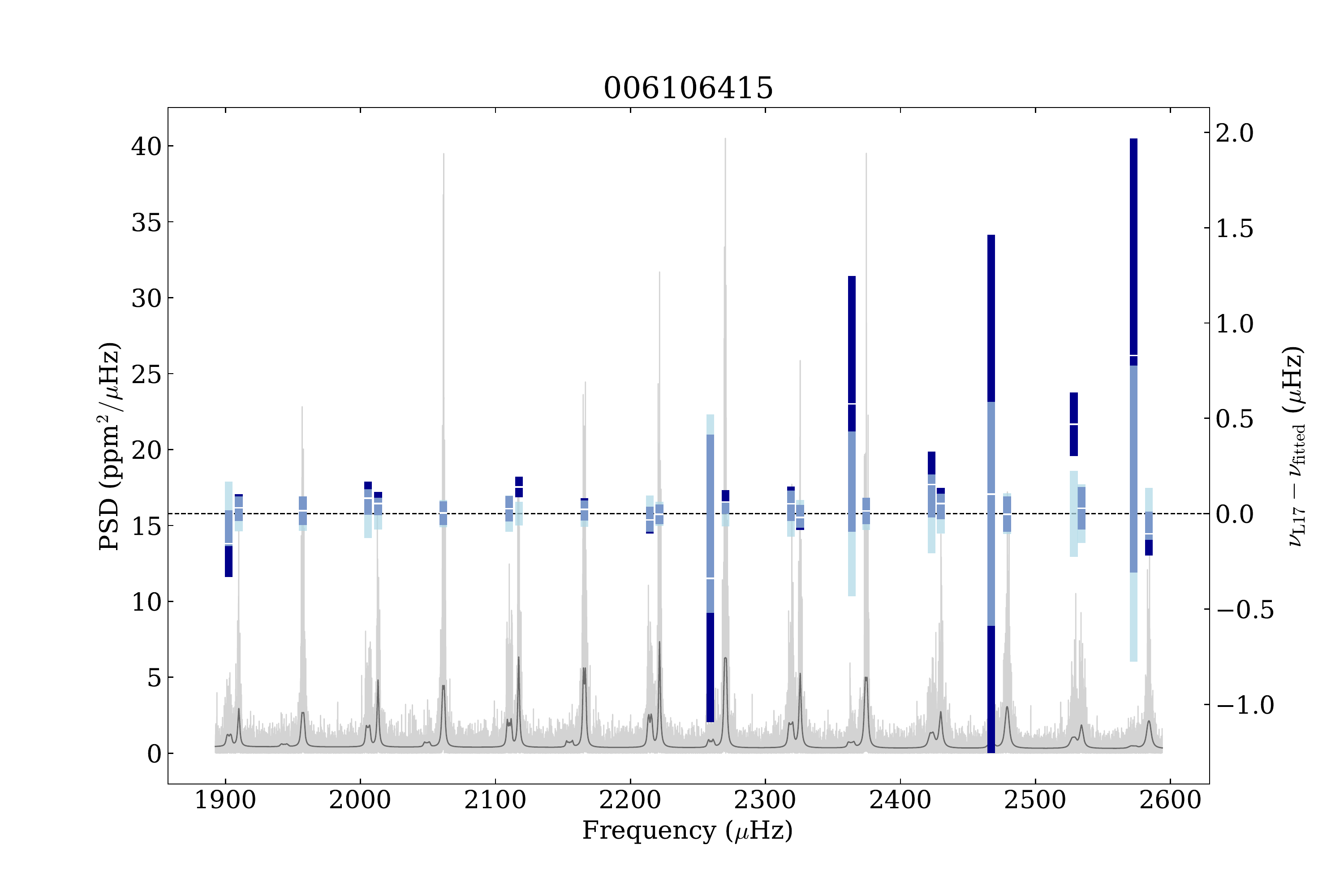}} 
  \subfigure{\includegraphics[width=.48\textwidth]{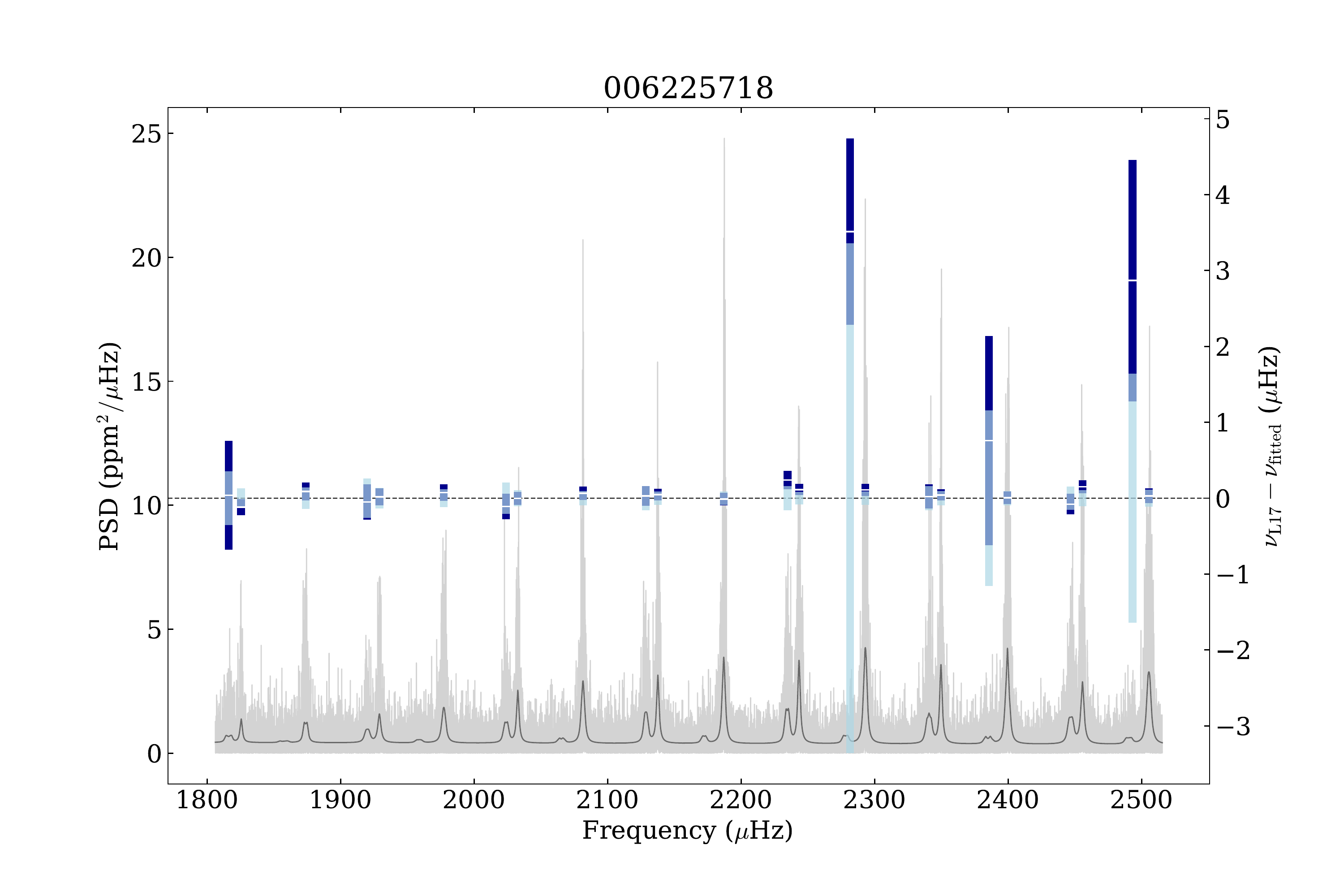}}
  \subfigure{\includegraphics[width=.48\textwidth]{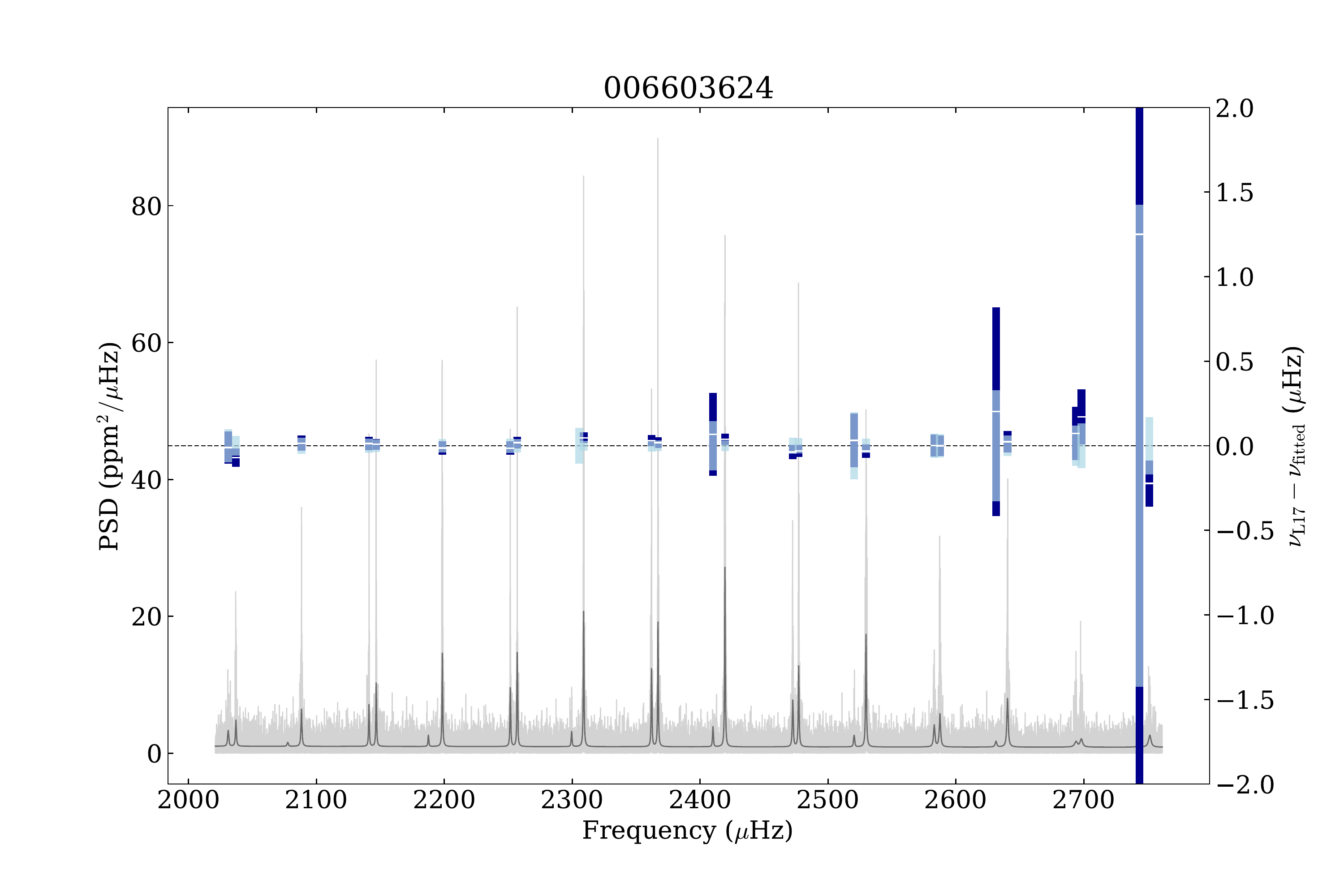}} 
  \subfigure{\includegraphics[width=.48\textwidth]{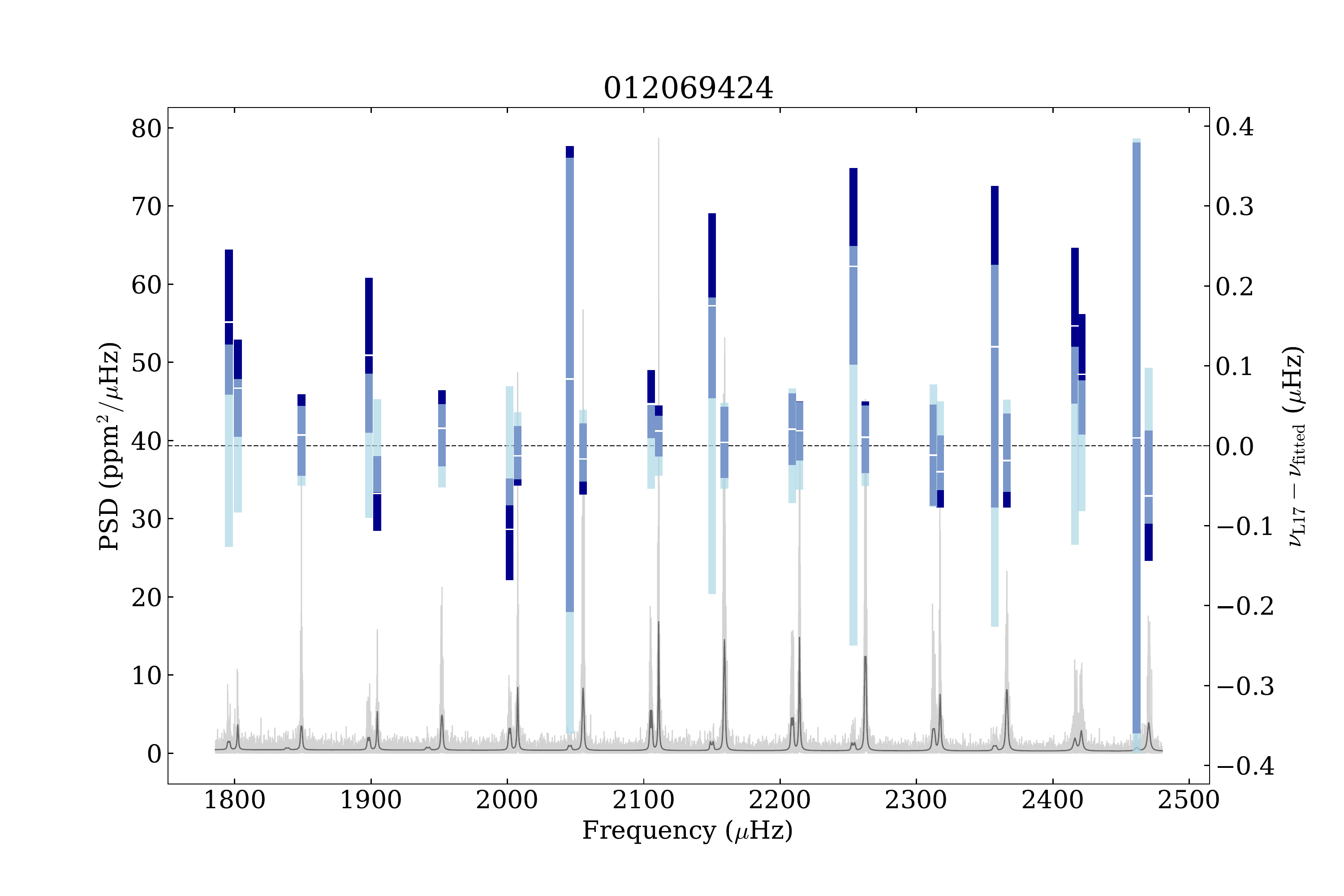}}
  \subfigure{\includegraphics[width=.48\textwidth]{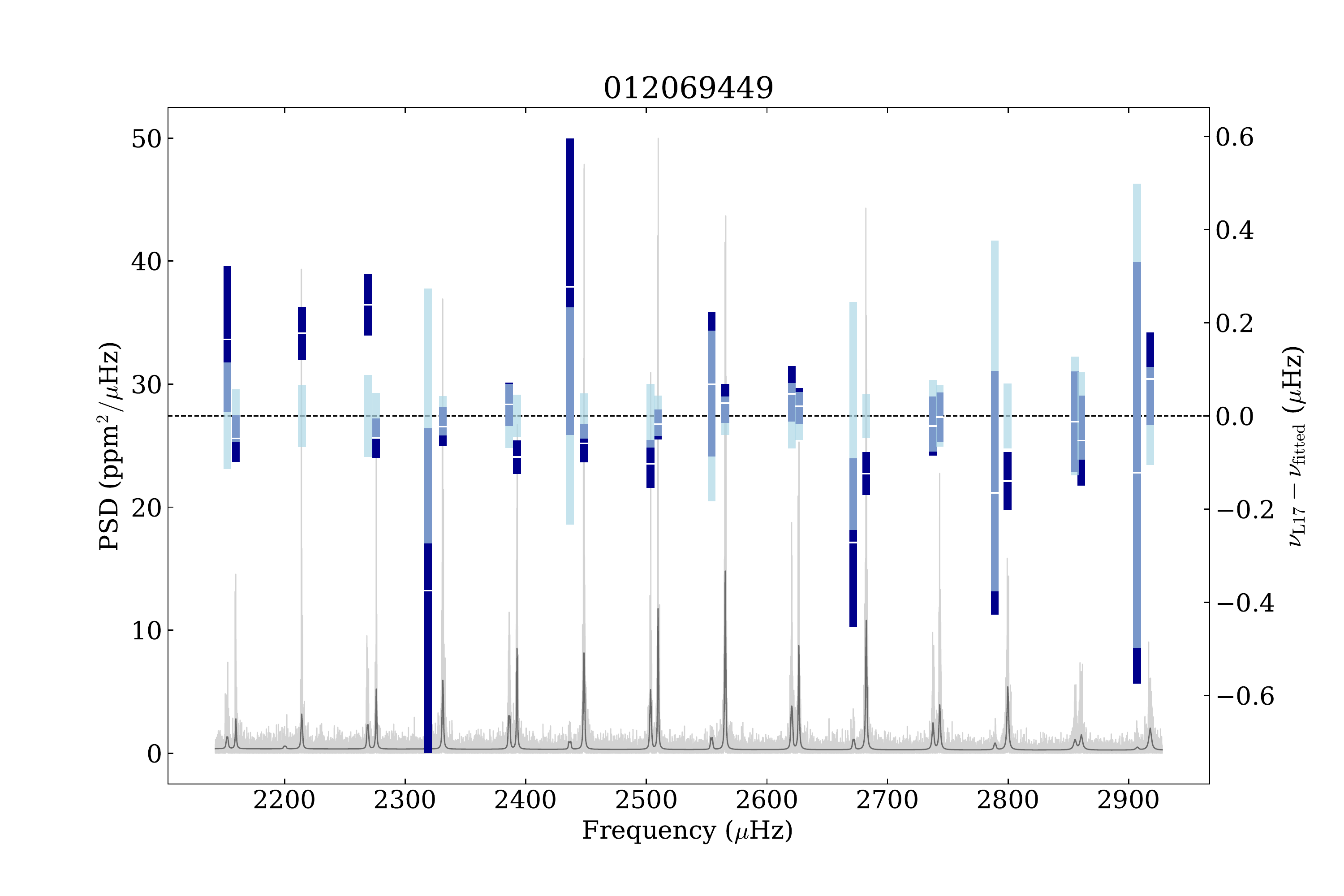}}
\caption{Comparison of p-mode frequencies yielded by \apollinaire (dark blue) and reference values from L17 (light blue). The error bars overlapping intervals are represented in medium blue. The KEPSEISMIC PSD are represented in light grey while the \apollinaire fitted model is shown in dark grey.}
\label{fig:frequency_stellar_benchmark}
\end{figure*}

\begin{figure*}[ht!]
 \centering
  \subfigure{\includegraphics[width=.48\textwidth]{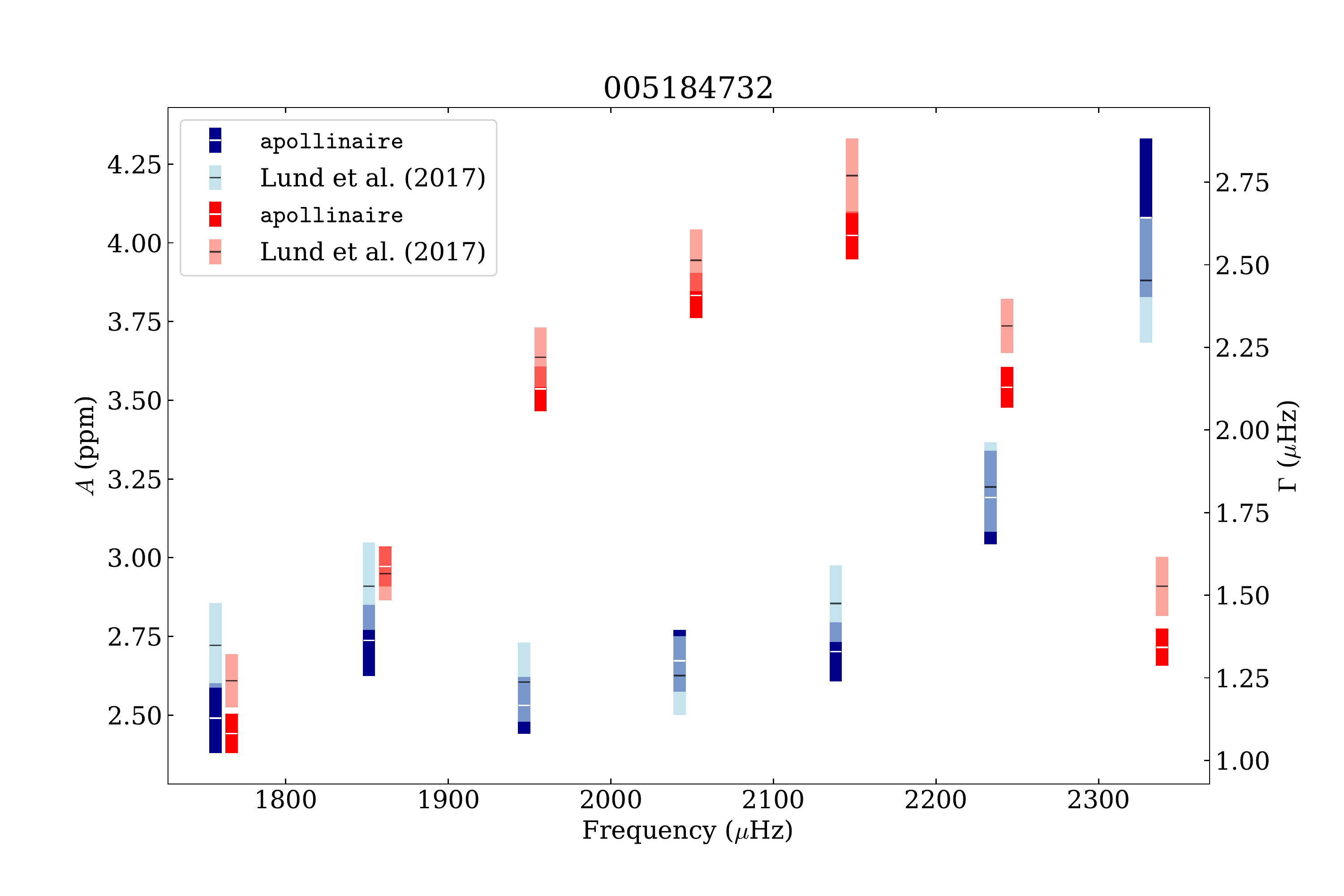}}
  \subfigure{\includegraphics[width=.48\textwidth]{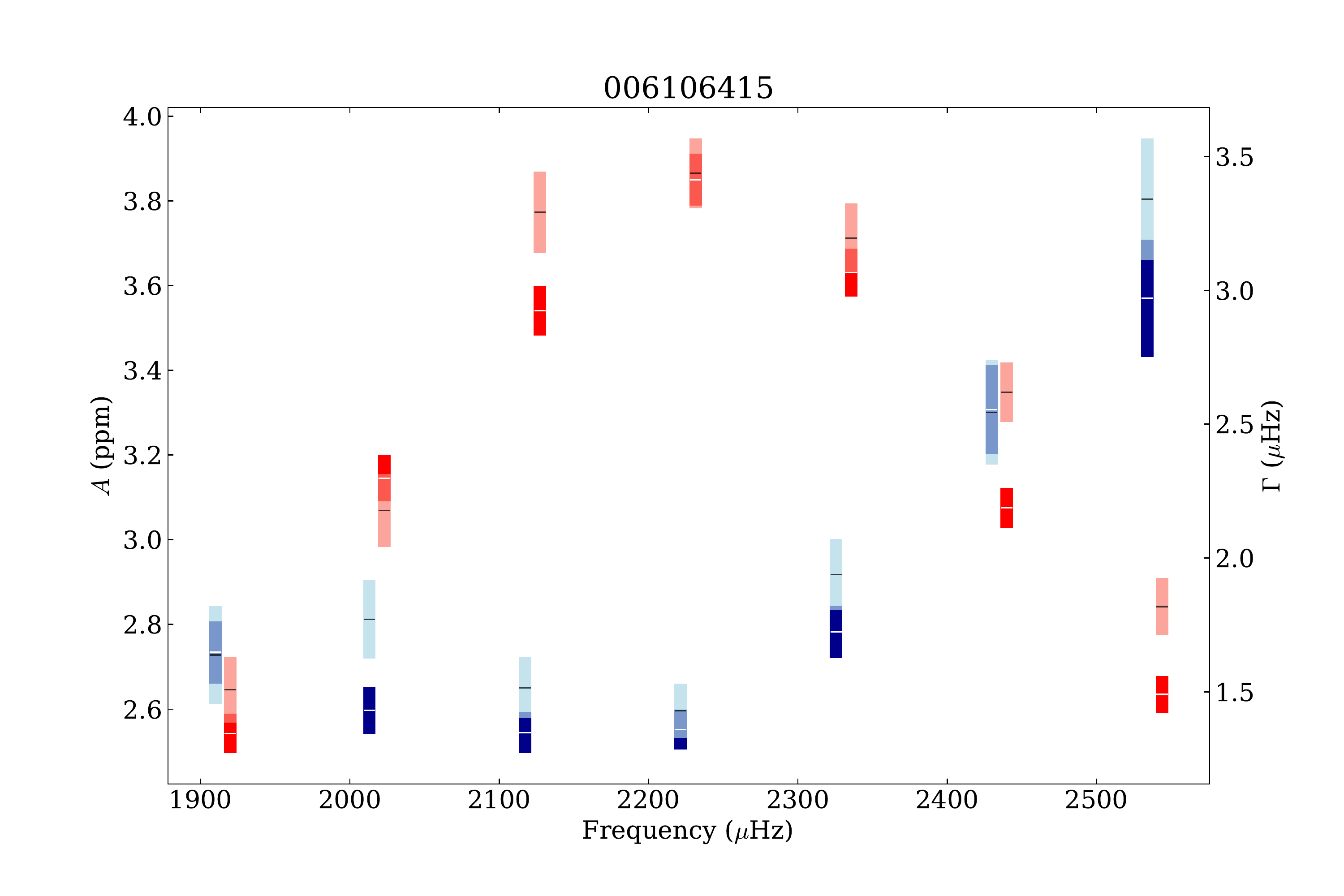}} 
  \subfigure{\includegraphics[width=.48\textwidth]{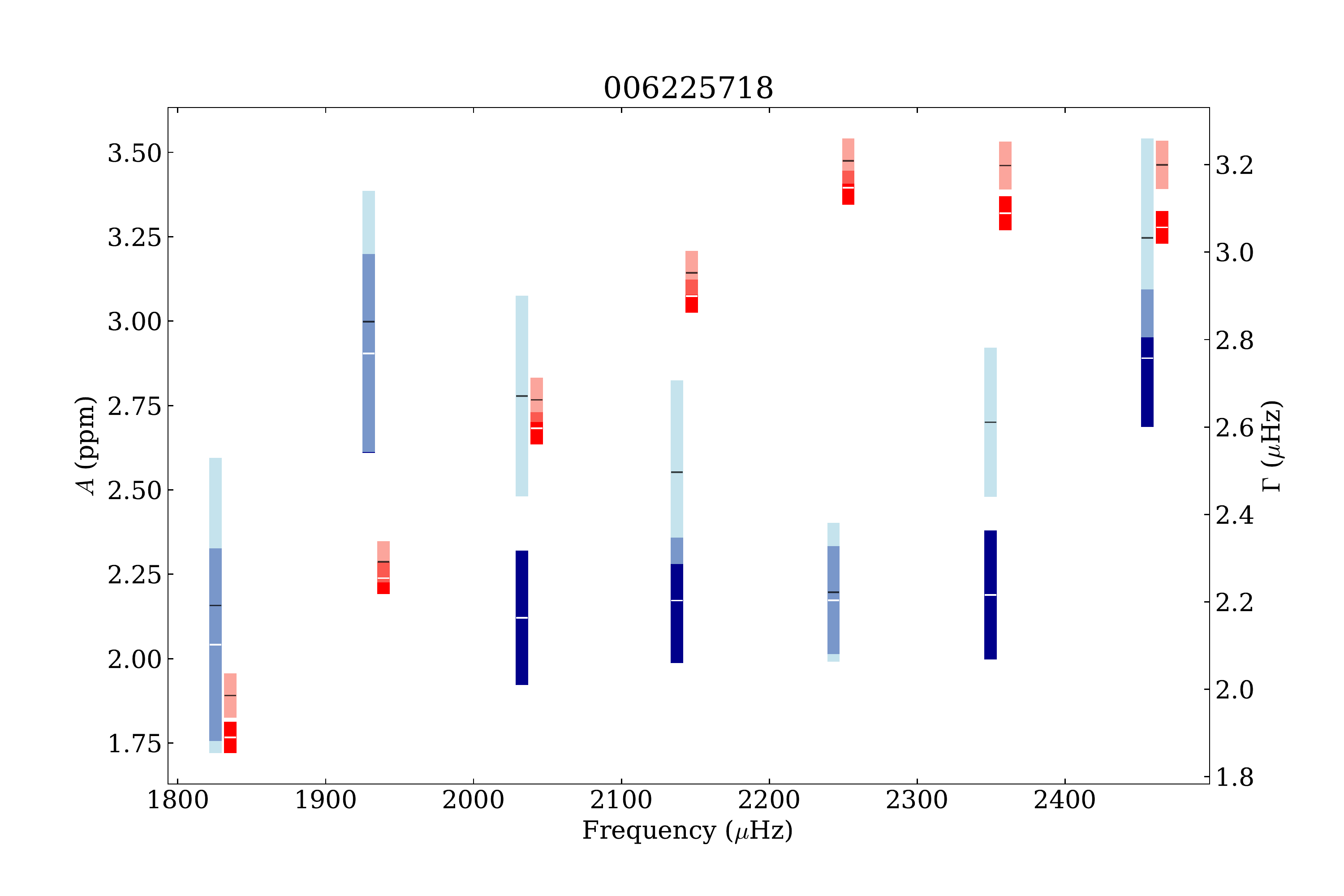}}
  \subfigure{\includegraphics[width=.48\textwidth]{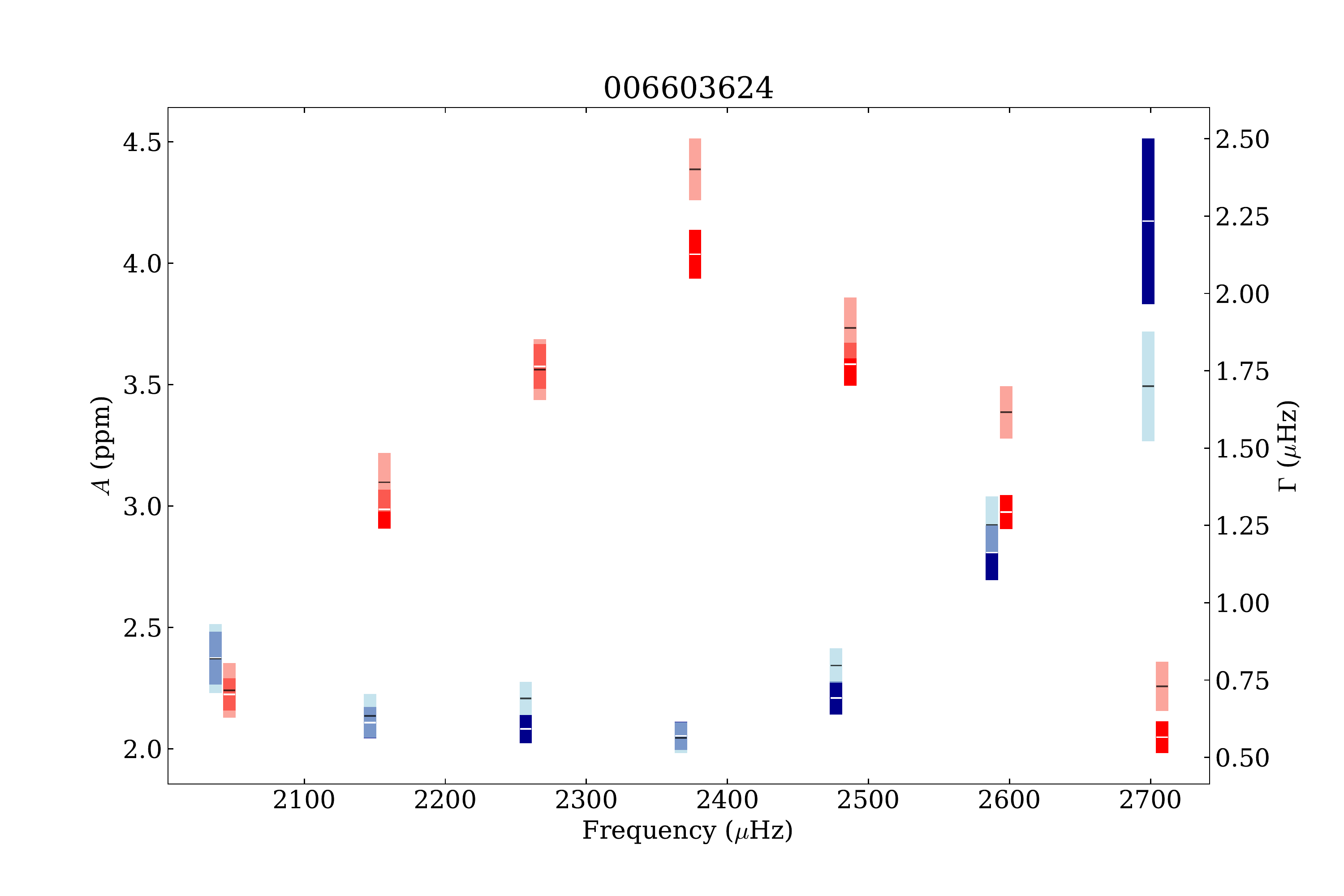}} 
  \subfigure{\includegraphics[width=.48\textwidth]{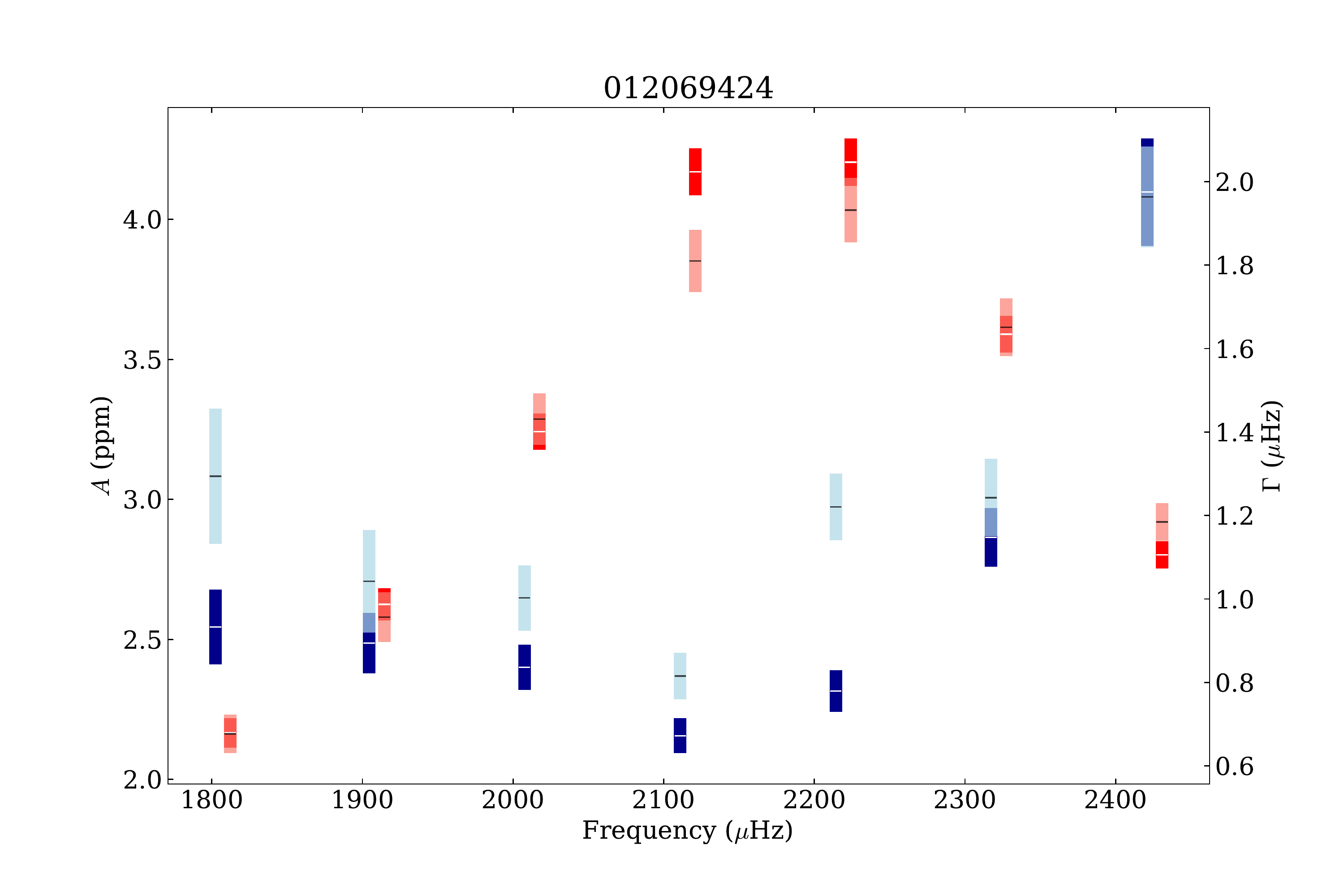}}
  \subfigure{\includegraphics[width=.48\textwidth]{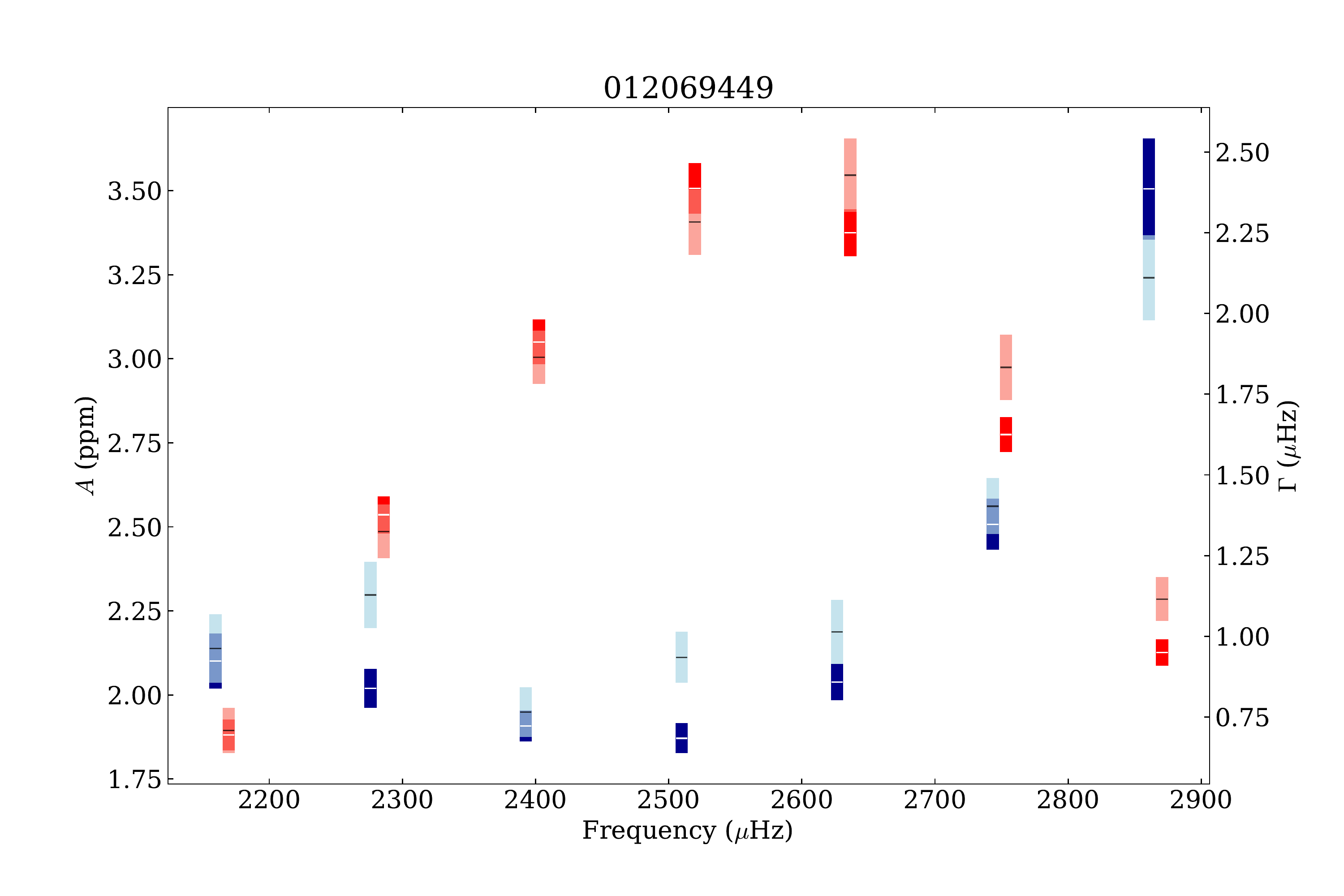}}
\caption{Comparison of p-mode FWHMs and amplitudes yielded by \apollinaire (dark blue and red, respectively) and reference values from L17 (light blue and salmon respectively). The overlapping error bars intervals are represented in medium blue and dark salmon, respectively.}
\label{fig:hw_stellar_benchmark}
\end{figure*}

For each fitted parameter, L17 specified two uncertainty values (in the same way \apollinaire allows computing $\sigma_+$ and $\sigma_-$). For sake of simplicity and clarity in the results comparison, we consider for each parameter only the larger uncertainty value. L17 epsilon $\epsilon$ values are compared to the results yielded by \apollinaire in Fig.~\ref{fig:epsilon}. The échelle diagrams obtained from the \apollinaire fits are provided in Fig.~\ref{fig:ed_stellar_benchmark} where we also represent the frequencies obtained from Eq.~\ref{eq:tassoul_2nd} and the corresponding pattern sampling. As it was already underlined in Sect.~\ref{section:pattern}, the échelle diagram clearly shows that, as we constrained the parameters from Eq.~\ref{eq:tassoul_2nd} only on the three central orders, the predicted value we obtain for mode far from $\nu_\mathrm{max}$ does not follow the actual node ridge in the échelle diagrams. Fitted frequencies are compared with the values from L17 in Fig.~\ref{fig:frequency_stellar_benchmark}. The same type of comparison is performed for mode amplitudes and FWHMs in Fig.~\ref{fig:hw_stellar_benchmark}. 
The \apollinaire fitted values and reference from L17 can be found in Tables~\ref{tab:benchmark_global_parameter}, \ref{tab:stellar_hw} and \ref{tab:stellar_frequency}, along with the quality assurances values $\ln K$ computed according to the method presented in Sect.~\ref{section:quality_assurance}. 
The results are globally in agreement for frequencies. Considering again the $1\sigma$ and $3\sigma$ interval relative to the \texttt{apollinaire} value, over the 154 fitted frequencies, the L17 frequency is beyond $1\sigma$ for 39 modes and beyond $3\sigma$ for three modes. We find larger discrepancies for mode amplitudes and widths: over the 42 fitted widths (one per order), the L17 value is outside the $1\sigma$ interval 25 times and outside the $3\sigma$ interval eight times. Over the 42 fitted amplitudes, the L17 value is outside the $1\sigma$ interval 31 times and outside the $3\sigma$ interval 12 times. We find no systematic bias in the frequency, amplitude and width comparison.
Concerning the mode frequencies, the largest observed discrepancy between L17 and \apollinaire is found to be on the KIC~006603624 mode $M_{18,3}$. We have $\nu_{18,3,\mathtt{apn}} = 2299.52 \pm 0.20$ $\mu$Hz and $\nu_{18,3,\mathrm{L17}} = 2305.47 \pm 0.13$ $\mu$Hz. However, inspecting the KIC~006603624 spectrum and the corresponding échelle diagram, it seems more credible to us that the correct position of this mode is the one yielded by \texttt{apollinaire}.  

The discrepancies in the values obtained for fitted parameters can have multiple explanations.  
While we used KEPSEISMIC data, L17 exploited light curves corrected with the KASOC filter \citep{2014MNRAS.445.2698H}. The data calibration influences the power redistribution in the PSD frequency bins and may therefore be responsible for some of the observed discrepancies between the mode widths and amplitudes, and, to a lesser extent, for frequency discrepancies. However, when comparing peakbagging codes, it is found that the main discrepancies are induced by the differences in the methods used to fit the background \citep{2014A&A...566A..20A}. 
It should also be underlined that there are several differences in the models we use, when compared to what L17 did. The first main difference resides in the fact that L17 included the $V_\ell$ parameters for which posterior probability distribution are sampled while these parameters were fixed in our model (see Sect.~\ref{section:visibilities_ratios}). The second difference is that, as we fit order by order, the inclination angle $i$ is poorly constrained with this choice of strategy.
It is also important to keep in mind that the uncertainty values yielded by a MCMC process are only formal uncertainties related to the variance of the parameters inside a given model confronted to the data. It has already been demonstrated that an inaccurate description (for example by neglecting the $\ell={4,5}$ contribution or the wing power of the modes outside the fitting window) of the mode profiles could bias the estimation of the real values by several times the uncertainty values \citep{2008MNRAS.389.1780J}.

Concerning $\epsilon$ values, L17 and \citet{2011ApJ...742L...3W} already noted that a strong anti-correlation existed between $\Delta\nu$ and $\epsilon$. We also note that for most $\nu_\mathrm{max}$, $\Delta\nu$, and $\epsilon$ the \apollinaire uncertainty values are much larger than for the L17 values. For KIC~006225718, for example, \apollinaire yielded $\Delta\nu = 106.78 \pm 0.43$ $\mu$Hz while L17 gave $\Delta\nu = 105.695 \pm 0.018$ $\mu$Hz.
This can be explained by the fact that, even if our Eq.~\ref{eq:tassoul_2nd} is really close to their Eq.~27, the strategy exploited to sample the parameter distribution is totally different. We use Eq.~\ref{eq:tassoul_2nd} together with Eq.~\ref{eq:height_glob} and a common $\Gamma$ value in order to compute a mode profile for the three central orders and directly constrain our parameters with the PSD. On the contrary, L17 used their Eq.~27 to constrain the pattern parameters from the individual mode frequencies they had already obtained from the PSD analysis. 

\begin{table*}[ht]
    \centering
    \caption{Comparison between $\nu_\mathrm{max}$, $\Delta\nu$ and $\epsilon$ values yielded by \apollinaire and reference values from L17.}
    \label{tab:benchmark_global_parameter}
\begin{tabular}{lrrrrrr}
\hline \hline
KIC &   $\nu_{\mathrm{max}, \mathtt{apn}}$ ($\mu$Hz) & $\nu_\mathrm{max, L17}$ ($\mu$Hz) & $\Delta\nu_\mathtt{apn}$ ($\mu$Hz) & $\Delta\nu_\mathrm{L17}$ ($\mu$Hz) & $\epsilon_\mathtt{apn}$ & $\epsilon_\mathrm{L17}$ \\
\hline
5184732  &  2164 $\pm$  35 & 2089.3 $\pm$ 4.4 &  95.94 $\pm$  0.15 & 95.545 $\pm$  0.024 & 1.287 $\pm$ 0.034 & 1.374 $\pm$ 0.005 \\
6106415  &  2215 $\pm$  50 & 2248.6 $\pm$ 4.6 & 103.36 $\pm$  0.37 & 104.074 $\pm$ 0.026 & 1.499 $\pm$ 0.079 & 1.343 $\pm$ 0.005 \\
6225718  &  2349 $\pm$  54 & 2364.2 $\pm$ 4.9 & 106.78 $\pm$  0.43 & 105.695 $\pm$ 0.018 & 1.005 $\pm$ 0.087 & 1.225 $\pm$ 0.004 \\
6603624  &  2386 $\pm$  24 & 2384.0 $\pm$ 5.6 & 110.05 $\pm$  0.03 & 110.128 $\pm$ 0.012 & 1.509 $\pm$ 0.006 & 1.492 $\pm$ 0.002 \\
12069424 &  2192 $\pm$  22 & 2188.5 $\pm$ 4.6 & 103.35 $\pm$  0.39 & 103.277 $\pm$ 0.021 & 1.426 $\pm$ 0.008 & 1.437 $\pm$ 0.004 \\
12069449 &  2625 $\pm$  93 & 2561.3 $\pm$ 5.6 & 116.87 $\pm$  0.87 & 116.929 $\pm$ 0.013 & 1.473 $\pm$ 0.007 & 1.461 $\pm$ 0.013 \\
\hline
\end{tabular}
\end{table*}

\begin{figure}[ht!]
    \centering
    \includegraphics[width=0.48\textwidth]{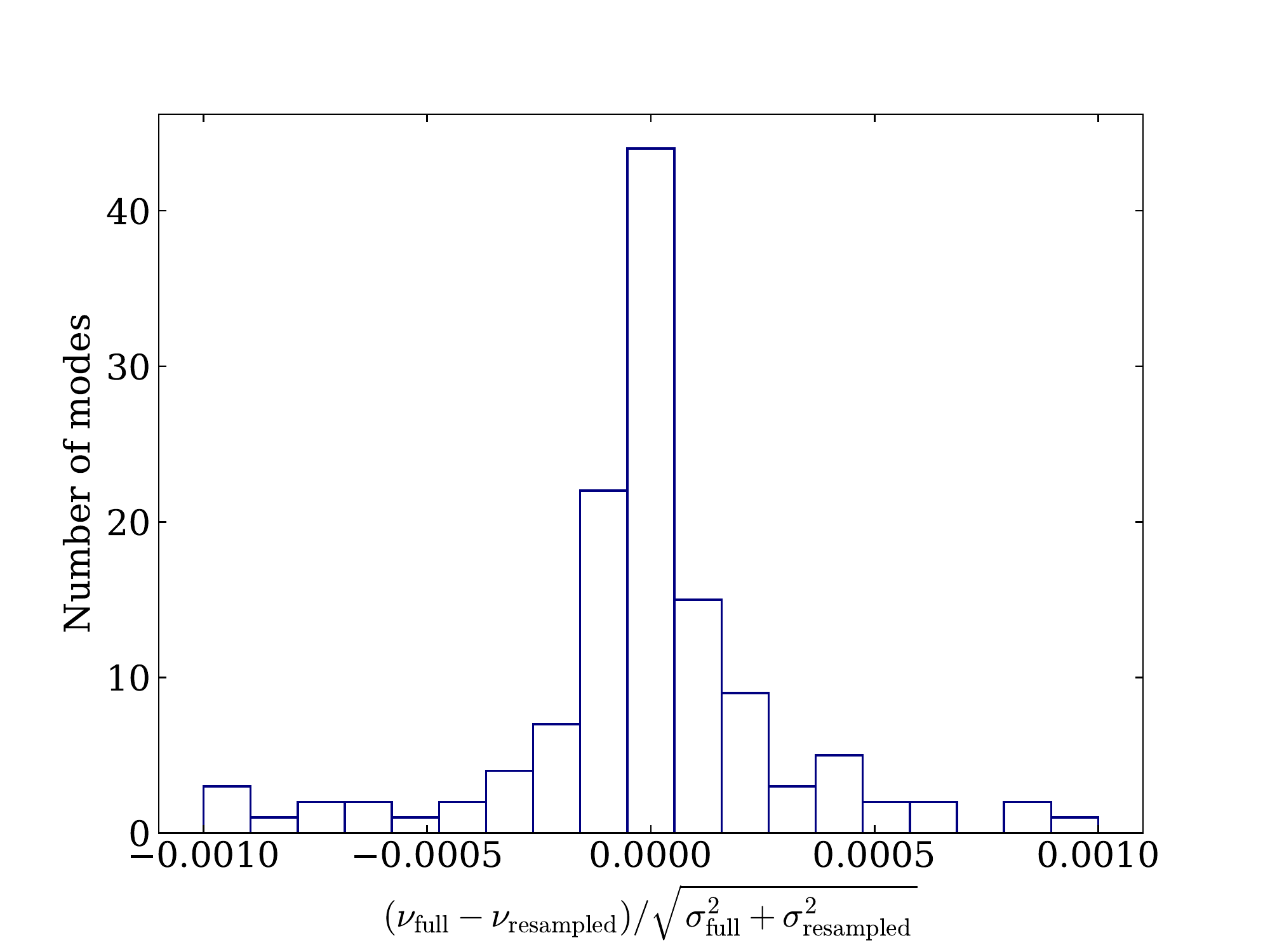}
    \caption{$(\nu_\mathrm{full} - \nu_\mathrm{resampled}) / \sqrt{\sigma_\mathrm{full}^2 + \sigma_\mathrm{resampled}^2}$ distribution for the six targets considered in the LEGACY sample.}
    \label{fig:comparison_quickfit_noquickfit}
\end{figure}

Finally, we represent in Fig.~\ref{fig:comparison_quickfit_noquickfit} the $(\nu_\mathrm{full} - \nu_\mathrm{resampled}) / \sqrt{\sigma_\mathrm{full}^2 + \sigma_\mathrm{resampled}^2}$ distribution, where  $\nu_\mathrm{full}$ and $\sigma_\mathrm{full}$ are the frequencies and uncertainties obtained by considering the complete PSD for the background fit, and $\nu_\mathrm{resampled}$ and $\sigma_\mathrm{resampled}$ are the frequencies and uncertainties obtained after using the background fitted on the resampled PSD. This shows that, for mode frequencies, the discrepancies between the two methods are negligible with respect to the fitted uncertainties, which means that frequencies obtained with the PSD resampling method can be used without risk as input for modelling codes. This resampling strategy is made relevant in the optic of the PLATO mission preparation, with the necessity to perform the peakbagging analysis for tens of thousands of stars in an optimised computing time. 

\section{Discussion \label{section:discussion}}

The possibilities offered by MCMC sampling methods deserve a brief discussion, as well as the flexibility of the possible strategies allowed by \texttt{apollinaire}. 
Another question that arise when using MCMC is the hyper-parameters value that must be chosen in order to correctly dimension the problem. We decided in this paper to use 500 walkers for all the sampling we perform, which is the typical value that is used in several peakbagging papers \citep[see e.g.][]{2016MNRAS.456.2183D,2017ApJ...835..172L}. Choosing the optimal number of steps to perform when sampling the chain and the proportion of steps that should be burned-in at the beginning of the run represents a complex issue that has not a universal solution. This can only be done through test-and-trial, especially when the initial guesses appears to be quite far from the posterior probability optimum. 
One way to check that the chain is fully converged is to split it in half after having removed the burned-in elements and to verify that the shapes of the two sub-distribution are identical.
It is also possible to estimate the auto-correlation time $\tau$ of the chain to assess its convergence. However, obtaining a reliable estimate of $\tau$ requires to run the sampling at least for 50$\tau$ steps, which can prove extremely computationally extensive. Having run a sampling experiment with KIC~6603624 as reference, we find that, when the posterior is data dominated, $\tau$ values are typically of a few hundreds of steps. Parameters for which the posterior is prior dominated typically have longer auto-correlation times.

It is also important to keep in mind that some fitting strategies will have to be preferred considering the target to analyse. For example, for high-resolution solar data we will favour a fit by pair, with the height and the FWHM of each mode set to vary freely. Mode asymmetries should be fitted for modes above 2450 $\mu$Hz. In stellar data, especially with low-resolution low-SNR data, mode will be fitted globally or by order, with a constraint on the mode relative amplitudes. It will be possible to constrain the stellar inclination angle only with a good SNR, in this case it will be necessary to perform a global fit in order to consider the m-height ratio of the largest possible number of modes at the same time. Mode asymmetries are usually not considered when fitting stellar data. The user may keep in mind that when fitting all the modes at once with a global method, 
the MCMC will converge slower due to the increased number of dimensions in the parameter space to explore. For power spectra obtained from short time series, because of the mode stochastic excitation, it may sometimes yield better results to let the mode FWHM and height vary freely than to use the ratios specified in Table~\ref{tab:amplitude_ratio}. 

\section{Conclusion \label{section:conclusion}}

In this paper, we introduced the \apollinaire module, an open-source \texttt{Python} 3 package designed for helio- and asteroseismic disk-integrated peakbagging. 
The module implements a set of function which are able to extract background, p-mode global pattern, and p-mode individual parameters through MCMC sampling implemented with \texttt{emcee}. The implementation of these functions have been thought to provide a flexible framework. They can be used independently or combined depending on the task the user wants to achieve.

We analysed data from the GOLF instrument and the \textit{Kepler} LEGACY sample in order to compare \apollinaire results against values available in the published literature. We find a good agreement between our results and the reference values. The discrepancies can be explained by difference in data calibrations, in fitting strategies and in adopted models.

The code is actively developed and the online repository is regularly updated. An online documentation of the package is available and includes several tutorials that should allow any interested user to quickly start working with \texttt{apollinaire} and use the different functions to analyse they own set of data.
With the increasing number of TESS targets of asteroseismic interest and the preparation of the PLATO mission, we hope that \texttt{apollinaire} will become in the future a widely-used tool in the community.

\begin{acknowledgements}
The authors thanks the anonymous referee for constructive comments and precious suggestions that greatly enriched and completed the material of the paper. 
S.N.B. and R.A.G acknowledge the support from PLATO and GOLF CNES grants. J.B. acknowledges the support from CNES. V.D. acknowledges the support of the IAC (Tenerife, Spain). The authors are deeply grateful to G.R. Davies who originally designed the a2z data structures in collaboration with R.A.G. They also thank L.~Bugnet, M.~Delorme, A.~Jim\'enez, S.~Mathur, and V.~Prat for fruitful discussions. A special thank to A.~Finley for his precious suggestions concerning the name of the paper. 
This paper includes data collected by the \textit{Kepler} mission. Funding for the \textit{Kepler} mission is provided by the NASA Science Mission Directorate. 
This research has made use of the VizieR catalogue access tool, CDS, Strasbourg, France (DOI : 10.26093/cds/vizier). The original description of the VizieR service was published in 2000, A\&AS 143, 23.

\\

\textit{Software:} \texttt{Python} \citep{10.5555/1593511}, \texttt{numpy} \citep{numpy,Harris_2020}, \texttt{pandas} \citep{reback2020pandas, mckinney-proc-scipy-2010}, \texttt{matplotlib} \citep{4160265}, \texttt{emcee} \citep{2013PASP..125..306F}, \texttt{scipy} \citep{2020SciPy-NMeth}, \texttt{corner} \citep{Foreman-Mackey2016}, \texttt{astropy }\citep{astropy:2013, astropy:2018}, \texttt{h5py} \citep{collette_python_hdf5_2014}, \texttt{george} \citep{hodlr}. The previous references should be cited together with \texttt{apollinaire} when using the module.
\\
Packaged version of \texttt{apollinaire} can be installed through \texttt{pip} and \texttt{conda-forge} \citep{conda_forge_community_2015_4774216}.
The source code and most recent versions of \texttt{apollinaire} can be found at \url{gitlab.com/sybreton/apollinaire} while the module documentation is hosted at \url{https://apollinaire.readthedocs.io/}.
\end{acknowledgements}

\bibliographystyle{aa} 
\bibliography{biblio.bib} 

\appendix
\vfill

\section{PSD resampling for background fit \label{appendix:background}}

We included in \texttt{apollinaire} several resampling strategy in order to perform quicker background fits (the computing time of the MCMC process being mainly constrained by the length of the vector on which the posterior probability is computed at each step). We describe here the resampling methods implemented in the module. The user has the choice between two resampling strategy, \texttt{reboxing} and \texttt{advanced\_reboxing}. 

The \texttt{reboxing} strategy uses the \texttt{numpy.logspace} function to create a new frequency vector with logarithmic frequency spacing and a number of data points specified by the user. The logarithmic spacing is preferred over a linear spacing in order to increase the proportion of low-frequency data-point in the fit. 
Each bin of the original PSD is then attributed to a box considering the closest frequency bin in the new frequency vector. 
The median of each box is then considered in order to compute the resampled PSD. 

In order to apply a distinct resampling to the p-mode region, the \texttt{advanced\_reboxing} extends this approach with a more subtle resampling techniques. In this case, the user specifies a target number $N$ of data points in the resampled PSD outside the $\nu_\mathrm{max}$ region and a number $M$ of data points in the $\nu_\mathrm{max}$ region. Considering consecutive frequency bins $\nu_k$ and $\nu_{k+1}$, they are directly put in the resampled frequency vector under the condition
\begin{equation}
    \frac{\nu_{k+1}}{\nu_k} > 10^{\frac{1}{N}} \; ,
\end{equation}
We define $k_0$ as the frequency bin verifying:
\begin{equation}
k_0 = \min_k \left\{ \frac{\nu_{k+1}}{\nu_k}  < 10^{\frac{1}{N}} \right\} \; ,
\end{equation} 
and three frequency intervals $I_j = [\nu_a, \nu_b]$ such as
\begin{align}
    I_1 &= \left[\nu_{k_0}, \frac{2}{3}\nu_\mathrm{max}\right] \; , \\
    I_2 &= \left[\frac{2}{3}\nu_\mathrm{max}, \frac{3}{2}\nu_\mathrm{max}\right] \; , \\
    I_3 &= \left[\frac{3}{2}\nu_\mathrm{max}, \nu_\mathrm{Nyquist}\right] \; .
\end{align}
For each frequency interval, the box size parameter $S_j$ is
\begin{equation}
    S_j = \left(\frac{\nu_b}{\nu_a}\right)^{\frac{1}{N_j}} \; ,
\end{equation}
with 
\begin{align}
    N_1 &=  \frac{N}{C} \log \left( \frac{2\nu_\mathrm{max}}{3\nu_{k_0}} \right) \; , \\
    N_2 &=  M \; , \\
    N_3 &=  \frac{N}{C} \log \left( \frac{3\nu_\mathrm{Nyquist}}{2\nu_\mathrm{max}} \right) \; ,
\end{align}
and the parameter $C$ is given by
\begin{equation}
    C = \log \left( \frac{4}{9}\frac{\nu_\mathrm{Nyquist}}{\nu_{k_0}} \right) \; .
\end{equation}
The bounds of the consecutive boxes in the $I_j$ interval are then $[\nu_a, S_j \nu_a]$, $[S_j \nu_a, S_j^2 \nu_a]$, ..., $[S_j^{N_j-1} \nu_a, \nu_b]$. 
Inside each box, the new PSD bin is taken as the median of the PSD values in the box while the corresponding new frequency bin is computed by considering the geometric means of the frequencies inside the box
\begin{equation}
    \nu_\mathrm{box} = 10^{\frac{1}{n_\mathrm{bin}}\sum\limits_{i=1}^{n_\mathrm{bin}}{\log \nu_i}} \; ,
\end{equation}
with $n_\mathrm{bin}$ the number of frequency bins inside the box.
\section{Subgiant mixed-mode fitting recipe \label{appendix:subgiant_fit}}


Although \texttt{apollinaire} was not initially designed to perform the automatic analysis of subgiant stars, it is possible to exploit the \texttt{global} fitting strategy in order to characterise stars with an important number of mixed modes. In this appendix we propose a recipe to analyse such stars. 
The user who would like to use this recipe should be aware that in its current version, \texttt{apollinaire} provides no method to automatically compute guesses for mixed-mode parameters. Such methods were presented, for example, in \citet{2015A&A...584A..50M} or \citet{2020A&A...642A.226A}. 

\begin{figure*}[!htb]
    \centering
    \includegraphics[width = 1 \textwidth]{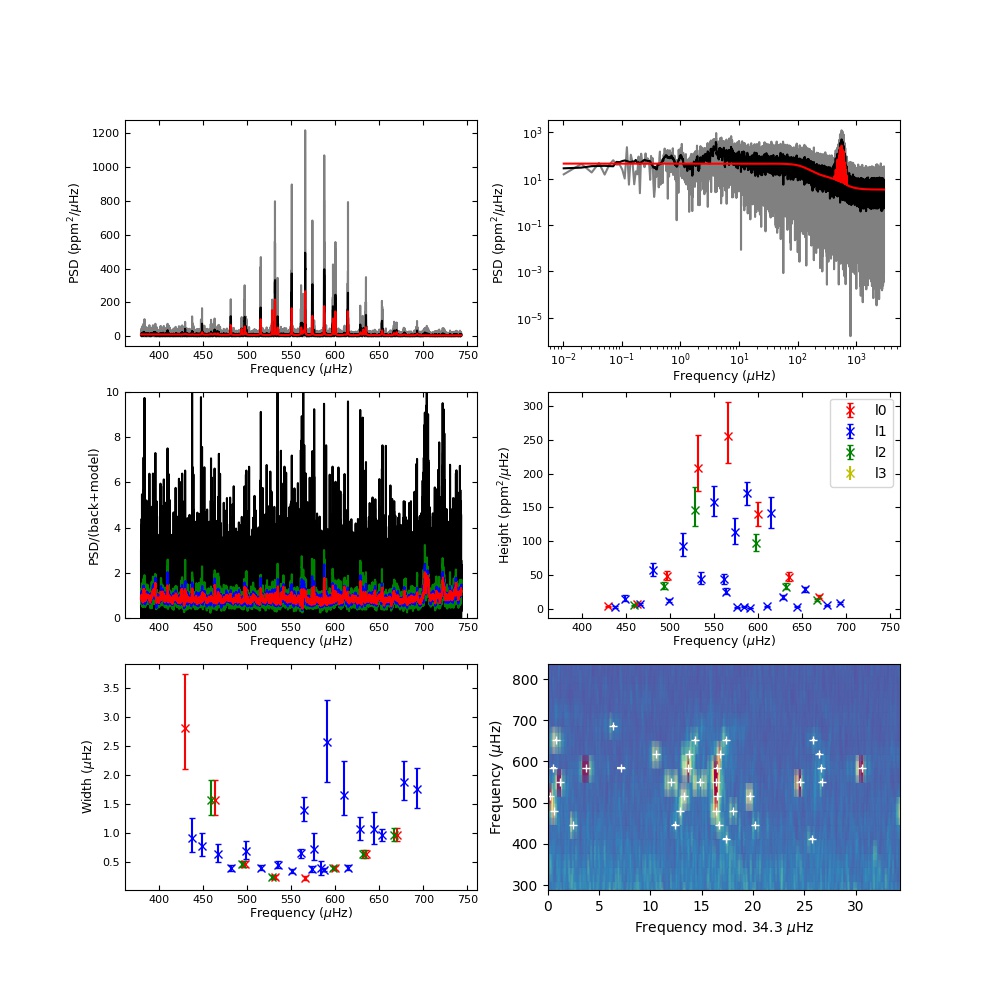}
    \caption{Summary of the peakbagging of the subgiant KIC~5723165. \textit{Top left:} original PSD as a function of frequency in a linear scale around $\nu_{\rm{max}}$ (grey), a smoothed PSD (black), and the resultant fit (red). \textit{Top right:} Same three PSDs in a log-log scale. \textit{Central left:} Residual of PSD (in S/N) after removing the fitted model. The colour lines represent different levels of smoothing. \textit{Central right:} Fitted mode heights as a function of the frequency. \textit{Bottom left:} Fitted mode widths as a function of Frequency. \textit{Bottom right:} Echelle diagram with a $\Delta\nu$ of 34.3 $\mu$Hz. Crosses indicate the fitted frequencies. This figure is has been produced as one of \texttt{apollinaire} standard outputs.}
    \label{fig:appB}
\end{figure*}

We provide here an example\footnote{This example is available on the same repository than the benchmark presented in Section~\ref{section:benchmark}: \url{https://gitlab.com/sybreton/benchmark_peakbagging_apollinaire/-/tree/master/subgiant_recipe/5723165}} of such a fit performed on KIC~5723165, a subgiant star observed by \textit{Kepler} in short cadence in Q1, Q5, and continuously from Q7 to Q13 for a total of $\sim$760 days \citep[e.g.][]{2020A&A...642A.226A}, which has not been peakbagged yet. The \texttt{apollinaire}  KIC~5723165 peakbagging summary plot is shown in Fig.~\ref{fig:appB}. 

The sequence of steps to be followed in order to apply the recipe is therefore:
\begin{enumerate}
    \item We start by fitting the background with the \texttt{explore\_distribution\_background} function, including the fitting of a Gaussian function to take into account the power hump of the p and mixed modes. 
    \item The presence of mixed modes can perturb the analysis of the mode pattern and the calculation of $\Delta\nu$. Therefore, after dividing the PSD by the background to work in S/N, we remove the region of the odd modes from the PSD prior to fit the universal pattern using only the even modes. To do so, we provide a Jupyter notebook in the documentation with an easy procedure to select the regions of the even modes by directly clicking on the PSD or by providing a list of frequencies determining the boudaries of these bands. Then, a median clipping of the n orders selected by the user around $\nu_{\rm{max}}$ is done. To help performing this manual selection, the user can use $\nu_{\rm{max}}$ obtained from the fitting of the Gaussian when the background is determined. A first estimation of $\Delta\nu$ can then be obtained using the seismic scaling relations.
    \item After median clipping the region of the odd modes, the fitting of the universal pattern can be performed using the masked PSD (see Fig.\ref{fig:appB2}). In general, if the masking is well done, the standard guesses of the universal pattern provided by \texttt{apollinaire} would be good enough. However, the user can try to change these original guesses and bounds for $\delta\nu_{0,2}$, as they are too low for subgiants. A good choice could be $\sim$4~$\mu$Hz, and the upper bound should be higher than 6~$\mu$Hz. The upper bound for the mode width $\Gamma$ could also be lowered. Indeed, it is important to note that having a small $\delta\nu_{0,2}$ guess value and a large one for $\Gamma$ often leads to an inaccurate fit, as the $\ell=0$ and $\ell=2$ are considered as part of the same mode. 
   
   \begin{figure}[!htb]
    \centering
    \includegraphics[width = 0.48 \textwidth]{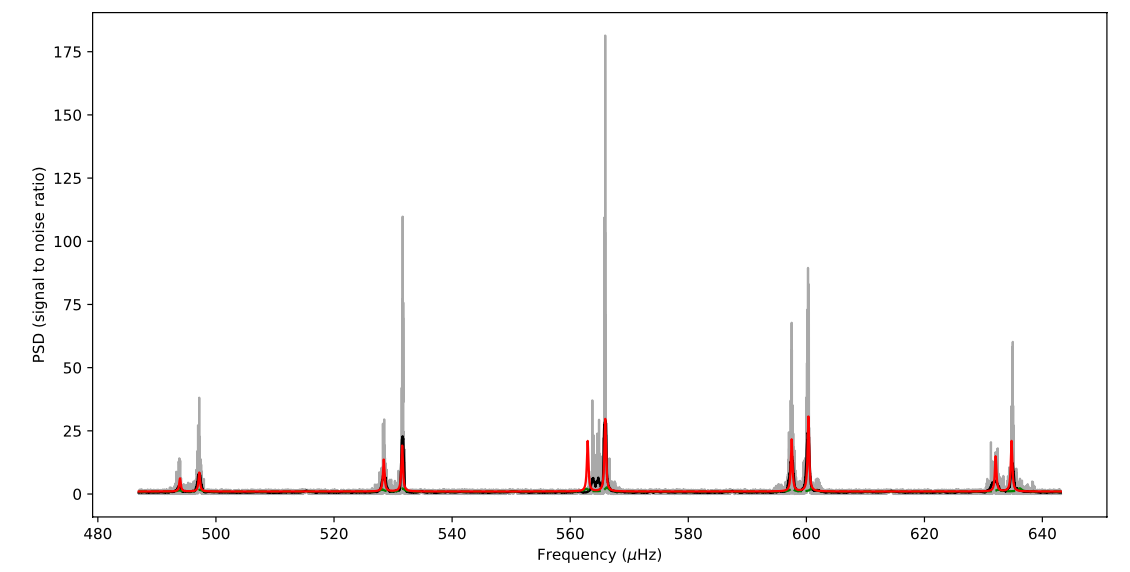}
    \caption{PSD of KIC~5723165 with the regions of the odd modes masked (grey line) and a smoothed version of this PSD in black. The red line corresponds to the result of fitting the \texttt{Universal} pattern. The structure between the modes $\ell$=2 and $\ell$=0 around 565 $\mu$Hz is a mixed mode.}
    \label{fig:appB2}
\end{figure}

    \item After fitting the \texttt{Universal} pattern, the code provides an a2z file with the guesses for the even modes to be fitted by the peak bagging module. A first characterization of those modes can be directly done if we are not interested in the odd ones.
    \item To fit the odd modes (see  Fig.~\ref{fig:appB}), the \texttt{a2z} guess file needs to be modified manually. The only way to do the fitting is \texttt{global} and not by \texttt{order} as the mixed modes cannot be assigned to any particular order. In the \texttt{a2z} guess file, we suggest to write the radial order n of each mixed mode starting by 100. The order is not important and these values are given only as an indication to the user that the corresponding modes are mixed modes. We recommend that each mixed mode is fitted with an individual frequency, width, and height. 
\end{enumerate}

The Table~\ref{tab:subgiant_fit} summarises the parameters obtained for the fitted modes. 

\begin{table}
\centering
\caption{Parameters fitted for KIC~5723165 modes. Mixed modes order $n$ and degrees $\ell$ are not inferred in the analysis and therefore not specified.}
\label{tab:subgiant_fit}
\begin{tabular}{lllll}
\toprule
  n &  $\ell$ & $\nu$ ($\mu$Hz) & $H$ (ppm$^2$/$\mu$Hz) & $\Gamma$ ($\mu$Hz) \\
\midrule
 11 &  0 &     $429.46_{-0.25}^{+0.23}$ &             $3.68_{-0.72}^{+0.93}$ &          $2.82_{-0.70}^{+0.93}$ \\
  - &  - &     $437.88_{-0.19}^{+0.23}$ &             $2.78_{-0.93}^{+1.44}$ &          $0.92_{-0.25}^{+0.34}$ \\
  - &  - &     $448.93_{-0.07}^{+0.07}$ &            $14.76_{-3.21}^{+4.86}$ &          $0.78_{-0.18}^{+0.22}$ \\
 11 &  2 &     $458.84_{-0.22}^{+0.21}$ &             $4.97_{-0.74}^{+0.96}$ &          $1.58_{-0.26}^{+0.34}$ \\
 12 &  0 &     $463.15_{-0.10}^{+0.10}$ &             $7.09_{-1.06}^{+1.38}$ &          $1.58_{-0.26}^{+0.34}$ \\
  - &  - &     $466.63_{-0.11}^{+0.12}$ &             $7.05_{-1.68}^{+2.12}$ &          $0.65_{-0.14}^{+0.17}$ \\
  - &  - &     $481.37_{-0.03}^{+0.03}$ &           $57.69_{-8.52}^{+10.44}$ &          $0.40_{-0.05}^{+0.06}$ \\
 12 &  2 &     $493.68_{-0.05}^{+0.05}$ &            $34.18_{-4.53}^{+5.27}$ &          $0.46_{-0.05}^{+0.06}$ \\
 13 &  0 &     $497.17_{-0.03}^{+0.03}$ &            $48.82_{-6.46}^{+7.53}$ &          $0.46_{-0.05}^{+0.06}$ \\
  - &  - &     $498.83_{-0.07}^{+0.08}$ &            $11.92_{-2.50}^{+3.24}$ &          $0.69_{-0.12}^{+0.18}$ \\
  - &  - &     $515.34_{-0.02}^{+0.03}$ &          $92.54_{-14.05}^{+19.36}$ &          $0.41_{-0.05}^{+0.05}$ \\
 13 &  2 &     $528.44_{-0.03}^{+0.03}$ &         $146.16_{-24.50}^{+33.48}$ &          $0.25_{-0.03}^{+0.03}$ \\
 14 &  0 &     $531.62_{-0.02}^{+0.02}$ &         $208.80_{-34.99}^{+47.84}$ &          $0.25_{-0.03}^{+0.03}$ \\
  - &  - &     $534.81_{-0.03}^{+0.03}$ &           $43.74_{-6.98}^{+10.05}$ &          $0.45_{-0.06}^{+0.06}$ \\
  - &  - &     $550.63_{-0.02}^{+0.02}$ &         $158.32_{-21.76}^{+23.46}$ &          $0.34_{-0.04}^{+0.04}$ \\
  - &  - &     $561.49_{-0.06}^{+0.06}$ &            $43.43_{-6.48}^{+8.21}$ &          $0.65_{-0.08}^{+0.09}$ \\
  - &  - &     $564.32_{-0.09}^{+0.09}$ &            $25.49_{-3.91}^{+4.68}$ &          $1.40_{-0.19}^{+0.22}$ \\
 15 &  0 &     $565.90_{-0.02}^{+0.02}$ &         $254.75_{-39.71}^{+50.90}$ &          $0.22_{-0.03}^{+0.03}$ \\
  - &  - &     $574.01_{-0.03}^{+0.03}$ &         $113.13_{-17.11}^{+21.66}$ &          $0.38_{-0.05}^{+0.05}$ \\
  - &  - &     $576.16_{-0.14}^{+0.94}$ &             $2.46_{-0.68}^{+0.76}$ &          $0.72_{-0.19}^{+0.28}$ \\
  - &  - &     $584.32_{-0.24}^{+0.19}$ &             $3.27_{-0.81}^{+1.34}$ &          $0.39_{-0.11}^{+0.13}$ \\
  - &  - &     $587.54_{-0.02}^{+0.02}$ &         $171.19_{-17.31}^{+16.61}$ &          $0.37_{-0.03}^{+0.03}$ \\
  - &  - &     $590.94_{-0.39}^{+0.29}$ &             $1.44_{-0.24}^{+0.37}$ &          $2.57_{-0.68}^{+0.73}$ \\
 15 &  2 &     $597.49_{-0.03}^{+0.03}$ &          $97.60_{-11.73}^{+12.75}$ &          $0.39_{-0.03}^{+0.04}$ \\
 16 &  0 &     $600.25_{-0.02}^{+0.02}$ &         $139.42_{-16.76}^{+18.21}$ &          $0.39_{-0.03}^{+0.04}$ \\
  - &  - &     $610.44_{-0.15}^{+0.15}$ &             $3.43_{-0.88}^{+1.21}$ &          $1.65_{-0.34}^{+0.59}$ \\
  - &  - &     $614.38_{-0.03}^{+0.03}$ &         $141.08_{-21.09}^{+24.45}$ &          $0.40_{-0.04}^{+0.05}$ \\
  - &  - &     $628.73_{-0.06}^{+0.06}$ &            $16.91_{-3.05}^{+3.76}$ &          $1.06_{-0.18}^{+0.22}$ \\
 16 &  2 &     $631.93_{-0.06}^{+0.06}$ &            $32.59_{-4.11}^{+5.24}$ &          $0.64_{-0.07}^{+0.07}$ \\
 17 &  0 &     $634.92_{-0.04}^{+0.04}$ &            $46.55_{-5.88}^{+7.49}$ &          $0.64_{-0.07}^{+0.07}$ \\
  - &  - &     $644.58_{-0.18}^{+0.19}$ &             $3.28_{-0.76}^{+0.98}$ &          $1.08_{-0.26}^{+0.29}$ \\
  - &  - &     $653.27_{-0.05}^{+0.05}$ &            $28.53_{-3.54}^{+4.13}$ &          $0.96_{-0.10}^{+0.12}$ \\
 17 &  2 &     $666.81_{-0.08}^{+0.08}$ &            $12.52_{-1.48}^{+1.83}$ &          $0.98_{-0.11}^{+0.12}$ \\
 18 &  0 &     $669.80_{-0.08}^{+0.08}$ &            $17.89_{-2.11}^{+2.62}$ &          $0.98_{-0.11}^{+0.12}$ \\
  - &  - &     $678.35_{-0.14}^{+0.14}$ &             $5.07_{-0.79}^{+0.93}$ &          $1.89_{-0.32}^{+0.35}$ \\
  - &  - &     $693.22_{-0.13}^{+0.14}$ &             $8.04_{-1.23}^{+1.41}$ &          $1.76_{-0.32}^{+0.38}$ \\
\bottomrule
\end{tabular}
\end{table}

\section{Detailed results from \textit{Kepler} LEGACY targets}

The detailed results of the p-mode individual parameters fitted for the benchmark on the six LEGACY targets performed in Sect.~\ref{section:benchmark_legacy} are presented here. Table~\ref{tab:stellar_hw} presents the comparison of $H$ and $\Gamma$ between \apollinaire and L17. As L17 chose to fit amplitude rather than heights, the corresponding mode heights and uncertainties have been recomputed through the proxy of their Eq.~6. Table~\ref{tab:stellar_frequency} presents the comparison of mode frequencies $\nu$ and quality assurance factors $\ln K$ between \apollinaire and L17.

\onecolumn
\begin{longtable}{rrllll}
\caption{Mode heights and widths obtained for modes fitted both in L17 and with \texttt{apollinaire}.}\label{tab:stellar_hw}\\
\toprule
      KIC &   n & $H_\mathtt{apn}$ (ppm$^2/\mu$Hz) & $H_\mathrm{L17}$ (ppm$^2/\mu$Hz) & $\Gamma_\mathtt{apn}$ ($\mu$Hz) & $\Gamma_\mathrm{L17}$ ($\mu$Hz) \\
\midrule
\endhead
\midrule
\multicolumn{6}{r}{{Continued on next page}} \\
\midrule
\endfoot

\bottomrule
\endlastfoot
  5184732 &  17 &        $2.44 \pm 0.06$ &        $2.61 \pm 0.08$ &                 $1.13 \pm 0.11$ &                 $1.35 \pm 0.13$ \\
  5184732 &  18 &        $2.97 \pm 0.06$ &        $2.95 \pm 0.08$ &                 $1.36 \pm 0.11$ &                 $1.53 \pm 0.13$ \\
  5184732 &  19 &        $3.54 \pm 0.07$ &        $3.64 \pm 0.09$ &                 $1.17 \pm 0.09$ &                 $1.24 \pm 0.12$ \\
  5184732 &  20 &        $3.83 \pm 0.07$ &        $3.94 \pm 0.10$ &                 $1.30 \pm 0.09$ &                 $1.26 \pm 0.12$ \\
  5184732 &  21 &        $4.02 \pm 0.08$ &        $4.21 \pm 0.12$ &                 $1.33 \pm 0.09$ &                 $1.47 \pm 0.12$ \\
  5184732 &  22 &        $3.54 \pm 0.06$ &        $3.74 \pm 0.09$ &                 $1.80 \pm 0.14$ &                 $1.83 \pm 0.14$ \\
  5184732 &  23 &        $2.72 \pm 0.06$ &        $2.91 \pm 0.09$ &                 $2.64 \pm 0.24$ &                 $2.45 \pm 0.19$ \\
  6106415 &  17 &        $2.54 \pm 0.05$ &        $2.65 \pm 0.08$ &                 $1.65 \pm 0.12$ &                 $1.64 \pm 0.18$ \\
  6106415 &  18 &        $3.14 \pm 0.05$ &        $3.07 \pm 0.09$ &                 $1.43 \pm 0.09$ &                 $1.77 \pm 0.15$ \\
  6106415 &  19 &        $3.54 \pm 0.06$ &        $3.77 \pm 0.10$ &                 $1.35 \pm 0.08$ &                 $1.52 \pm 0.11$ \\
  6106415 &  20 &        $3.85 \pm 0.06$ &        $3.87 \pm 0.08$ &                 $1.36 \pm 0.07$ &                 $1.43 \pm 0.10$ \\
  6106415 &  21 &        $3.63 \pm 0.06$ &        $3.71 \pm 0.08$ &                 $1.72 \pm 0.10$ &                 $1.94 \pm 0.13$ \\
  6106415 &  22 &        $3.08 \pm 0.05$ &        $3.35 \pm 0.07$ &                 $2.55 \pm 0.17$ &                 $2.54 \pm 0.20$ \\
  6106415 &  23 &        $2.63 \pm 0.04$ &        $2.84 \pm 0.07$ &                 $2.97 \pm 0.22$ &                 $3.34 \pm 0.23$ \\
  6225718 &  16 &        $1.77 \pm 0.05$ &        $1.89 \pm 0.07$ &                 $2.10 \pm 0.22$ &                 $2.19 \pm 0.34$ \\
  6225718 &  17 &        $2.24 \pm 0.05$ &        $2.29 \pm 0.06$ &                 $2.77 \pm 0.23$ &                 $2.84 \pm 0.30$ \\
  6225718 &  18 &        $2.68 \pm 0.05$ &        $2.77 \pm 0.07$ &                 $2.16 \pm 0.15$ &                 $2.67 \pm 0.23$ \\
  6225718 &  19 &        $3.07 \pm 0.05$ &        $3.14 \pm 0.06$ &                 $2.20 \pm 0.14$ &                 $2.50 \pm 0.21$ \\
  6225718 &  20 &        $3.40 \pm 0.05$ &        $3.47 \pm 0.07$ &                 $2.20 \pm 0.12$ &                 $2.22 \pm 0.16$ \\
  6225718 &  21 &        $3.32 \pm 0.05$ &        $3.46 \pm 0.07$ &                 $2.22 \pm 0.15$ &                 $2.61 \pm 0.17$ \\
  6225718 &  22 &        $3.28 \pm 0.05$ &        $3.46 \pm 0.07$ &                 $2.76 \pm 0.16$ &                 $3.03 \pm 0.23$ \\
  6603624 &  17 &        $2.22 \pm 0.07$ &        $2.24 \pm 0.11$ &                 $0.82 \pm 0.09$ &                 $0.82 \pm 0.11$ \\
  6603624 &  18 &        $2.99 \pm 0.08$ &        $3.10 \pm 0.12$ &                 $0.61 \pm 0.05$ &                 $0.63 \pm 0.07$ \\
  6603624 &  19 &        $3.58 \pm 0.09$ &        $3.56 \pm 0.12$ &                 $0.59 \pm 0.05$ &                 $0.69 \pm 0.05$ \\
  6603624 &  20 &        $4.04 \pm 0.10$ &        $4.39 \pm 0.13$ &                 $0.57 \pm 0.05$ &                 $0.56 \pm 0.05$ \\
  6603624 &  21 &        $3.58 \pm 0.09$ &        $3.73 \pm 0.13$ &                 $0.69 \pm 0.05$ &                 $0.80 \pm 0.06$ \\
  6603624 &  22 &        $2.98 \pm 0.07$ &        $3.39 \pm 0.11$ &                 $1.16 \pm 0.09$ &                 $1.25 \pm 0.09$ \\
  6603624 &  23 &        $2.05 \pm 0.07$ &        $2.26 \pm 0.10$ &                 $2.23 \pm 0.27$ &                 $1.70 \pm 0.18$ \\
 12069424 &  16 &        $2.17 \pm 0.05$ &        $2.16 \pm 0.07$ &                 $0.93 \pm 0.09$ &                 $1.29 \pm 0.16$ \\
 12069424 &  17 &        $2.63 \pm 0.06$ &        $2.58 \pm 0.09$ &                 $0.89 \pm 0.07$ &                 $1.04 \pm 0.12$ \\
 12069424 &  18 &        $3.24 \pm 0.07$ &        $3.29 \pm 0.09$ &                 $0.84 \pm 0.05$ &                 $1.00 \pm 0.08$ \\
 12069424 &  19 &        $4.17 \pm 0.08$ &        $3.85 \pm 0.11$ &                 $0.67 \pm 0.04$ &                 $0.82 \pm 0.06$ \\
 12069424 &  20 &        $4.20 \pm 0.09$ &        $4.03 \pm 0.12$ &                 $0.78 \pm 0.05$ &                 $1.22 \pm 0.08$ \\
 12069424 &  21 &        $3.59 \pm 0.07$ &        $3.61 \pm 0.10$ &                 $1.15 \pm 0.07$ &                 $1.24 \pm 0.09$ \\
 12069424 &  22 &        $2.80 \pm 0.05$ &        $2.92 \pm 0.07$ &                 $1.97 \pm 0.13$ &                 $1.96 \pm 0.12$ \\
 12069449 &  17 &        $1.88 \pm 0.05$ &        $1.89 \pm 0.07$ &                 $0.92 \pm 0.09$ &                 $0.96 \pm 0.11$ \\
 12069449 &  18 &        $2.54 \pm 0.06$ &        $2.49 \pm 0.08$ &                 $0.84 \pm 0.06$ &                 $1.13 \pm 0.10$ \\
 12069449 &  19 &        $3.05 \pm 0.07$ &        $3.00 \pm 0.08$ &                 $0.72 \pm 0.05$ &                 $0.77 \pm 0.08$ \\
 12069449 &  20 &        $3.51 \pm 0.08$ &        $3.41 \pm 0.10$ &                 $0.68 \pm 0.05$ &                 $0.94 \pm 0.08$ \\
 12069449 &  21 &        $3.37 \pm 0.07$ &        $3.55 \pm 0.11$ &                 $0.86 \pm 0.06$ &                 $1.01 \pm 0.10$ \\
 12069449 &  22 &        $2.77 \pm 0.05$ &        $2.97 \pm 0.10$ &                 $1.35 \pm 0.08$ &                 $1.40 \pm 0.09$ \\
 12069449 &  23 &        $2.13 \pm 0.04$ &        $2.28 \pm 0.07$ &                 $2.39 \pm 0.16$ &                 $2.11 \pm 0.13$ \\
\end{longtable}

\begin{longtable}{rrrllll} 
\caption{Modes frequencies and quality assurances factor $\ln K$ obtained for modes fitted both in L17 and with \texttt{apollinaire}.} \label{tab:stellar_frequency} \\
\toprule
      KIC &   n &  $\ell$ & $\nu_\mathtt{apn}$ ($\mu$Hz) & $\nu_\mathrm{L17}$ ($\mu$Hz) & $\ln K_\mathtt{apn}$ & $\ln K_\mathrm{L17}$ \\
\midrule
\endhead
\midrule
\multicolumn{7}{r}{{Continued on next page}} \\
\midrule
\endfoot

\bottomrule
\endlastfoot
  5184732 &  16 &  2 &           $1750.31 \pm 0.16$ &           $1750.26 \pm 0.19$ &                            > 6 &         > 6 \\
  5184732 &  17 &  0 &           $1756.69 \pm 0.06$ &           $1756.69 \pm 0.07$ &                            > 6 &         > 6 \\
  5184732 &  17 &  1 &           $1800.54 \pm 0.07$ &           $1800.50 \pm 0.08$ &                            > 6 &         > 6 \\
  5184732 &  17 &  2 &           $1844.66 \pm 0.12$ &           $1844.74 \pm 0.15$ &                            > 6 &         > 6 \\
  5184732 &  18 &  0 &           $1851.08 \pm 0.07$ &           $1851.15 \pm 0.08$ &                            > 6 &         > 6 \\
  5184732 &  18 &  1 &           $1895.61 \pm 0.07$ &           $1895.59 \pm 0.06$ &                            > 6 &         > 6 \\
  5184732 &  18 &  2 &           $1940.39 \pm 0.09$ &           $1940.45 \pm 0.13$ &                            > 6 &         > 6 \\
  5184732 &  18 &  3 &           $1981.70 \pm 0.85$ &           $1982.10 \pm 0.74$ &                            > 6 &        2.62 \\
  5184732 &  19 &  0 &           $1946.67 \pm 0.06$ &           $1946.65 \pm 0.06$ &                            > 6 &         > 6 \\
  5184732 &  19 &  1 &           $1991.62 \pm 0.06$ &           $1991.57 \pm 0.06$ &                            > 6 &         > 6 \\
  5184732 &  19 &  2 &           $2036.29 \pm 0.09$ &           $2036.35 \pm 0.10$ &                            > 6 &         > 6 \\
  5184732 &  19 &  3 &           $2076.94 \pm 0.85$ &           $2076.67 \pm 0.55$ &                            > 6 &        1.56 \\
  5184732 &  20 &  0 &           $2042.36 \pm 0.06$ &           $2042.31 \pm 0.06$ &                            > 6 &         > 6 \\
  5184732 &  20 &  1 &           $2087.47 \pm 0.06$ &           $2087.45 \pm 0.06$ &                            > 6 &         > 6 \\
  5184732 &  20 &  2 &           $2132.34 \pm 0.08$ &           $2132.34 \pm 0.10$ &                            > 6 &         > 6 \\
  5184732 &  20 &  3 &           $2172.49 \pm 0.63$ &           $2173.09 \pm 0.60$ &                            > 6 &         > 6 \\
  5184732 &  21 &  0 &           $2138.20 \pm 0.06$ &           $2138.21 \pm 0.06$ &                            > 6 &         > 6 \\
  5184732 &  21 &  1 &           $2183.39 \pm 0.07$ &           $2183.32 \pm 0.07$ &                            > 6 &         > 6 \\
  5184732 &  21 &  2 &           $2227.94 \pm 0.12$ &           $2227.82 \pm 0.13$ &                            > 6 &         > 6 \\
  5184732 &  21 &  3 &           $2269.33 \pm 0.77$ &           $2269.23 \pm 0.57$ &                            > 6 &        3.44 \\
  5184732 &  22 &  0 &           $2233.54 \pm 0.07$ &           $2233.51 \pm 0.09$ &                            > 6 &         > 6 \\
  5184732 &  22 &  1 &           $2279.06 \pm 0.08$ &           $2279.09 \pm 0.09$ &                            > 6 &         > 6 \\
  5184732 &  22 &  2 &           $2323.78 \pm 0.22$ &           $2324.03 \pm 0.20$ &                            > 6 &         > 6 \\
  5184732 &  22 &  3 &           $2365.14 \pm 1.07$ &           $2365.48 \pm 0.88$ &                        3.96081 &        3.31 \\
  5184732 &  23 &  0 &           $2328.96 \pm 0.13$ &           $2329.03 \pm 0.13$ &                            > 6 &         > 6 \\
  5184732 &  23 &  1 &           $2375.03 \pm 0.11$ &           $2375.03 \pm 0.12$ &                            > 6 &         > 6 \\
  6106415 &  16 &  2 &           $1902.71 \pm 0.18$ &           $1902.55 \pm 0.18$ &                            > 6 &         > 6 \\
  6106415 &  17 &  0 &           $1909.90 \pm 0.07$ &           $1909.94 \pm 0.09$ &                            > 6 &         > 6 \\
  6106415 &  17 &  1 &           $1957.30 \pm 0.08$ &           $1957.32 \pm 0.09$ &                            > 6 &         > 6 \\
  6106415 &  17 &  2 &           $2005.58 \pm 0.09$ &           $2005.66 \pm 0.13$ &                            > 6 &         > 6 \\
  6106415 &  18 &  0 &           $2013.01 \pm 0.06$ &           $2013.06 \pm 0.09$ &                            > 6 &         > 6 \\
  6106415 &  18 &  1 &           $2061.47 \pm 0.06$ &           $2061.47 \pm 0.07$ &                            > 6 &         > 6 \\
  6106415 &  18 &  2 &           $2110.17 \pm 0.07$ &           $2110.19 \pm 0.11$ &                            > 6 &         > 6 \\
  6106415 &  19 &  0 &           $2117.21 \pm 0.05$ &           $2117.35 \pm 0.07$ &                            > 6 &         > 6 \\
  6106415 &  19 &  1 &           $2165.88 \pm 0.06$ &           $2165.91 \pm 0.07$ &                            > 6 &         > 6 \\
  6106415 &  19 &  2 &           $2214.42 \pm 0.07$ &           $2214.39 \pm 0.10$ &                            > 6 &         > 6 \\
  6106415 &  19 &  3 &           $2259.49 \pm 0.75$ &           $2259.15 \pm 0.52$ &                            > 6 &        3.87 \\
  6106415 &  20 &  0 &           $2221.52 \pm 0.05$ &           $2221.52 \pm 0.06$ &                            > 6 &         > 6 \\
  6106415 &  20 &  1 &           $2270.33 \pm 0.06$ &           $2270.39 \pm 0.07$ &                            > 6 &         > 6 \\
  6106415 &  20 &  2 &           $2318.76 \pm 0.09$ &           $2318.82 \pm 0.12$ &                            > 6 &         > 6 \\
  6106415 &  20 &  3 &           $2363.38 \pm 0.67$ &           $2363.96 \pm 0.47$ &                            > 6 &         > 6 \\
  6106415 &  21 &  0 &           $2325.66 \pm 0.07$ &           $2325.64 \pm 0.09$ &                            > 6 &         > 6 \\
  6106415 &  21 &  1 &           $2374.53 \pm 0.07$ &           $2374.55 \pm 0.08$ &                            > 6 &         > 6 \\
  6106415 &  21 &  2 &           $2422.96 \pm 0.17$ &           $2423.11 \pm 0.21$ &                            > 6 &         > 6 \\
  6106415 &  21 &  3 &           $2467.07 \pm 1.36$ &           $2467.17 \pm 0.64$ &                            > 6 &         > 6 \\
  6106415 &  22 &  0 &           $2429.76 \pm 0.08$ &           $2429.81 \pm 0.11$ &                            > 6 &         > 6 \\
  6106415 &  22 &  1 &           $2478.93 \pm 0.09$ &           $2478.93 \pm 0.11$ &                            > 6 &         > 6 \\
  6106415 &  22 &  2 &           $2527.92 \pm 0.17$ &           $2528.38 \pm 0.23$ &                            > 6 &         > 6 \\
  6106415 &  22 &  3 &           $2571.93 \pm 1.14$ &           $2572.76 \pm 0.87$ &                            > 6 &         > 6 \\
  6106415 &  23 &  0 &           $2534.00 \pm 0.11$ &           $2534.02 \pm 0.15$ &                            > 6 &         > 6 \\
  6106415 &  23 &  1 &           $2583.95 \pm 0.12$ &           $2583.85 \pm 0.14$ &                            > 6 &         > 6 \\
  6225718 &  15 &  2 &           $1816.15 \pm 0.72$ &           $1816.19 \pm 0.36$ &                      0.0717307 &         > 6 \\
  6225718 &  16 &  0 &           $1825.53 \pm 0.11$ &           $1825.41 \pm 0.13$ &                            > 6 &         > 6 \\
  6225718 &  16 &  1 &           $1873.79 \pm 0.12$ &           $1873.88 \pm 0.14$ &                            > 6 &         > 6 \\
  6225718 &  16 &  2 &           $1920.02 \pm 0.23$ &           $1919.97 \pm 0.26$ &                        2.71076 &         > 6 \\
  6225718 &  17 &  0 &           $1929.03 \pm 0.11$ &           $1929.05 \pm 0.14$ &                            > 6 &         > 6 \\
  6225718 &  17 &  1 &           $1977.27 \pm 0.11$ &           $1977.35 \pm 0.12$ &                            > 6 &         > 6 \\
  6225718 &  17 &  2 &           $2023.91 \pm 0.17$ &           $2023.80 \pm 0.22$ &                            > 6 &         > 6 \\
  6225718 &  18 &  0 &           $2032.68 \pm 0.09$ &           $2032.68 \pm 0.11$ &                            > 6 &         > 6 \\
  6225718 &  18 &  1 &           $2081.51 \pm 0.09$ &           $2081.57 \pm 0.09$ &                            > 6 &         > 6 \\
  6225718 &  18 &  2 &           $2128.59 \pm 0.13$ &           $2128.62 \pm 0.16$ &                            > 6 &         > 6 \\
  6225718 &  19 &  0 &           $2137.54 \pm 0.08$ &           $2137.59 \pm 0.10$ &                            > 6 &         > 6 \\
  6225718 &  19 &  1 &           $2186.90 \pm 0.08$ &           $2186.89 \pm 0.09$ &                            > 6 &         > 6 \\
  6225718 &  19 &  2 &           $2234.46 \pm 0.12$ &           $2234.70 \pm 0.16$ &                            > 6 &         > 6 \\
  6225718 &  19 &  3 &           $2278.10 \pm 1.23$ &           $2281.61 \pm 3.36$ &                            > 6 &        3.01 \\
  6225718 &  20 &  0 &           $2243.30 \pm 0.07$ &           $2243.42 \pm 0.08$ &                            > 6 &         > 6 \\
  6225718 &  20 &  1 &           $2292.94 \pm 0.08$ &           $2293.05 \pm 0.09$ &                            > 6 &         > 6 \\
  6225718 &  20 &  2 &           $2340.61 \pm 0.16$ &           $2340.63 \pm 0.17$ &                            > 6 &         > 6 \\
  6225718 &  20 &  3 &           $2384.81 \pm 1.38$ &           $2385.57 \pm 1.16$ &                            > 6 &        3.94 \\
  6225718 &  21 &  0 &           $2349.60 \pm 0.07$ &           $2349.64 \pm 0.09$ &                            > 6 &         > 6 \\
  6225718 &  21 &  1 &           $2399.38 \pm 0.08$ &           $2399.39 \pm 0.10$ &                            > 6 &         > 6 \\
  6225718 &  21 &  2 &           $2446.78 \pm 0.14$ &           $2446.71 \pm 0.16$ &                            > 6 &         > 6 \\
  6225718 &  21 &  3 &           $2490.21 \pm 1.59$ &           $2493.08 \pm 1.64$ &                            > 6 &        3.66 \\
  6225718 &  22 &  0 &           $2455.54 \pm 0.09$ &           $2455.69 \pm 0.11$ &                            > 6 &         > 6 \\
  6225718 &  22 &  1 &           $2505.31 \pm 0.10$ &           $2505.34 \pm 0.11$ &                            > 6 &         > 6 \\
  6603624 &  16 &  2 &           $2030.98 \pm 0.10$ &           $2030.97 \pm 0.10$ &                            > 6 &         > 6 \\
  6603624 &  17 &  0 &           $2037.05 \pm 0.05$ &           $2036.98 \pm 0.06$ &                            > 6 &         > 6 \\
  6603624 &  17 &  1 &           $2088.30 \pm 0.04$ &           $2088.32 \pm 0.05$ &                            > 6 &         > 6 \\
  6603624 &  17 &  2 &           $2141.08 \pm 0.04$ &           $2141.09 \pm 0.04$ &                            > 6 &         > 6 \\
  6603624 &  18 &  0 &           $2146.75 \pm 0.03$ &           $2146.76 \pm 0.04$ &                            > 6 &         > 6 \\
  6603624 &  18 &  1 &           $2198.48 \pm 0.04$ &           $2198.47 \pm 0.04$ &                            > 6 &         > 6 \\
  6603624 &  18 &  2 &           $2251.61 \pm 0.04$ &           $2251.60 \pm 0.04$ &                            > 6 &         > 6 \\
  6603624 &  18 &  3 &           $2299.52 \pm 0.20$ &           $2305.47 \pm 0.13$ &                            > 6 &         > 6 \\
  6603624 &  19 &  0 &           $2256.99 \pm 0.04$ &           $2257.01 \pm 0.04$ &                            > 6 &         > 6 \\
  6603624 &  19 &  1 &           $2308.95 \pm 0.03$ &           $2309.00 \pm 0.03$ &                            > 6 &         > 6 \\
  6603624 &  19 &  2 &           $2362.00 \pm 0.03$ &           $2362.03 \pm 0.04$ &                            > 6 &         > 6 \\
  6603624 &  19 &  3 &           $2410.02 \pm 0.25$ &           $2410.08 \pm 0.15$ &                            > 6 &         > 6 \\
  6603624 &  20 &  0 &           $2367.06 \pm 0.03$ &           $2367.08 \pm 0.03$ &                            > 6 &         > 6 \\
  6603624 &  20 &  1 &           $2419.40 \pm 0.03$ &           $2419.44 \pm 0.03$ &                            > 6 &         > 6 \\
  6603624 &  20 &  2 &           $2472.42 \pm 0.04$ &           $2472.38 \pm 0.05$ &                            > 6 &         > 6 \\
  6603624 &  20 &  3 &           $2520.40 \pm 0.16$ &           $2520.44 \pm 0.20$ &                            > 6 &         > 6 \\
  6603624 &  21 &  0 &           $2477.07 \pm 0.04$ &           $2477.04 \pm 0.04$ &                            > 6 &         > 6 \\
  6603624 &  21 &  1 &           $2529.69 \pm 0.04$ &           $2529.66 \pm 0.04$ &                            > 6 &         > 6 \\
  6603624 &  21 &  2 &           $2583.03 \pm 0.07$ &           $2583.03 \pm 0.07$ &                            > 6 &         > 6 \\
  6603624 &  21 &  3 &           $2631.27 \pm 0.62$ &           $2631.48 \pm 0.35$ &                         4.0125 &        1.89 \\
  6603624 &  22 &  0 &           $2587.51 \pm 0.06$ &           $2587.51 \pm 0.07$ &                            > 6 &         > 6 \\
  6603624 &  22 &  1 &           $2640.33 \pm 0.06$ &           $2640.36 \pm 0.07$ &                            > 6 &         > 6 \\
  6603624 &  22 &  2 &           $2693.81 \pm 0.16$ &           $2693.89 \pm 0.12$ &                            > 6 &         > 6 \\
  6603624 &  22 &  3 &           $2742.30 \pm 5.26$ &           $2743.55 \pm 1.43$ &                        2.55786 &        1.14 \\
  6603624 &  23 &  0 &           $2698.06 \pm 0.16$ &           $2698.23 \pm 0.13$ &                            > 6 &         > 6 \\
  6603624 &  23 &  1 &           $2751.56 \pm 0.14$ &           $2751.34 \pm 0.18$ &                            > 6 &         > 6 \\
 12069424 &  15 &  2 &           $1795.69 \pm 0.09$ &           $1795.84 \pm 0.13$ &                            > 6 &         > 6 \\
 12069424 &  16 &  0 &           $1802.28 \pm 0.06$ &           $1802.35 \pm 0.08$ &                            > 6 &         > 6 \\
 12069424 &  16 &  1 &           $1849.00 \pm 0.05$ &           $1849.01 \pm 0.06$ &                            > 6 &         > 6 \\
 12069424 &  16 &  2 &           $1898.29 \pm 0.10$ &           $1898.40 \pm 0.11$ &                            > 6 &         > 6 \\
 12069424 &  17 &  0 &           $1904.58 \pm 0.05$ &           $1904.52 \pm 0.06$ &                            > 6 &         > 6 \\
 12069424 &  17 &  1 &           $1951.99 \pm 0.05$ &           $1952.01 \pm 0.05$ &                            > 6 &         > 6 \\
 12069424 &  17 &  2 &           $2001.69 \pm 0.06$ &           $2001.59 \pm 0.09$ &                            > 6 &         > 6 \\
 12069424 &  17 &  3 &           $2045.77 \pm 0.29$ &           $2045.85 \pm 0.38$ &                            > 6 &         > 6 \\
 12069424 &  18 &  0 &           $2007.55 \pm 0.04$ &           $2007.54 \pm 0.05$ &                            > 6 &         > 6 \\
 12069424 &  18 &  1 &           $2055.51 \pm 0.04$ &           $2055.49 \pm 0.05$ &                            > 6 &         > 6 \\
 12069424 &  18 &  2 &           $2105.32 \pm 0.04$ &           $2105.37 \pm 0.06$ &                            > 6 &         > 6 \\
 12069424 &  18 &  3 &           $2149.88 \pm 0.12$ &           $2150.06 \pm 0.22$ &                            > 6 &         > 6 \\
 12069424 &  19 &  0 &           $2110.93 \pm 0.03$ &           $2110.95 \pm 0.04$ &                            > 6 &         > 6 \\
 12069424 &  19 &  1 &           $2159.14 \pm 0.04$ &           $2159.15 \pm 0.05$ &                            > 6 &         > 6 \\
 12069424 &  19 &  2 &           $2208.91 \pm 0.04$ &           $2208.93 \pm 0.07$ &                            > 6 &         > 6 \\
 12069424 &  19 &  3 &           $2253.57 \pm 0.12$ &           $2253.80 \pm 0.25$ &                            > 6 &         > 6 \\
 12069424 &  20 &  0 &           $2214.21 \pm 0.04$ &           $2214.23 \pm 0.05$ &                            > 6 &         > 6 \\
 12069424 &  20 &  1 &           $2262.55 \pm 0.04$ &           $2262.56 \pm 0.05$ &                            > 6 &         > 6 \\
 12069424 &  20 &  2 &           $2312.52 \pm 0.06$ &           $2312.50 \pm 0.08$ &                            > 6 &         > 6 \\
 12069424 &  20 &  3 &           $2357.37 \pm 0.20$ &           $2357.50 \pm 0.23$ &                            > 6 &         > 6 \\
 12069424 &  21 &  0 &           $2317.31 \pm 0.05$ &           $2317.28 \pm 0.06$ &                            > 6 &         > 6 \\
 12069424 &  21 &  1 &           $2366.26 \pm 0.06$ &           $2366.24 \pm 0.06$ &                            > 6 &         > 6 \\
 12069424 &  21 &  2 &           $2416.10 \pm 0.10$ &           $2416.25 \pm 0.12$ &                            > 6 &         > 6 \\
 12069424 &  21 &  3 &           $2461.44 \pm 0.37$ &           $2461.45 \pm 0.38$ &                            > 6 &         > 6 \\
 12069424 &  22 &  0 &           $2420.85 \pm 0.08$ &           $2420.94 \pm 0.08$ &                            > 6 &         > 6 \\
 12069424 &  22 &  1 &           $2470.29 \pm 0.08$ &           $2470.23 \pm 0.10$ &                            > 6 &         > 6 \\
 12069449 &  16 &  2 &           $2152.35 \pm 0.16$ &           $2152.52 \pm 0.11$ &                            > 6 &         > 6 \\
 12069449 &  17 &  0 &           $2159.55 \pm 0.05$ &           $2159.50 \pm 0.06$ &                            > 6 &         > 6 \\
 12069449 &  17 &  1 &           $2214.16 \pm 0.06$ &           $2214.33 \pm 0.07$ &                            > 6 &         > 6 \\
 12069449 &  17 &  2 &           $2268.87 \pm 0.07$ &           $2269.11 \pm 0.10$ &                            > 6 &         > 6 \\
 12069449 &  17 &  3 &           $2319.33 \pm 0.35$ &           $2318.96 \pm 0.31$ &                            > 6 &         > 6 \\
 12069449 &  18 &  0 &           $2276.00 \pm 0.04$ &           $2275.95 \pm 0.06$ &                            > 6 &         > 6 \\
 12069449 &  18 &  1 &           $2331.19 \pm 0.04$ &           $2331.16 \pm 0.04$ &                            > 6 &         > 6 \\
 12069449 &  18 &  2 &           $2386.23 \pm 0.05$ &           $2386.25 \pm 0.07$ &                            > 6 &         > 6 \\
 12069449 &  18 &  3 &           $2436.50 \pm 0.32$ &           $2436.78 \pm 0.28$ &                            > 6 &         > 6 \\
 12069449 &  19 &  0 &           $2392.73 \pm 0.04$ &           $2392.64 \pm 0.05$ &                            > 6 &         > 6 \\
 12069449 &  19 &  1 &           $2448.24 \pm 0.04$ &           $2448.18 \pm 0.05$ &                            > 6 &         > 6 \\
 12069449 &  19 &  2 &           $2503.51 \pm 0.05$ &           $2503.41 \pm 0.07$ &                            > 6 &         > 6 \\
 12069449 &  19 &  3 &           $2554.11 \pm 0.16$ &           $2554.18 \pm 0.19$ &                            > 6 &         > 6 \\
 12069449 &  20 &  0 &           $2509.70 \pm 0.03$ &           $2509.68 \pm 0.04$ &                            > 6 &         > 6 \\
 12069449 &  20 &  1 &           $2565.40 \pm 0.04$ &           $2565.43 \pm 0.05$ &                            > 6 &         > 6 \\
 12069449 &  20 &  2 &           $2620.51 \pm 0.06$ &           $2620.56 \pm 0.07$ &                            > 6 &         > 6 \\
 12069449 &  20 &  3 &           $2671.86 \pm 0.18$ &           $2671.59 \pm 0.28$ &                            > 6 &         > 6 \\
 12069449 &  21 &  0 &           $2626.44 \pm 0.04$ &           $2626.46 \pm 0.05$ &                            > 6 &         > 6 \\
 12069449 &  21 &  1 &           $2682.37 \pm 0.05$ &           $2682.25 \pm 0.05$ &                            > 6 &         > 6 \\
 12069449 &  21 &  2 &           $2737.73 \pm 0.06$ &           $2737.71 \pm 0.08$ &                            > 6 &         > 6 \\
 12069449 &  21 &  3 &           $2789.17 \pm 0.26$ &           $2789.00 \pm 0.38$ &                            > 6 &         > 6 \\
 12069449 &  22 &  0 &           $2743.32 \pm 0.05$ &           $2743.32 \pm 0.07$ &                            > 6 &         > 6 \\
 12069449 &  22 &  1 &           $2799.75 \pm 0.06$ &           $2799.61 \pm 0.07$ &                            > 6 &         > 6 \\
 12069449 &  22 &  2 &           $2855.52 \pm 0.11$ &           $2855.51 \pm 0.13$ &                            > 6 &         > 6 \\
 12069449 &  22 &  3 &           $2907.03 \pm 0.45$ &           $2906.90 \pm 0.50$ &                            > 6 &         > 6 \\
 12069449 &  23 &  0 &           $2860.73 \pm 0.10$ &           $2860.68 \pm 0.10$ &                            > 6 &         > 6 \\
 12069449 &  23 &  1 &           $2917.81 \pm 0.10$ &           $2917.89 \pm 0.12$ &                            > 6 &         > 6 \\
\end{longtable}

\twocolumn

\section{Input and output files \label{section:format}}

The \apollinaire module is built around a certain number of input and output files which are described in the subsequent sections.

\subsection{The a2z file and the a2z DataFrame}

The syntax of the a2z files has been designed to specify some pieces of information the procedure has to be aware of when dealing with individual mode parameters extraction. A line corresponds to a parameter. The file contains nine columns: order $n$, degree $\ell$, \texttt{name} of the parameter, \texttt{extent}, \texttt{value}, uncertainty $\sigma_\mathrm{sym}$ (see Sect.~\ref{section:sampling_MCMC}),\texttt{fixed} key, \texttt{low bound} and \texttt{up bound} of the parameter.

If a parameter has to apply to every order or degrees, the value to specify in the corresponding column is $a$. The possible parameters \texttt{name} are \texttt{freq, height, width, asym, angle, split}. Parameters with name \texttt{freq, height} or \texttt{width} cannot have $n=a$, parameters with name \texttt{freq} cannot have $\ell=a$. The \texttt{extent} columns reminds the extent of application of a given parameter: \texttt{mode}, \texttt{pair}, \texttt{order}, or \texttt{global}. The \texttt{pair} keyword can only be used when fitting modes by pair. Only \texttt{angle} or \texttt{split} parameters can be set as \texttt{global}. For an input a2z file, the \texttt{value} column specify the value around where the sampler will initialise the walkers that will sample the MCMC. The \texttt{uncertainty} column is relevant only for output a2z files. It has to be read as the $\sigma_y$ values explicited in Eq.~\ref{eq:sigma}. The \texttt{fixed} key column values must be set to 0 or 1 and are managed by the \apollinaire function. Parameters with \texttt{fixed} key 0 at a given step will be fitted while parameters with 1-value will be read as frozen parameters. In input a2z files, the \texttt{low bound} and \texttt{up bound} columns specify the limit values inside which the posterior probability will be sampled. Obviously, the \texttt{up bound} must be greater than the \texttt{low bound}, and the \texttt{value} term must lay inside the defined interval.  

An example of input for solar data is given in Table~\ref{table:example_golf_a2z}. In \texttt{apollinaire}, a2z files are read as a2z DataFrame (\texttt{pandas} DataFrame with the structure specified in the previous paragraph) with the function \texttt{read\_a2z}. 

\begin{table*}
\centering
\caption{An example of helioseismic a2z input file corresponding to priors for the $n=20$ GOLF order.}
\begin{tabular}{llllrrrrrrr}
\hline\hline
 $n$ &  $\ell$ &  name & extent & value & $\sigma_\mathrm{sym}$ & fixed & low bound & up bound \\ 
 \hline
 19 &  2 &  height &    mode &    0.0146 & 0.0000 & 0 &    0.0000 &    0.0732 \\
 19 &  2 &   width &    mode &    0.8585 & 0.0000 & 0 &    0.0000 &    8.0000 \\
 19 &  2 &    asym &    mode &   -0.0082 & 0.0000 & 0 &   -0.2000 &    0.2000 \\
 19 &  2 &   split &    mode &    0.4000 & 0.0000 & 0 &    0.1000 &    0.8000 \\
 20 &  0 &    freq &    mode & 2899.0861 & 0.0000 & 0 & 2897.0861 & 2901.0861 \\
 20 &  0 &  height &    mode &    0.0163 & 0.0000 & 0 &    0.0000 &    0.0814 \\
 20 &  0 &   width &    mode &    0.8585 & 0.0000 & 0 &    0.0000 &    8.0000 \\
 20 &  0 &    asym &    mode &   -0.0082 & 0.0000 & 0 &   -0.2000 &    0.2000 \\
 a &  a &   angle &  global &   90.0000 & 0.0000 & 1 &    0.0000 &   90.0000 \\
\hline
\end{tabular}
\label{table:example_golf_a2z}
\end{table*}

\subsection{The pkb file and the pkb arrays}

In pkb files, each line correspond to a given mode of order $n$ and degree $\ell$. The file contains 14 columns: order $n$, degree $\ell$, mode frequency $\nu$, uncertainty over frequency $\sigma_{\mathrm{approx}, \nu}$, mode height $H$, uncertainty over height $\sigma_{\mathrm{approx}, H}$, mode FWHM $\Gamma$, uncertainty over FWHM $\sigma_{\mathrm{approx}, \Gamma}$, stellar angle $i$, uncertainty over stellar angle $\sigma_{\mathrm{approx}, i}$, mode splitting $s$, uncertainty over mode splitting $\sigma_{\mathrm{approx}, s}$, mode asymmetry $\alpha$, uncertainty over mode asymmetry $\sigma_{\mathrm{approx}, \alpha}$. The way $\sigma_\mathrm{sym}$ are computed is described in Sect.~\ref{section:sampling_MCMC}.

An example of pkb files is presented in Table~\ref{table:example_golf_pkb}.
In \apollinaire procedures, pkb arrays (\texttt{numpy} arrays with the shape specified in the previous paragraph) are usually created by converting a2z DataFrame with the \texttt{a2Z\_to\_pkb} function. Those pkb arrays are used to build the models $S_\mathrm{SN}(\nu)$ which are necessary to compute the likelihood and the posterior probability. 

\begin{table*}
\caption{An example of helioseismic pkb file corresponding to fitted parameters for the $n=20$ GOLF order.}
\centering
\begin{adjustbox}{width=1.\textwidth}
\begin{tabular}{cccccccccccccc}
\hline\hline
 $n$ &  $l$ &  $\nu$ & $\sigma_\mathrm{sym,\nu}$ & $H$ & $\sigma_{\mathrm{sym},H}$ & $\Gamma$ & $\sigma_\mathrm{sym,\Gamma}$  & $i$ & $\sigma_{\mathrm{sym},i}$  & $s$ & $\sigma_{\mathrm{sym},s}$  & $\alpha$ & $\sigma_{\mathrm{sym},\alpha}$  \\ 
  &  & ($\mu$Hz) & ($\mu$Hz) & ($(m/s)^2/\mu$Hz) & ($(m/s)^2/\mu$Hz) & ($\mu$Hz) & ($\mu$Hz) & ($^{\circ}$) & ($^{\circ}$) & ($\mu$Hz) & ($\mu$Hz) &  &  \\
 \hline
20 &  2 &  2889.57 &  0.07 &  2.10e-2 &  5.07e-3 &  0.827 &  0.110 &  90 &  0 &  0.386 &  0.034 & -0.005 &  0.008 \\
 20 &  0 &  2898.97 &  0.07 &  1.27e-2 &  2.73e-3 &  1.107 &  0.128 &  90 &  0 &  0 &  0 & -0.023 &  0.010 \\
\hline
\end{tabular}
\end{adjustbox}
\label{table:example_golf_pkb}
\end {table*}

\subsection{The extended pkb array}

The structure of the extended pkb array is close to the classical pkb array structure, except that it contains 20 columns. Instead of $\sigma_\mathrm{sym}$, two uncertainties values are provided for each parameters: $\sigma_-$ and $\sigma_+$ (see Sect.~\ref{section:sampling_MCMC}).

\end{document}